\documentclass[10pt]{article}

\usepackage{amssymb,amsmath,amsthm} 
\usepackage[margin=1in]{geometry}
\usepackage{graphicx,ctable,booktabs}
\usepackage{verbatim}
\usepackage{enumerate}
\usepackage{float}
\usepackage{overpic}

\usepackage{epstopdf}
\usepackage{url}
\usepackage{xcolor}

\usepackage{booktabs} 
\usepackage{multicol}
\usepackage{multirow}
\newcolumntype{P}[1]{>{\centering\arraybackslash}p{#1}}
\newcolumntype{M}[1]{>{\centering\arraybackslash}m{#1}}

\begin{document}

\title{Controlled Comparison of Simulated Hemodynamics across Tricuspid and Bicuspid Aortic Valves}
\author{Alexander D. Kaiser$^{1,2,3}$, Rohan Shad$^{3,4}$, Nicole Schiavone$^{5}$, \vspace{5pt} \\ 
William Hiesinger$^{3,4}$, Alison L. Marsden$^{1,2,3,5,6}$}
\date{
{\small
$^{1}$Institute for Computational and Mathematical Engineering, Stanford University; \\
$^{2}$Department of Pediatrics (Cardiology), Stanford University; 
$^{3}$Stanford Cardiovascular Institute; \\
$^{4}$Department of Cardiothoracic Surgery, Stanford University; \\ 
$^{5}$Department of Mechanical Engineering, Stanford University; 
$^{6}$Department of Bioengineering, Stanford University 
}
\\
\vspace{10pt}
\today}

\maketitle

\thispagestyle{empty}

\begin{abstract}

Bicuspid aortic valve is the most common congenital heart defect, affecting 1-2\% of the global population. 
Patients with bicuspid valves frequently develop dilation and aneurysms of the ascending aorta. 
Both hemodynamic and genetic factors are believed to contribute to dilation, yet the precise mechanism underlying this progression remains under debate. 
Controlled comparisons of hemodynamics in patients with different forms of bicuspid valve disease are challenging because of confounding factors, and simulations offer the opportunity for direct and systematic comparisons.
Using fluid-structure interaction simulations, we simulate flows through multiple aortic valve models in a patient-specific geometry.
The aortic geometry is based on a healthy patient with no known aortic or valvular disease, which allows us to isolate the hemodynamic consequences of changes to the valve alone. 
Four fully-passive, elastic model valves are studied: a tricuspid valve and bicuspid valves with fusion of the left- and right-, right- and non-, and non- and left-coronary cusps. 
The resulting tricuspid flow is relatively uniform, with little secondary or reverse flow, and little to no pressure gradient across the valve. 
The bicuspid cases show localized jets of forward flow, excess streamwise momentum, elevated secondary and reverse flow, and clinically significant levels of stenosis. 
Localized high flow rates correspond to locations of dilation observed in patients, with the location related to which valve cusps are fused. 
Thus, the simulations support the hypothesis that chronic exposure to high local flow contributes to localized dilation and aneurysm formation.

\end{abstract}


\section{Introduction}

Bicuspid aortic valve, in which the aortic valve has two functional cusps rather than the normal three, is the most common congenital heart defect and is estimated to occur in 1.3\% of the global population \cite{Verma_nejm_bicuspid_review}. 
The presence of a bicuspid valve dramatically changes the hemodynamics in the aorta and may cause stenosis, even in the absence of calcification of the valve leaflets.  
Additionally, approximately half of adult patients with bicuspid aortic valves have pathological dilation of the ascending aorta, which may progress to an ascending aortic aneurysm \cite{Verma_nejm_bicuspid_review}. 
The morphology and presentation of the disease is diverse, with fusion of the left- and right-coronary cusps (LC/RC fusion) in the majority of cases, fusion of the right- and non-coronary cusps (RC/NC fusion) in the minority of cases, and rarely fusion of the non- and left-coronary cusps (NC/LC fusion) \cite{2Schaefer1634}.

Dilation of the ascending aorta associated with bicuspid aortic valve disease is caused by a combination of the chronic accumulation of aberrant hemodynamics and heightened genetic susceptibility, but the precise mechanism of dilation remains poorly understood \cite{Verma_nejm_bicuspid_review}. 
Dilation is frequently located asymmetrically on the greater curvature of the aorta \cite{Girdauskas2011}. 
The phenotype of dilation is not uniform, however, instead varying with the phenotype of cusp fusion. 
Patients with LC/RC fusion typically have dilation throughout the greater curvature of the ascending aorta, commonly with involvement of the aortic root. 
In contrast, patients with RC/NC fusion tend to have more distal disease: along the ascending aorta with little dilation of the aortic root \cite{2Schaefer1634,3losenno2012bicuspid}. 

Attempts to define the  association of hemodynamic features with the phenotype of aortic dilation using 4D-Magnetic Resonance Imaging (MRI) have yielded conflicting results \cite{4DuxSantoy_wss_dilation,5barker2012bicuspid}. 
Multiple factors increase the difficulty in identifying the mechanism of dilation. 
Patients with bicuspid valves are typically recruited into studies after aneurysmal dilation and pathological remodeling have already manifested, and baseline scans for prior to dilation are usually not available. 
Furthermore, since bicuspid valves are present at birth and there are other confounding factors, it is difficult to isolate the effects of bicuspid valve morphology on downstream dilation via such approaches.

Several simulation studies have examined the hemodynamics associated with bicuspid aortic valve morphology.  
Studies compared multiple bicuspid phenotypes but did not include a patient-specific aortic geometry \cite{cao2017simulations} or compared a tricuspid and one bicuspid phenotype \cite{gilmanov2016comparative}. 
Another studied a single patient with a bicuspid valve but did not compare results across valve phenotypes \cite{emendi2021patient}.
Others did not include an anatomical aorta \cite{marom2013fully,Lavon2018} or model leaflet motion \cite{kimura2017patient,youssefi2017patient}.

Experimental studies on adult patients with bisucpid aortic valves based on 4D MRI have shown that hemodynamics, specifically wall shear stress, are associated with aortopathy. 
One study showed that regions of elevated local wall shear stress are associated with dilation growth rate \cite{guala2022wall}. 
Another study in patients with advanced, operable, aortopathy showed that locally elevated wall shear stress was associated with elastin degradation and extracellular matrix dysregulation \cite{guzzardi2015valve}. 
One simulation study showed substantial changes in quantities derived from hemodynamics in a patient with a bicuspid valve and ascending aortic aneursym \cite{de2020deciphering}, and one study combining simulation and experiment suggested links between hemodynamics and aneurysm growth \cite{jayendiran2020computational}. 
These studies, however, were conducted after somatic growth and possible pathological remodeling had occurred over the patients' lifetimes, and thus are subject to the confounding factors discussed above.

In this work, we use computational modeling and fluid-structure interaction (FSI) simulations to compare the hemodynamics of tricuspid and all common variants of bicuspid aortic valves and its relationship to aortic dilation. 
Our simulation setup serves to remove confounding factors discussed above that are necessarily present in studies based on medical imaging of adult patients with bicuspid valves.
The aortic geometry is patient-specific, modeled from the CT scan of an adult patient with a normal tricuspid aortic valve and aortic geometry. 
By using a single geometry, we isolate the hemodynamic effects caused by changes in valve morphology prior to the onset of remodeling.
A fully-passive, elastic, fiber-based model aortic valve is incorporated, which proved highly robust in previous FSI studies \cite{kaiser2020designbased}  and has been validated by direct comparisons to experimental data \cite{kaiser2021validation}. 
Four models were constructed: a healthy, normal tricuspid valve, and three bicuspid valve morphologies: a bicuspid valve with LC/RC fusion, RC/NC fusion and NC/LC fusion. 
All models have the same geometry and material properties, barring cusp fusion, which is modeled by mathematically forcing points along the free edge of the leaflets to coincide. 
A qualitative and quantitative analysis of hemodynamics is presented. 
Connections between hemodynamics and the physiology and etiology of aortic dilation and aneurysm formulation are discussed. 
Code for model generation and FSI simulations is available at \url{github.com/alexkaiser/heart_valves}.
To our knowledge, this is the first systematic comparison of hemodynamics in a patient-specific model generated by a normal tricuspid valve compared to all common bicuspid valve variants, keeping all other conditions fixed.

\section{Methods}
\label{Methods}

We discuss construction of the model valve in Section \ref{model_construction}, image segmentation and creation of the model aorta in Section \ref{aorta}, FSI and simulation setup in Section \ref{fsi}, and integral metrics for flow analysis in Section \ref{Integral_metrics}.

\subsection{Construction of the model aortic valve}
\label{model_construction}

The model aortic valves were constructed using a design-based approach to elasticity, as previously introduced in \cite{kaiser2020designbased}. 
We specified that the tension in the leaflets must support a pressure, then derived an associated system of partial differential equations. 
The solution of this system specified the loaded model geometry and tensions needed to support the specified pressure load. 
Using this information, a reference configuration and constitutive law were derived. 
This configuration automatically included local fiber structure and heterogeneous material properties that would be challenging to measure experimentally. 
By tuning free parameters in these differential equations, this construction allowed us to design a model that is consistent with known anatomy and material properties. 
This method was previously applied to create realistic and effective mitral valve models for FSI simulations \cite{kaiser2019modeling,thesis}.

We created four model aortic valves. 
The first is trileaflet and serves as a control, and is identical to the model derived in  \cite{kaiser2020designbased}. 
Three are bicuspid, representing LC/RC, RC/NC and NC/LC fusion. 
Aside from fusion at the free edge, all four models have identical material properties and leaflet reference geometry.

We represented the valve leaflets as an unknown parametric surface in $\mathbb R^{3}$, 
\begin{align}  
\mathbf X(u,v) : \Omega \subset \mathbb{R}^{2} \to \mathbb{R}^{3} . 
\end{align}
The parametrization was assumed to conform to the fiber and cross fiber directions, meaning that curves of constant $v$ ran circumferentially in the leaflets and curves of constant $u$ ran radially. 
Let subscripts denote partial derivatives. 
The unit tangents in the circumferential and radial directions in the leaflets were defined as 
\begin{align}
\frac{\mathbf X_{u}}{ | \mathbf X_{u} |} \quad \text{ and } \quad \frac{\mathbf X_{v}}{ | \mathbf X_{v} |}. 
\end{align}
The leaflet exerted tensions $S$ and $T$ in the circumferential and radial directions, respectively. 
Shear tensions were assumed to be identically zero. 
Let $p$ denote the pressure supported by the leaflets. 
Consider the equilibrium of pressure and tension on an arbitrary patch of leaflet corresponding to $[u_{1},u_{2}] \times [v_{1},v_{2}]$. 
This dictated that tension integrated over on the boundary of the leaflet and pressure integrated with respect to area sum to zero, or  
\begin{align} 
&0 = \int_{v_{1}}^{v_{2}}   \int_{u_{1}}^{u_{2}}    p  \left(  \mathbf X_{u}(u,v) \times \mathbf X_{v}(u,v) \right)    du dv  \\ 
&\hspace{-3pt}+ \int_{v_{1}}^{v_{2}}  \left(  S(u_{2}, v) \frac{ \mathbf X_{u} (u_{2}, v) }{|  \mathbf X_{u} (u_{2}, v) |} - S(u_{1},v)   \frac{ \mathbf X_{u} (u_{1}, v) }{|  \mathbf X_{u} (u_{1}, v) |}   \right)    dv \nonumber  \\ 
&\hspace{-3pt}+ \int_{u_{1}}^{u_{2}} \left(  T(u, v_{2})  \frac{\mathbf X_{v} (u, v_{2})}{|\mathbf X_{v} (u, v_{2})|} - T(u,v_{1})  \frac{\mathbf X_{v} (u, v_{1}) }{|\mathbf X_{v} (u, v_{1})|} \right)  du . \nonumber
\end{align}
We then applied the fundamental theorem of calculus, differentiating and integrating to convert tensions integrated on the boundary of the patch to derivatives of tension integrated on the entire patch. 
The order of integration was swapped formally and the integrals were combined to obtain 
\begin{align} 
0 = &\int_{v_{1}}^{v_{2}}  \int_{u_{1}}^{u_{2}} \bigg(   p  (  \mathbf X_{u} \times \mathbf X_{v} )    \\ 
&+  \frac{\partial}{\partial u}  \left( S \frac{ \mathbf X_{u} }{ |\mathbf X_{u}| } \right) +  \frac{\partial}{\partial v}  \left( T \frac{ \mathbf X_{v} }{|\mathbf X_{v}|} \right)    \bigg)  \;  du dv . \nonumber
\end{align}
Since the patch of leaflet is arbitrary, the integrals can be dropped. 
This process gave the following system of partial differential equations for equilibrium of the leaflets: 
\begin{align} 
0 = p  (  \mathbf X_{u} \times \mathbf X_{v} )  +   \frac{\partial}{\partial u}  \left( S \frac{ \mathbf X_{u} }{ |\mathbf X_{u}| } \right)  +  \frac{\partial}{\partial v}  \left( T \frac{ \mathbf X_{v} }{|\mathbf X_{v}|} \right).    
\label{eq_eqns}
\end{align}

The equations \eqref{eq_eqns} have three components and five unknowns, and so are not closed. 
To close them we temporarily specified that 
\begin{align}
S(u,v) &= \alpha \left( 1 - \frac{1}{1 + |\mathbf X_{u}|^{2} / a^{2} } \right), \label{dec_tension} \\ 
T(u,v) &= \beta \left( 1 - \frac{1}{1 + |\mathbf X_{v}|^{2} / b^{2} } \right). \nonumber
\end{align}
The values $\alpha,\beta,a,b$ are tunable free parameters that are not required to be constants. 
The parameters $\alpha, \beta$ specify the maximum tension that each fiber can achieve. 
The parameters $a,b$ can be tuned to influence spacing of fibers and achieve realistic gross morphology in the closed configuration of the valve. 
Equations \eqref{dec_tension} are not meant to represent a physical constitutive law, rather they provided a way to arrive at a closed configuration for the valve and the heterogeneous tensions it supported in its loaded state without specifying either a reference configuration or physical constitutive law. 
 Based on preliminary experiments in prior work \cite{kaiser2019modeling}, we expect the leaflet geometry to be fairly insensitive to the specific functional form of equations \eqref{dec_tension}, and that other smooth functions varying from zero to the maximum allowed tension would work similarly.
The emergent leaflet geometry, however, depends strongly on the selected coefficients.

Incorporating the tensions \eqref{dec_tension}, the equilibrium equations \eqref{eq_eqns} were discretized with centered finite differences. 
Let $\mathbf X^{j,k}$ denote an arbitrary node in the discretized leaflets. 
The nonlinear system of equations that corresponded to this node was given 
\begin{align}
&0 =  p  \left(  \frac{(\mathbf X^{j+1,k} - \mathbf X^{j-1,k} )}{2\Delta u}  \times \frac{ ( \mathbf X^{j,k+1} - \mathbf X^{j,k-1} ) }{2\Delta v} \right)  \label{equilbrium_eqn_discrete}  \\ 
            &+  
             \frac{\alpha}{\Delta u} 
	     \left(  1 - 
	     \dfrac{1}{1 + \dfrac{ | \mathbf X^{j+1,k}  -  \mathbf X^{j,k} |^{2}}{  a^{2} (\Delta u)^{2} } }  \right)            
             \frac{ \mathbf X^{j+1,k}  -  \mathbf X^{j,k}  }{ | \mathbf X^{j+1,k}  -  \mathbf X^{j,k} |  }  
              -    
             \frac{\alpha}{\Delta u} 
	     \left( 1 - 
	     \dfrac{1}{ 1 + \dfrac{ | \mathbf X^{j,k}  -  \mathbf X^{j-1,k} |^{2}}{  a^{2} (\Delta u)^{2}  } } \right) 
              \frac{ \mathbf X^{j,k}  -  \mathbf X^{j-1,k}  }{ | \mathbf X^{j,k}  -  \mathbf X^{j-1,k} |  }     \nonumber   \\   
             &  + 
             \frac{\beta}{\Delta v} 
	     \left( 1 -
	     \dfrac{1}{ 1 + \dfrac{ | \mathbf X^{j,k+1}  -  \mathbf X^{j,k} |^{2}}{  b^{2} (\Delta v)^{2}  } } \right) 
             \frac{ \mathbf X^{j,k+1}  -  \mathbf X^{j,k}  }{ | \mathbf X^{j,k+1}  -  \mathbf X^{j,k} |  } 
               -               
              \frac{\beta}{\Delta v} 
	     \left( 1 -
	     \dfrac{1}{ 1 + \dfrac{ | \mathbf X^{j,k}  -  \mathbf X^{j,k-1} |^{2}}{  b^{2} (\Delta v)^{2}  } } \right)
             \frac{ \mathbf X^{j,k}  -  \mathbf X^{j,k-1}  }{ | \mathbf X^{j,k}  -  \mathbf X^{j,k-1} |  }    .  \nonumber 
\end{align}
The annulus and commissure positions were prescribed as Dirichlet boundary conditions. 
The free edges were treated with homogeneous Neumann (zero tension) boundary conditions. 
The nonlinear system of equations was solved with Newton's method with line search.

Next, we used the solution to equations \eqref{eq_eqns}, which represents the closed, loaded configuration of the valve, to specify a reference configuration and constitutive law. 
We prescribed uniform, experimentally measured  circumferential strains of $E_{u}$ = 0.15 and radial strains of $E_{v}$ = 0.54 radially \cite{yap2009dynamic}. 
(In-membrane shear strain was not reported in \cite{yap2009dynamic}. We do not include it here nor expect it to contribute substantially were it to be included.)
The equation $E = L/R - 1$ was solved for the reference length $R$, where $E$ denotes engineering strain and $L$ is the current length. 
(Note that stretch ratio, $L/R$, could be used and is equivalent; use of engineering strains is not meant to denote that deformations are small.)
The tension/strain relation for each link was taken to be exponential through the origin with exponential rates of $57.46$ circumferentially and $22.40$ radially \cite{may2009hyperelastic}.
We solved for the stiffness coefficient $\kappa$ in the equation 
\begin{align} 
t = \kappa (e^{ \lambda E} - 1) 
\end{align} 
to scale stiffness for each link to match tension $t$ at the relevant strain $E$. 

This process produces material properties that are comparable to those measured in in vitro testing.
The tension in the solution to equations \eqref{eq_eqns} is heterogeneous, so each link in the model has its own stiffness coefficient. 
This process thus created heterogeneous material properties. 
We estimated the tangent modulus at the loaded circumferential strains of $E_{u}$ and radial strains of $E_{v}$, and compared to an experimental study on human data that tested a small patch on the central or belly region of the leaflet \cite{pham2017quantification}. 
 The circumferential tangent modulus has minimum, maximum and mean values of $6.6\cdot 10^{7}$, $1.7 \cdot 10^{9}$ and $1.4 \cdot 10^{8}$ dynes/cm$^{2}$, respectively, compared to an experimental estimate of $9.9 \pm 1.8 \cdot 10^{7}$ dynes/cm$^{2}$ \cite{pham2017quantification}. 
The radial tangent modulus has minimum, maximum and mean values of $2.4\cdot 10^{5}$, $3.2 \cdot 10^{7}$ and $5.7 \cdot 10^{6}$ dynes/cm$^{2}$, respectively, compared to an experimental estimate of $2.3 \pm 0.4 \cdot 10^{7}$ dynes/cm$^{2}$\cite{pham2017quantification}. 
At the belly of the leaflet, we estimated the pointwise tangent modulus to be $9.4 \cdot 10^{7}$ dynes/cm$^{2}$ circumferentially and $8.5 \cdot 10^{6}$ dynes/cm$^{2}$ radially. 
Thus, at the belly of the leaflet, our model has circumferential tangent modulus that agrees with this experiment to within standard error. 
Further, along the midline of the leaflet, the radial tangent modulus varies from a maximum of $3.0 \cdot 10^{7}$ dynes/cm$^{2}$ at the attachment of the leaflet and annulus to a minimum of $4.0\cdot 10^{5}$  dynes/cm$^{2}$ at the free edge. 
The radial stiffness of our model on the centerline varies from just above their experimentally reported range at the annulus, through the range for a portion of leaflet, then is lower at the free edge, where the leaflet is expected to be more compliant radially. 
Thus, the tangent moduli in our model agree with measured data in certain specific, relevant locations. 
However, the tangent moduli are heterogenous and thus not equal everywhere. 
Studies have found substantial variety in the material properties of leaflet tissue, especially near the commissures \cite{billiar2000biaxial}. 
We observe local increases in the circumferential tangent modulus near the commissures (see \cite{kaiser2019modeling}), which explains slightly elevated mean values compared to those of \cite{pham2017quantification}. 

The resulting fiber orientations, which run circumferentially, are comparable to experimental data, especially in the belly region of the leaflet, as discussed further in \cite{kaiser2019modeling}.  
The model exerts tension only in the circumferential and radial directions, and we do not include in-plane coupling terms. 
We are aware of little data that addresses the material response to in-plane shear, but one study showed that the changes to the closing kinematics of the leaflet were minimal with or without such terms on a similar material model \cite{hammer2011mass}. 
In future work, these radial and circumferential tensions and strains could be fit to a variety of constitutive laws including those with in-plane coupling or other general hyperelastic laws.
One minor limitation of these models is that they lack bending rigidity (beyond what emerges from the thickening process described below), and hence may not capture fluttering behavior in the leaflets accurately. 
We do not alter material properties in the bicuspid leaflets from those of the tricuspid case, as to avoid confounding factors associated with pathological remodeling of bicuspid valves and further isolate the effects of changes to leaflet geometry alone. 
Thus, the material properties prescribed and derived for our model agree with experimental data in terms of tangent modulus, shape of the tension/strain curves and fiber orientation.

\begin{figure}[th!]
\centering
\setlength{\tabcolsep}{2pt}
\begin{tabular}[t]{c c}
\includegraphics[width=.3\textwidth]{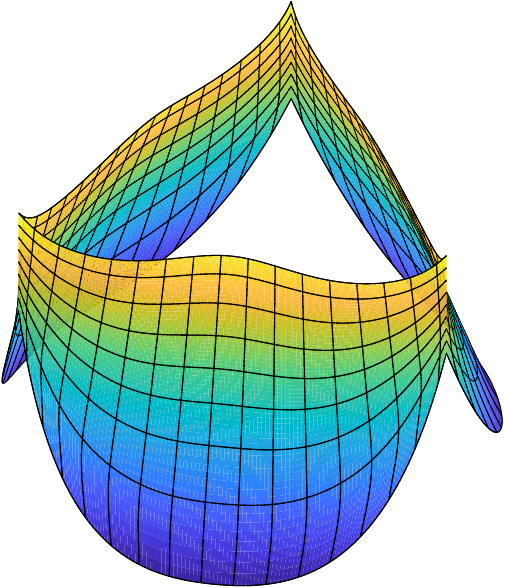} & 
\includegraphics[width=.3\textwidth]{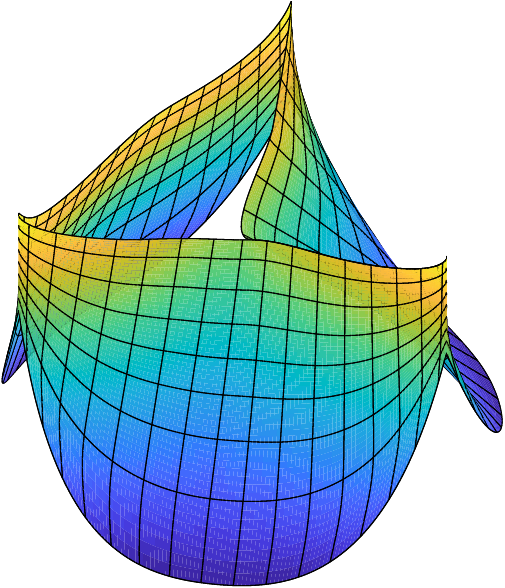}  
\end{tabular}
\caption{ The tricuspid (left) and bicuspid (right) aortic valve. This configuration is used to create initial conditions for FSI simulations.} 
\label{aortic_initial}
\end{figure}

To obtain initial conditions for FSI, we solved for open configurations of the tri- and bicuspid valve as follows. 
Using this newly generated constitutive law for tensions, we again solved the generic equations of equilibrium \eqref{eq_eqns} with zero pressure. 
The positions of the free edges were prescribed as a Dirichlet boundary condition to ensure an open configuration. 
Evenly spaced points were computed on a chord between the two commissure points on either side of this leaflet. 
Each free edge point was placed on a line between the annulus and the corresponding point on the chord. 
The distance from the annulus was set to be the total reference length in the radial direction associated with this radially-oriented curve  multiplied by $E_{v}$, the radial strain in the predicted loaded configuration. 
This distance reduced overall residual tension in the leaflets compared to putting the points directly on the chord or at the rest length from the annulus. 
To construct a bicuspid valve, two leaflets were prescribed to coincide exactly on half of their free edges.
For the fused half of the leaflets, the point was determined to lie at the reference length circumferentially from the commissure, and at the reference length multiplied by $E_{v}$ from each corresponding point on the annulus. 
The free half of the fused leaflets was set with a spline interpolant from the center point to the relevant commissure. 
The third leaflet, which has no fusion, is set identically to the tricuspid case. 
This process created a raphe in the center of the fused cusps. 
The reference configuration and constitutive law of the bicuspid valve are identical to those of the tricuspid valve; the models differ only in their initial configuration and coinciding points on the free edge. 
For all models, this process produced an initial configuration that is not the reference configuration; thus the initial condition included pre-strain and residual tension. 
Further, a configuration that is tension-free may not exist. 
Experiments have shown that the aortic valve has pre-strain and residual tension throughout the cardiac cycle, including when the valve is open, both in vitro \cite{yap2009dynamic} and in vivo \cite{aggarwal2016vivo}. 
The resultant configurations are shown in Figure \ref{aortic_initial}. 
Until this point we worked with a zero-thickness membrane and membrane tensions; next we thickened the model to a physiological thickness of  $0.044$ cm \cite{sahasakul1988age}. 
Two additional layers were placed normal to the membrane, and membrane stiffnesses were divided by three to obtain the relevant stiffness for each layer. 
Linear springs attached the layers together in the normal directions; these prevent the leaflets from separating from each other. 
Their stiffnesses were tuned empirically such that the layers undergo minimal movement from each other without causing further time step restrictions. 
This also served to mitigate the ``grid-aligned artifact'' that occurs with pressure discontinuities across thin, tensioned membranes in the IB method \cite{kaiser2019modeling,thesis}.

\subsection{Construction of the model aorta}
\label{aorta}

Image segmentations to construct a patient-specific model aortic geometry were constructed from CT scan data using SimVascular \cite{lan2018re}. 
The patient was a 50 year old male with no known complications in the aorta or aortic valve. 
A pathline for the approximate centerline of the vessel was created. 
Two-dimensional contours were segmented manually on slice views normal to the pathline. 
These contours were then lofted into a surface, and a triangular mesh of this surface was exported. 
The aorta model was held in an approximately constant position with target points and linear springs on each edge of the triangular mesh (Section \ref{fsi}). 
Keeping the aorta approximately rigid serves to further isolate the effects of changes to valve morphology alone, because in the bicuspid cases it is unclear whether to use healthy material properties to remove effects of material property changes due to disease, or material properties from adult bicuspid patients, which potentially be influenced by pathological remodeling. 
Because the fluid domain was a Cartesian box, flow extenders of length 1 cm were added to the inlets and outlets. 
This ensured that the normal to the vessel was aligned with the normal to the fluid domain at the inlet and outlet. 
An interpolant connected the aorta from immediately below the aortic annulus to the flow extender at the edge of the fluid domain. 
As with the valve, two additional layers were added by extruding the model in the normal direction at one and two times the structure mesh width, or $0.025$ and $0.05$ cm.

We positioned the valve manually in the aorta model as follows. 
We imported the aorta geometry into Paraview \cite{ahrens2005paraview}, then applied the slice operator and interactively manipulated its position to determine the origin and normal of the minimum plane of the annulus. 
A rigid body rotation and translation were applied to align the axis and origin of the valve with the axis and normal of the minimum annulus plane. 
We applied an additional rotation, computed by trial and error, around the axis of the valve to ensure that the commissures align with the aortic sinus.
The model has exactly a 120 degree angle between each of the commissures, and this aligned well with the patient's aortic sinus after segmentation. 
Expert clinicians confirmed the anatomical accuracy of this placement by visual inspection of the models.

\subsection{Fluid-structure interaction}
\label{fsi}

Simulations were run using the immersed boundary (IB) method for FSI \cite{ib_acta_numerica}. 
In the IB method, fluid quantities are represented with respect to the lab or Eulerian frame. 
Structure quantities, such as the aorta and valve, are represented with respect to a material or Lagrangian frame. 
The fluid occupies the entire domain of interest. 
The structure occupies some subset of this domain and is treated as neutrally buoyant, meaning that it has no additional mass beyond the mass of the local fluid. 
The two frames are coupled via convolutions with the Dirac-$\delta$ function, as described below. 
The structure influences the fluid through a body force computed via these convolutions, which is one distinctive feature of the IB method.

The fluid velocity and pressure are denoted $\mathbf u, p$ respectively.
The Eulerian frame force $\mathbf f$ represents force exerted by the structure onto the fluid. 
These functions take arguments of fixed spatial position $\mathbf x$ and time $t$. 
The constants $\rho$ and $\mu$ denote fluid density and dynamic viscosity. 
Let $\mathbf X(\mathbf s,t)$ denote the current configuration of the structure, where $\mathbf s$ labels a material point. 
The Lagrangian frame force $\mathbf F$ denotes the force exerted by the structure in the Lagrangian frame. 
(We previously used $u,v$ to label material points, but switch to $\mathbf s$ to avoid confusion with fluid velocity here.) 
The Dirac delta function is denoted as $\delta$. 
 
The governing equations of the IB method are 
{\allowdisplaybreaks
\begin{align}
\rho \bigg(  \frac{ \partial \mathbf u (\mathbf x, t)}{\partial t}  +  \mathbf u (\mathbf x, t) \cdot \nabla \mathbf u (\mathbf x, t) \bigg) &= \label{momentum} 
- \nabla p (\mathbf x, t) + \mu \Delta \mathbf u (\mathbf x, t) + \mathbf f (\mathbf x, t)   \\ 
\nabla \cdot \mathbf u(\mathbf x, t)  &= 0   		\label{mass}		\\
\mathbf F( \, \cdot \, , t) &=   \mathcal F (\mathbf X( \, \cdot \, ,t))  \label{nonlinear_force}  \\   
\frac{ \partial \mathbf X(\mathbf s,t)}{\partial t}&=  \mathbf u(\mathbf X(\mathbf s,t), t) 	 \label{interpolate} 	 \\
		&= \int \mathbf u(\mathbf x, t)   \delta (  \mathbf x  - \mathbf X(\mathbf s,t) )  \;  d  \mathbf x   \nonumber  \\ 
\mathbf f(\mathbf x, t)   &= \int  \mathbf F(\mathbf s,t)  \delta(  \mathbf x - \mathbf X(\mathbf s,t)  )   \; d\mathbf s   .       \label{spreading} 
\end{align}  
}

Equations \eqref{momentum} and \eqref{mass} are the Navier Stokes equations for conservation of momentum and volume of an incompressible Newtonian fluid, plus the Eulerian frame force $\mathbf f$. 
Equation \eqref{nonlinear_force} is a mapping from the configuration of the structure to the force it exerts on the fluid. 
The omitted argument indicates that the mapping $\mathcal F$ takes the entire configuration of the structure at time $t$ and produces the  Lagrangian frame force $\mathbf F$. 
Equations \eqref{interpolate} and \eqref{spreading} are interaction equations that couple the fluid and structure. 
The interpolation equation \eqref{interpolate} specifies that the structure moves at the local fluid velocity. 
The force spreading equation \eqref{spreading} determines the Eulerian frame force $\mathbf f$ from the Lagrangian frame force $\mathbf F$.

Simulations were conducted with the software library IBAMR (Immersed Boundary Adaptive Mesh Refinement) using a staggered Cartesian grid for the fluid \cite{griffith2010parallel,IBAMR}. 
Simulations were run for three cardiac cycles at 75 beats per minute, or 0.8 s per beat. 
The fluid density and viscosity were set to $\rho = 1.0$ g/cm$^{3}$ and 0.04 Poise, respectively.
The velocity was initialized to zero and the simulation began at early diastole. 
The second and third cardiac cycles appeared qualitatively similar in all simulations and sufficient to wash out initialization effects; no further cycles were run due to to computational expense and similarity of further cycles in previous validation tests. 
Simulations were run on Stanford University's Sherlock cluster on 48 Intel Xeon Gold 5118 cores across two nodes with a 2.30 GHz clock speed. 

The fluid mesh width was set to $\Delta x = $ 0.05 cm. 
The fluid domain was taken to be   $7.2\times4.8\times11.2$ cm, corresponding to $144\times96\times224$ points.  
The structure mesh width was targeted to half that of the fluid mesh width, or $\Delta s = \Delta x/2 = 0.025$ cm.
The lengths of individual links were determined as described in Section \ref{model_construction} and evolved according to the FSI dynamics. 
A scaffold was placed around the valve to ensure that there are no holes between the model valve and aorta. 
The aorta, scaffold and the annular edge of the leaflets were held in place by \emph{target points}, a penalty method that approximately enforces a fixed position. 
For a point $\mathbf X$ and its desired location $\mathbf X_{target}$, this is a force of the form $\mathbf F = -k (\mathbf X - \mathbf X_{target})$, representing a linear spring of zero rest length connecting the current and desired positions with $k = 5.8 \cdot 10^{4}$ dynes/cm. 
A 5-point discrete $\delta$ function was used \cite{IB5_arxiv}. 
At the outlet, open-boundary stabilization was applied with penalty coefficient $\eta = \rho / (4 \Delta t) $ dynes s/cm$^{4}$ \cite{CNM:CNM2918}. 
 Open boundary stabilization was necessary to prevent spurious, non-physical flows from emerging from the intersection of the aortic domain with the boundary of the fluid domain.
At the inlet, a flow-averaging force $ \mathbf f = -\eta (u - \bar u, v, w)$ was applied, where the $x$ direction is normal to the inlet and $\bar u$ is the mean inflow and $\eta = \rho / (4 \Delta t) $ dynes s/cm$^{4}$. 
This approximately enforced a uniform or plug profile to the inflow. 
Empirically, this was necessary to avoid numerical instabilities  in the form of rapidly increasing spurious flows at the boundary of the aortic inlet. 
This was the only stabilization applied at the inlet, but in the even of a local flow reversal, this stabilization applies a similar force to that of open boundary stabilization.  
Outside of the aorta, a force $ \mathbf f = -\eta \mathbf u$ was applied to damp flow to zero  with $\eta = \rho / (4 \Delta t) $ dynes /cm$^{4}$. 
In the first 0.1 s, this force was applied on the entire domain to damp out initial transients associated with residual stress in the initial conditions.

Pressure boundary conditions created a pressure gradient across the valve and aorta to drive the simulations. 
(Here the term ``pressure gradient'' denotes the difference in two local values of pressure in mmHg, as is standard in clinical literature.)
To prescribe the pressure boundary conditions, we prescribed zero tangential slip and applied a normal traction with value equal to the negative of the desired pressure. 
Applying the incompressibility constraint, zero tangential slip implies that this normal traction is then equal to the pressure locally.
We targeted aortic systolic, diastolic and mean pressures of 120, 80 and 96 mmHg respectively and a flow rate of 5.6 L/min. 
The aortic pressure was initialized to be equal to the initial ventricular pressure of 23 mmHg at $t = 0$ s, meaning that the initial pressure gradient was zero. 
The aortic pressure then ramped linearly to 94 mmHg at $t = 0.1$. 
For the remainder of the simulation, the pressure at the aorta side outlet was determined by an RCR lumped parameter network \cite{kim2009coupling}. 
The total resistance was computed as the ratio of the targeted mean pressure over the targeted mean flow. 
The ratio of the distal to proximal resistors was set to $0.064$ \cite{laskey1990estimation}. 
Capacitance was set using an exponential exact solution to the governing differential equations in the case of assumed zero flow in diastole. 
This procedure gave values of proximal resistance of $R_{p} = 83.67$ s dynes cm$^{-5}$, distal resistance of $R_{d} = 1287.65$ s dynes cm$^{-5}$ and capacitance of $C = 0.0017  \text{ dynes}^{-1} \text{ cm}^{5} $. 
The pressure on the left ventricular side was prescribed based on experimental measurements \cite{yellin_book}. 
During peak systole, a pressure loss was observed across the location of the flow averaging force, which removes energy from the flow locally. 
Thus, we added 15 mmHg to the systolic pressure upstream of the flow extender to obtain the desired physiological ventricular systolic pressure of 120-130 mmHg immediately proximal to the valve. 
In results, we report ventricular pressures taken from the pointwise fluid pressure proximal to the valve and downstream of the flow extender. 
Outside of the aorta, no slip boundary conditions were prescribed on faces that include the inlet and outlet, keeping these structures stationary through the simulation.  
Zero pressure boundary conditions were prescribed on all other faces.

To validate these methods, we conducted a physical experiment with a 3D-printed vascular geometry and prosthetic aortic valve and measured its flow field via 4D MRI, then, with the same methods used in this work, simulated the physical experiment  \cite{kaiser2021validation}. 
We achieved excellent qualitative agreement on the hemodynamics and valve kinematics between simulation and experiment. 
We computed the $L^{1}$ relative error of streamwise velocity and velocity magnitude on the entire flow domain and relevant two-dimensional slices, and achieved reasonable qualitative agreement of order 30\% during forward flow on all quantities. 
Other studies on flows through heart valves have performed qualitative comparisons or quantitative validation of leaflet kinematics, but we not are aware of previous studies have directly compared the three-dimensional velocity fields and achieved agreement in relative error. 
We also compared varying levels of resolution for the bicuspid aortic simulations shown in the current paper, which produced qualitatively and quantitatively similar results (see Appendix). 
Thus, we consider the hemodynamics that result from our method to be well-validated, and these comparisons ensure that downstream hemodynamics shown in this work are realistic.

\subsection{Integral metrics}
\label{Integral_metrics}

To quantify the nature of the flows, we computed a number of integral metrics on two-dimensional cross sections of the flow. 
These metrics were previously used to study flow in the lungs \cite{banko2016oscillatory} and through the pulmonary valve \cite{schiavone2021vitro}. 
The two-dimensional cross sections are denoted $S_{1}, S_{2} \dots S_{5}$, numbered streamwise, and located at the aortic annulus, the sinotubular junction and three additional cross sections in the ascending aorta. 

The metric $I_{1}$ represents the nondimensional streamwise momentum flux and is given as 
\begin{align}
I_{1} &= \left( \frac{ 1}{U_{T}^{2} A}  \iint_{S_{k}} (\mathbf u \cdot \mathbf n)^{2} \; dA  \right)^{1/2}  \label{I1_def}  
=  \left\| \frac{\mathbf u \cdot \mathbf n} {U_{T} A^{1/2}}  \right\|_{L^{2}(S_{k})} , 
\end{align}
where $A$ is the area of the cross section, $U_{T}$ is a constant velocity scale, and $\mathbf n$ is the unit normal to the cross section. 
This is the $L^{2}$ norm of the normal component of velocity, made non-dimensional by inclusion of the velocity scale and area. 
In the case of a constant or plug velocity profile with velocity equal to $U_{T}$, the velocity scale, $I_{1} = 1$. 

The metric $I_{2}$ represents secondary or tangential flow strength and is given as 
\begin{align}
I_{2} &= \left( \frac{ 1}{U_{T}^{2} A}  \iint_{S_{k}} \| \mathbf u - (\mathbf u \cdot \mathbf n) \mathbf n \|_{2}^{2} \; dA  \right)^{1/2} \label{I2_def} 
= \left\| \frac{\mathbf u - (\mathbf u \cdot \mathbf n) \mathbf n}{U_{T} A^{1/2}}  \right\|_{L^{2}(S_{k})} .  
\end{align}
Similarly, this is the $L^{2}$ norm of the tangential component of velocity on the cross section, nondimensionalized by the inclusion of the velocity scale and area. 

For both $I_{1}$ and $I_{2}$, the velocity scale was set to the spatial average of the maximum instantaneous flow rate, 
\begin{align}
U_{T} = \max_{t \in [0, 2.4]}  \left( \frac{1}{A} \iint_{S_{k}} \left( \mathbf u(\mathbf x, t) \cdot \mathbf n \right) \; dA \right), 
\end{align}
where we explicitly notate the space and time variables for clarity.
This value was computed for each case individually. 

Finally, we compute $I_{R}$, the fraction of reverse flow on the slice. 
The indicator or characteristic function of the set on which flow moves in the reverse direction is defined as 
\begin{align}
\text I_{ \{ \mathbf u \cdot \mathbf n < 0 \} }(\mathbf x) = 
\begin{cases}
1 & : \mathbf u \cdot \mathbf n < 0 \\ 
0 & : \mathbf u \cdot \mathbf n \geq 0 , \\ 
\end{cases}
\end{align}
and the metric is defined as 
\begin{align}
I_{R} &=  \frac{1}{A} \iint_{S_{k}} \text I_{ \{ \mathbf u \cdot \mathbf n < 0 \} }(\mathbf x) \; dA. 
\label{IR_def} 
\end{align}
This value represents the portion of cross sectional area in which flow moves towards the left ventricle, rather than up the aorta.

To evaluate the flow fields and integrals on the selected cross sections, the flow field was first restricted to the interior to the aorta model using PyVista \cite{sullivan2019pyvista}. 
Using Paraview filters \cite{ahrens2005paraview}, data were projected to the normal and tangential components using Gram Scimidt orthogonalization, two dimensional slices were extracted from these data, then integrated.

\section{Results}
\label{Results}

\subsection{Hemodynamics}

\begin{figure*}[p!]  

       {\Large A:}
       \hfill     
       \raisebox{0pt}{\includegraphics[width=.08\textwidth]{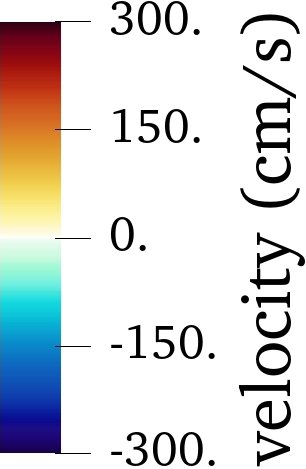}}
       
	\vspace{2pt}
    
        {
        \centering
        \setlength{\tabcolsep}{1.0pt}        
        \begin{tabular}{c | c | c | c | c}        
         & 
         tricuspid &
        LC/RC fusion  & 
        RC/NC fusion & 
        NC/LC fusion \\ 
        \rotatebox[origin=l]{90}{$t = 2.178 $ s} &
        \includegraphics[width=.24\textwidth]{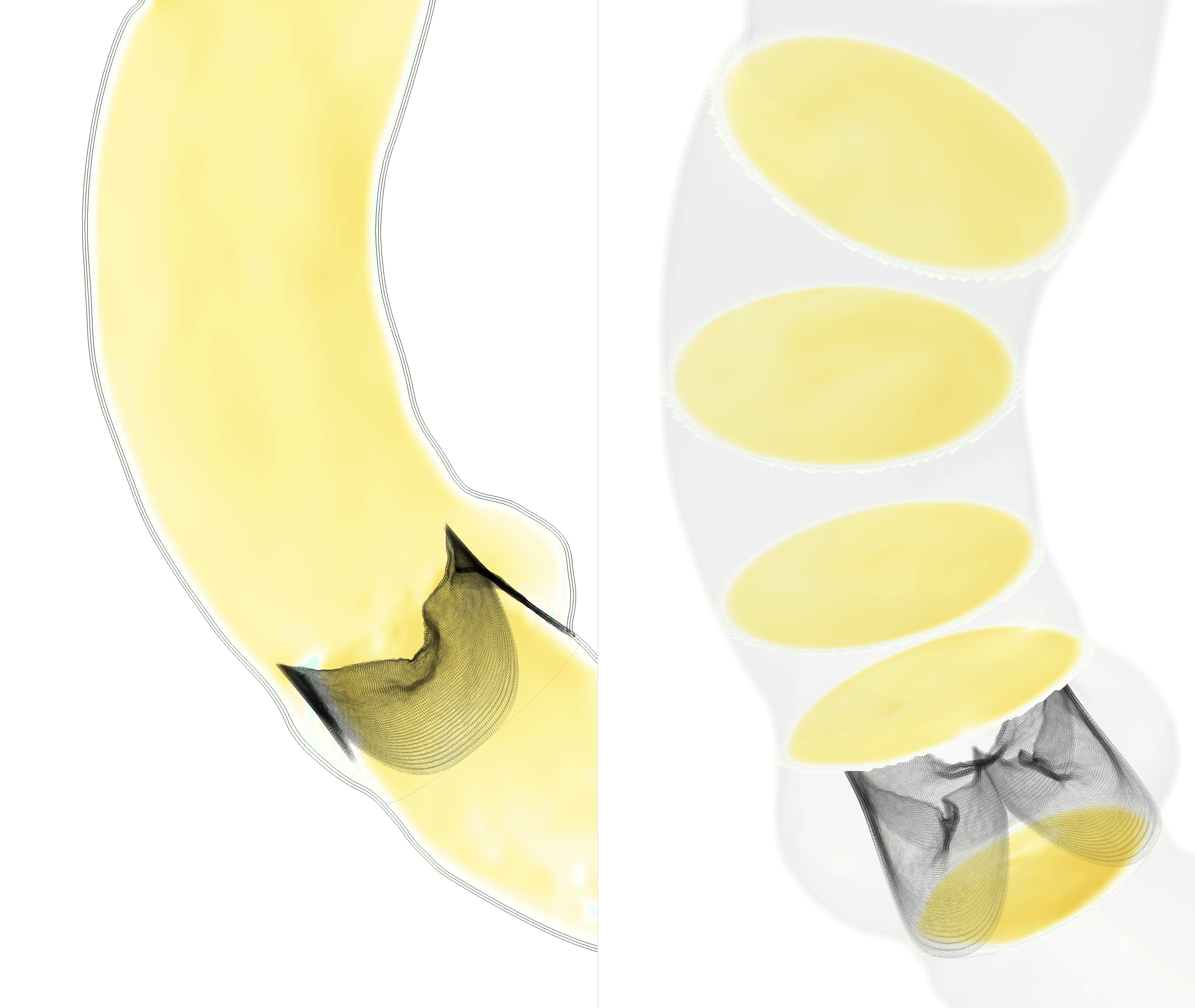}   &
        \includegraphics[width=.24\textwidth]{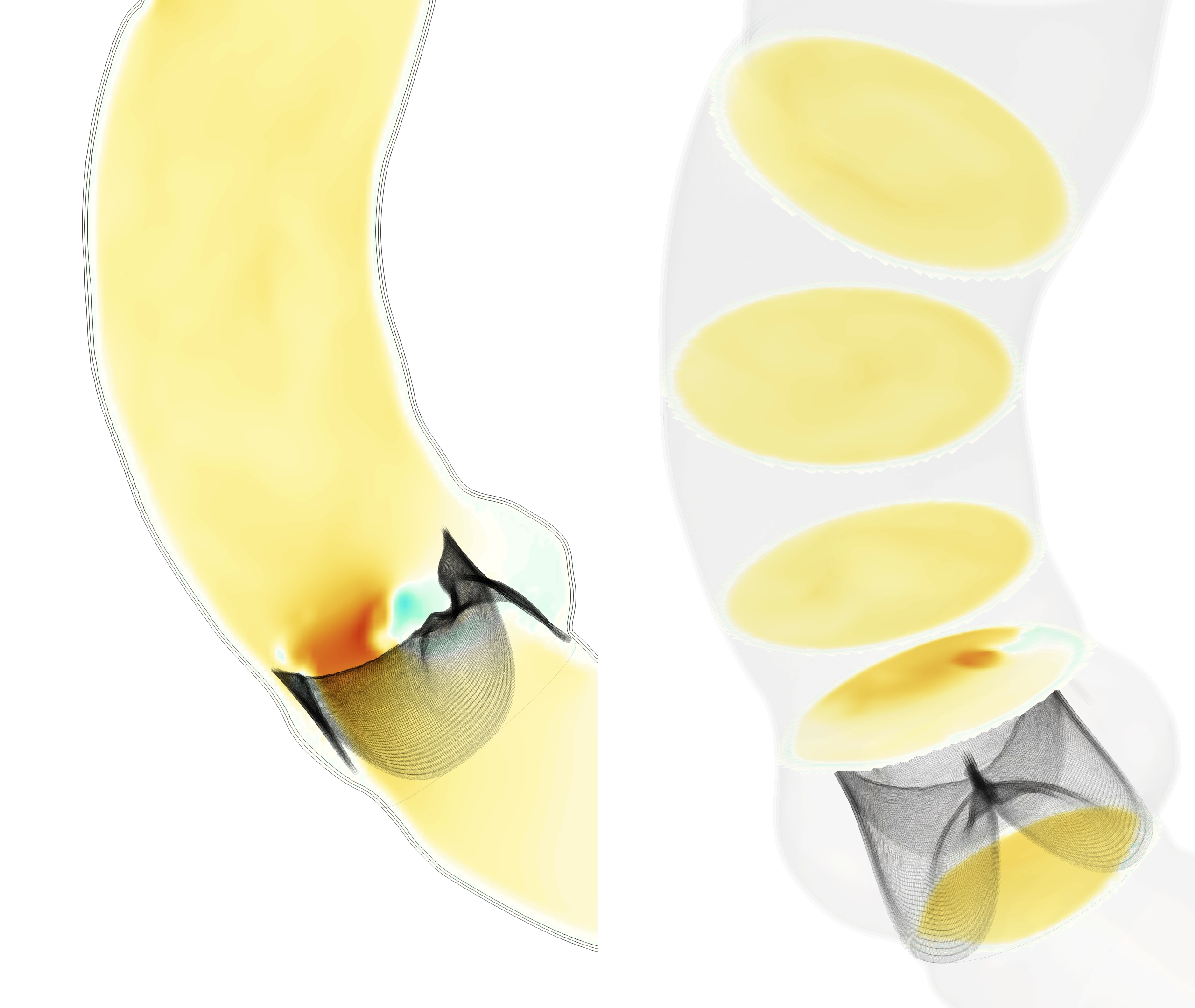}  &
        \includegraphics[width=.24\textwidth]{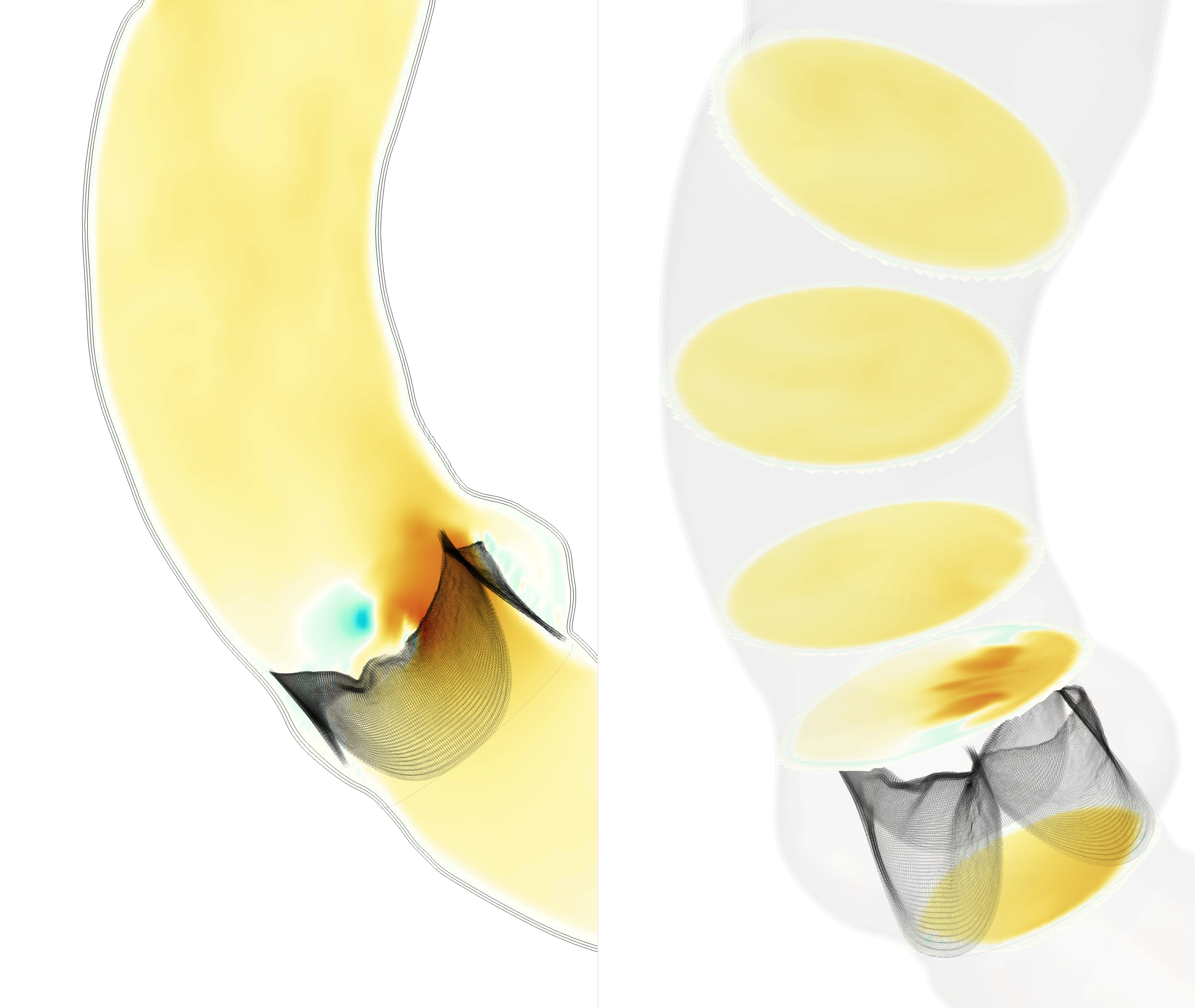}  & 
        \includegraphics[width=.24\textwidth]{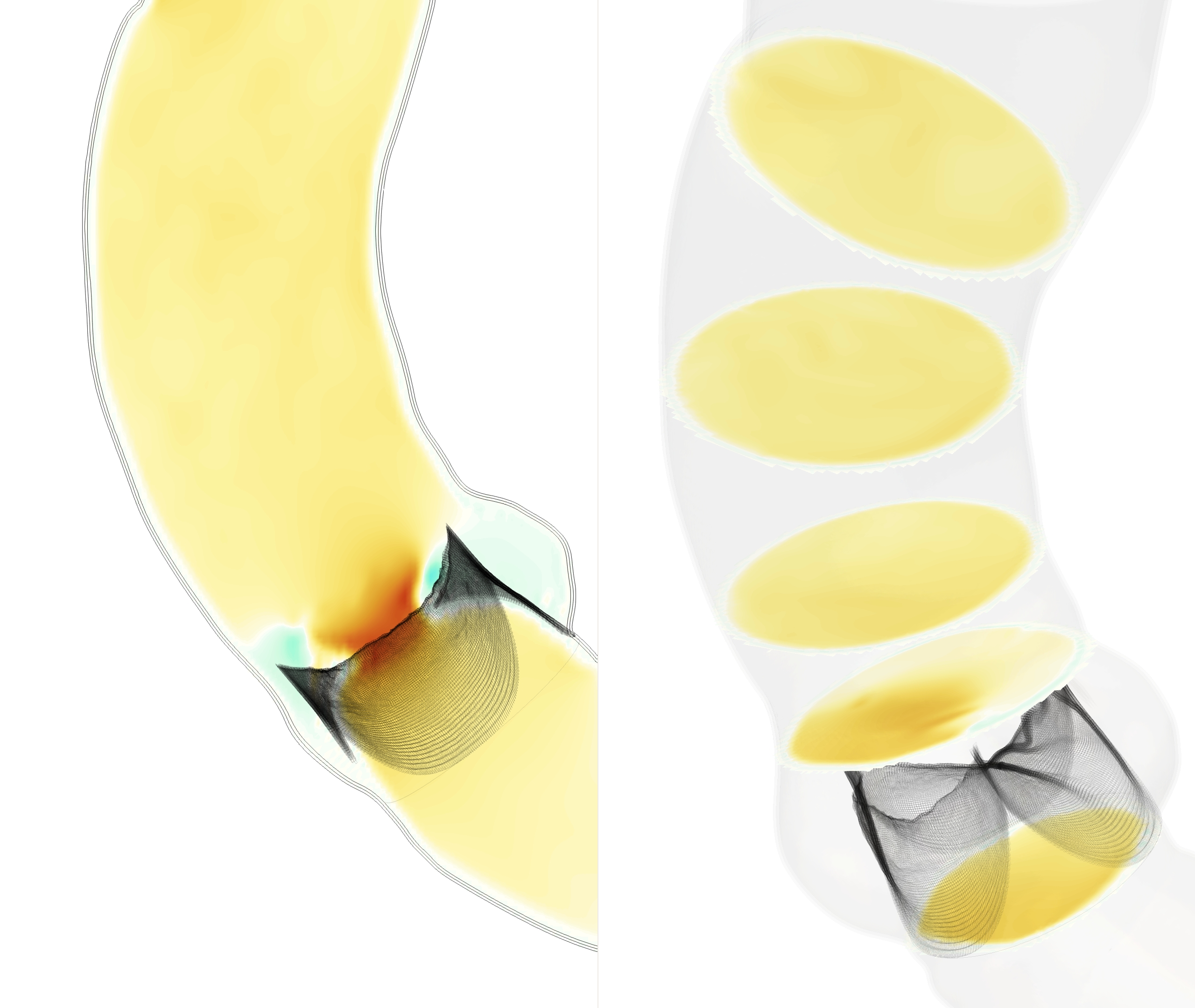} \\ 
        \rotatebox[origin=l]{90}{ $t = 2.231 $ s} &
        \includegraphics[width=.24\textwidth]{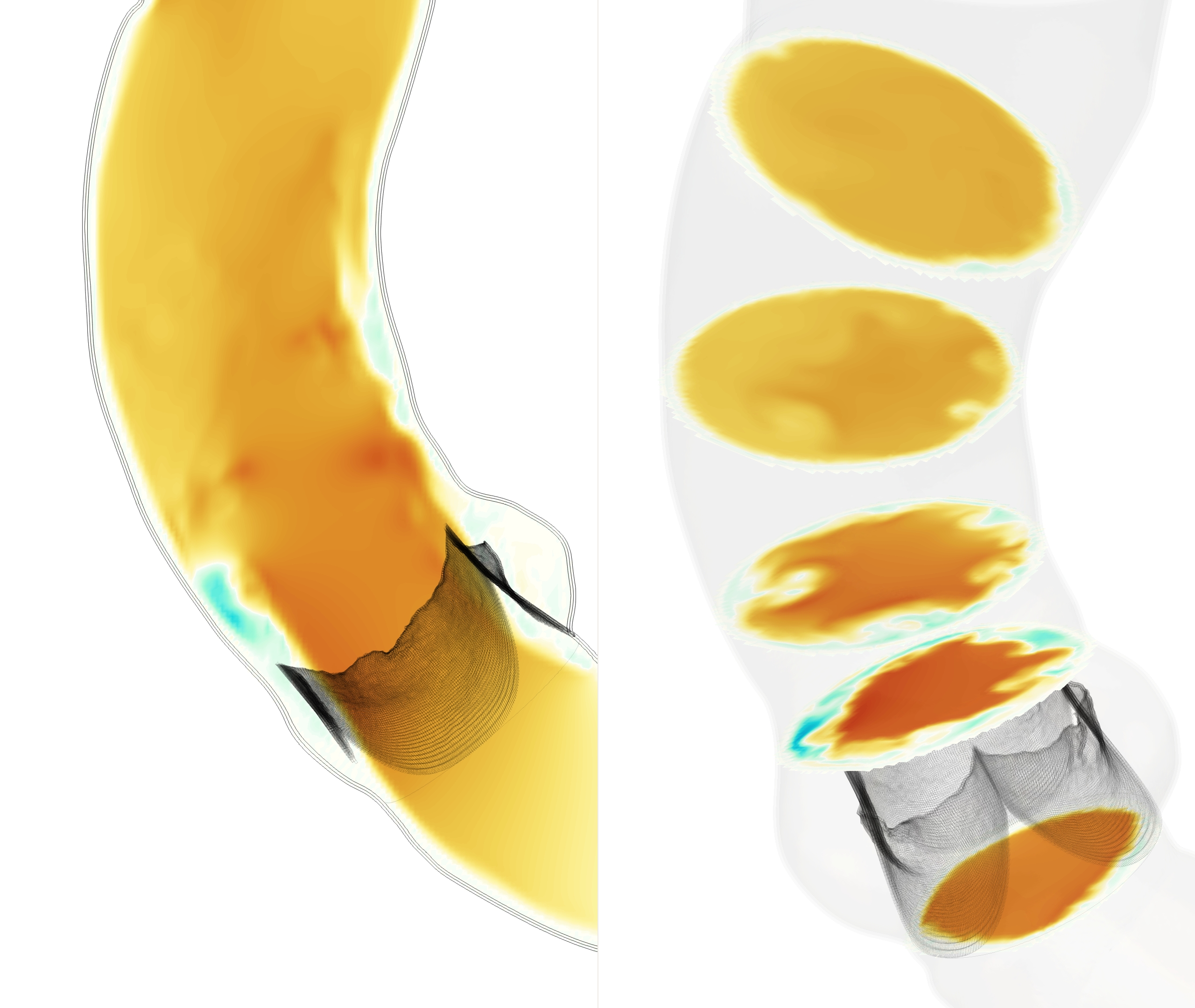}   &
        \includegraphics[width=.24\textwidth]{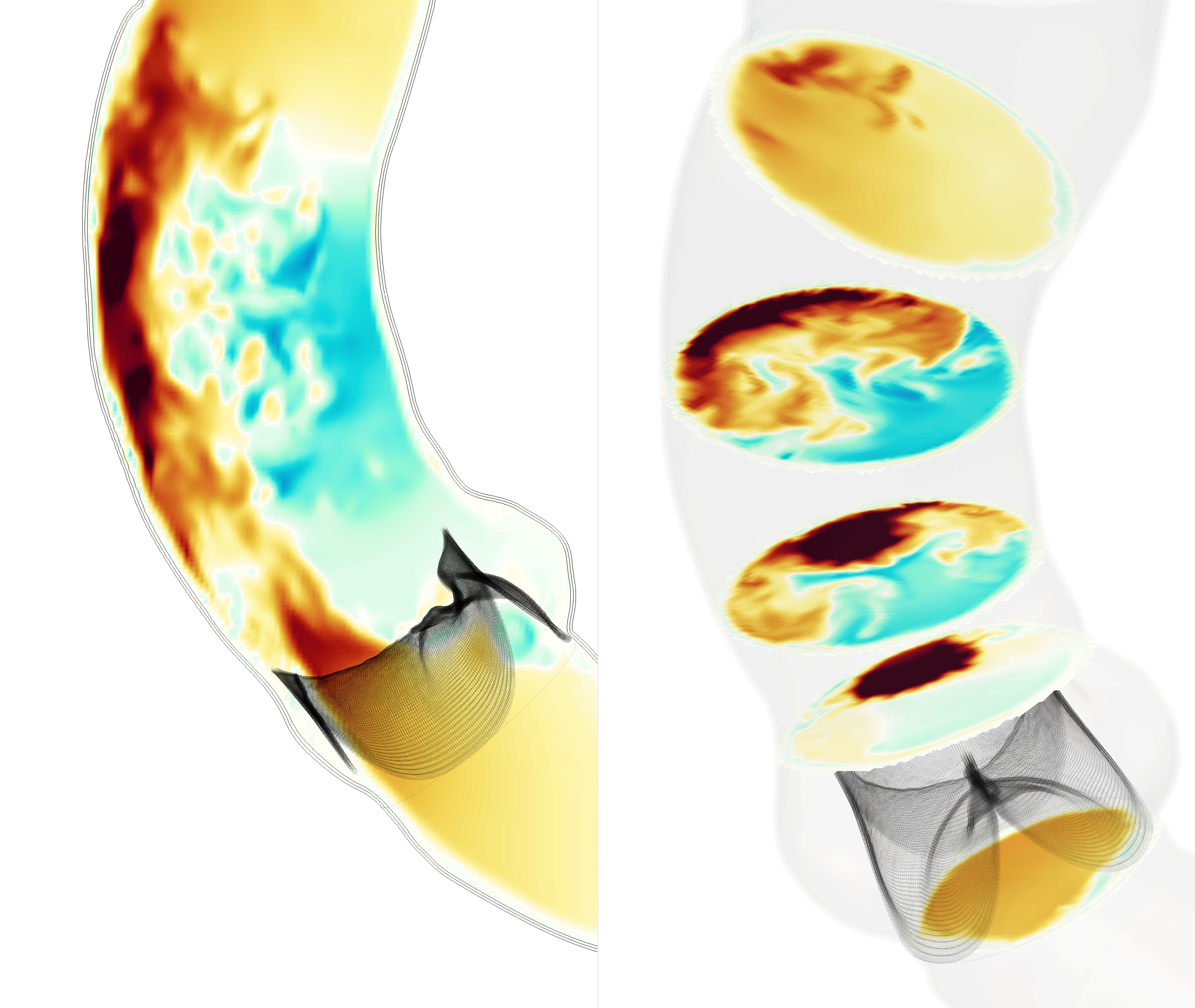}  &
        \includegraphics[width=.24\textwidth]{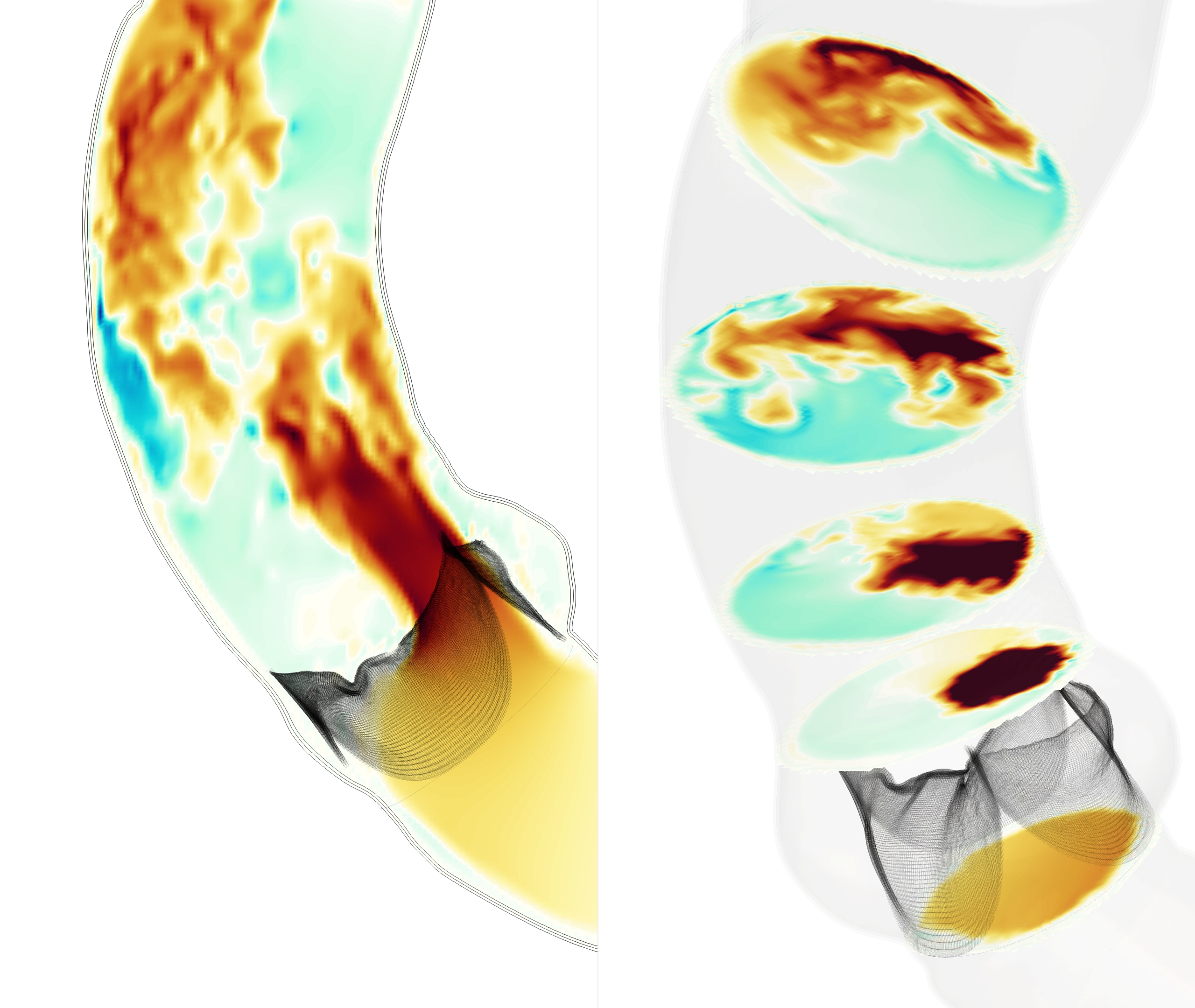}  & 
        \includegraphics[width=.24\textwidth]{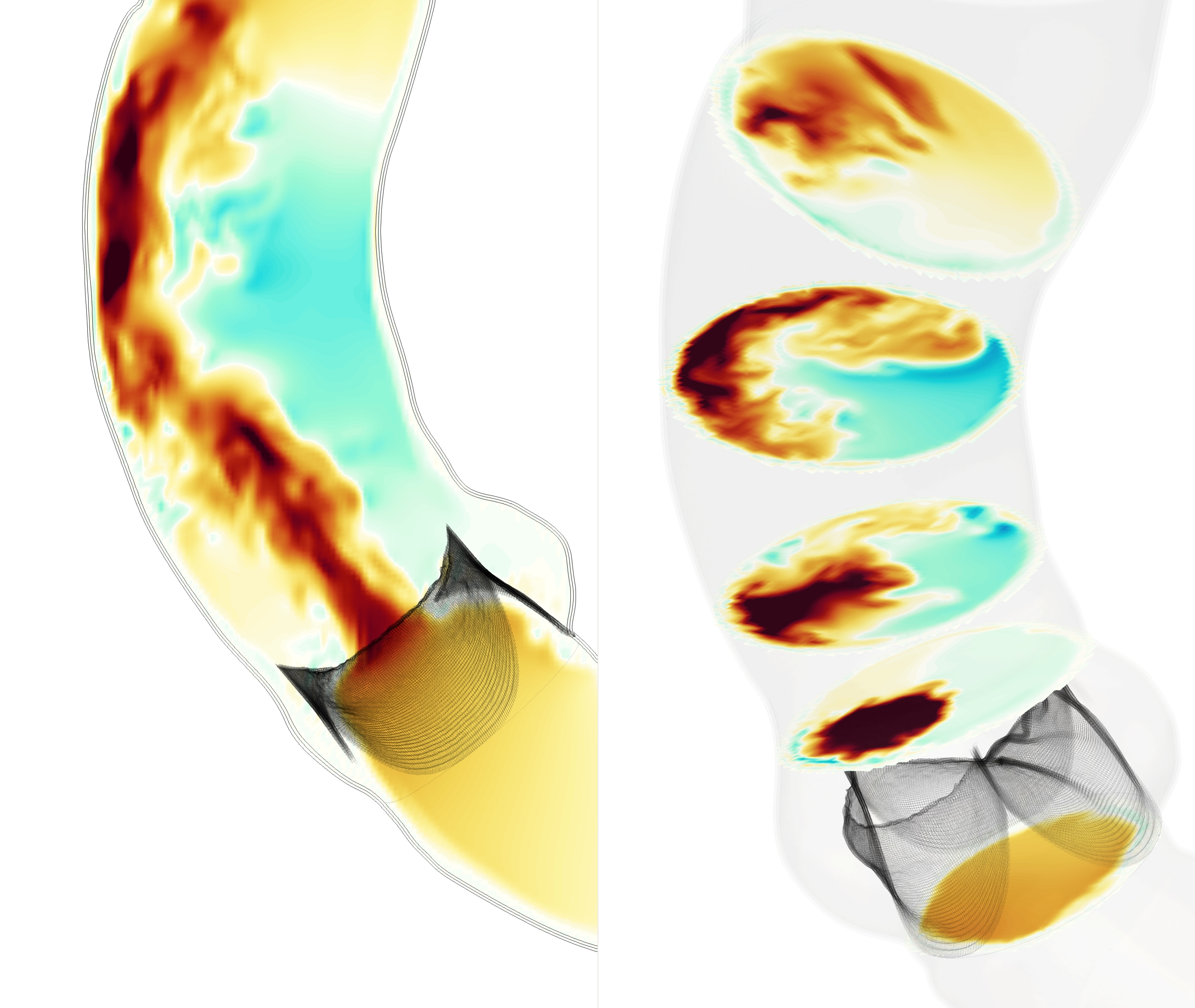} \\ 
        \rotatebox[origin=l]{90}{ $t = 2.271 $ s } &
        \includegraphics[width=.24\textwidth]{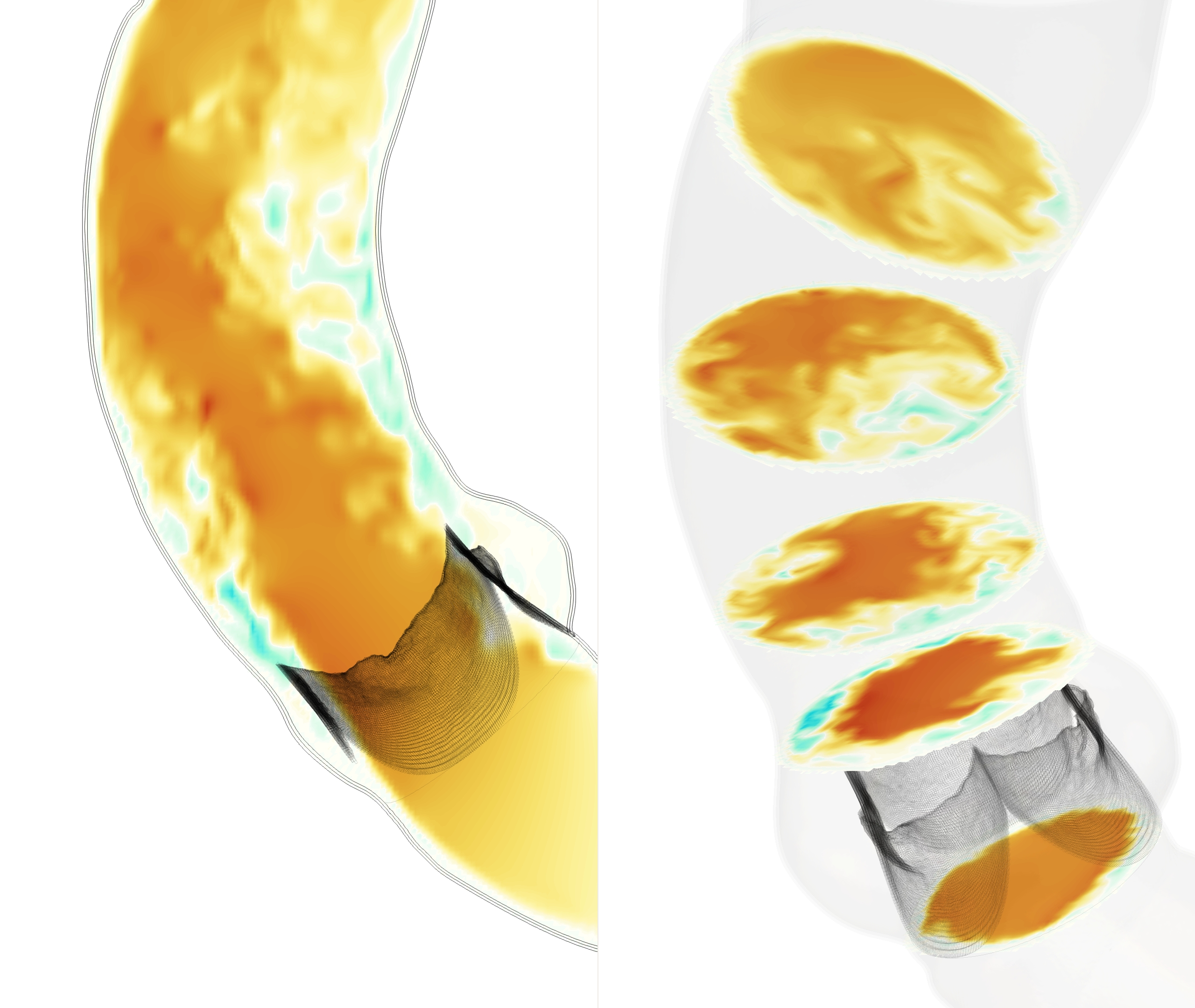}   &
        \includegraphics[width=.24\textwidth]{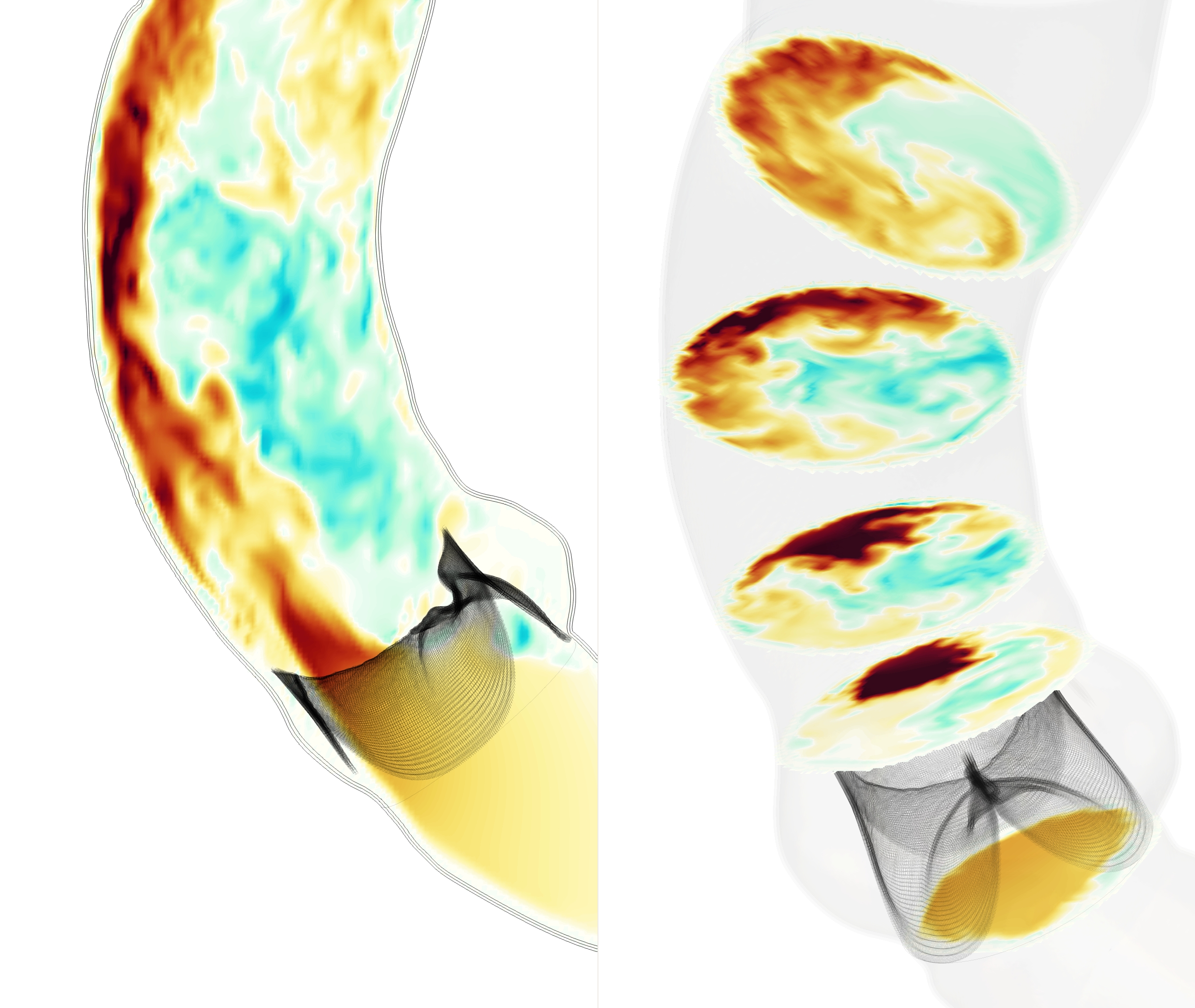}  &
        \includegraphics[width=.24\textwidth]{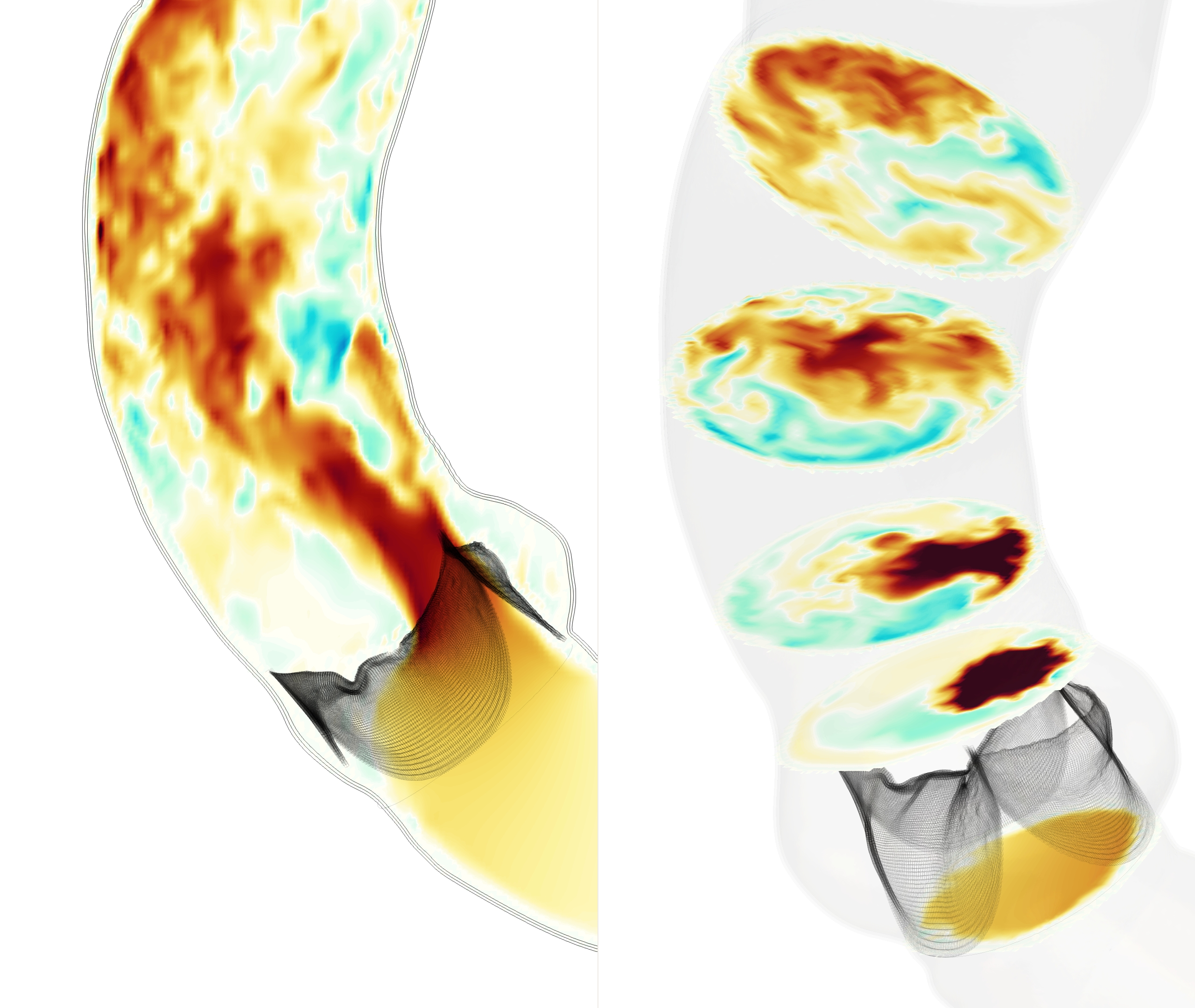}  & 
        \includegraphics[width=.24\textwidth]{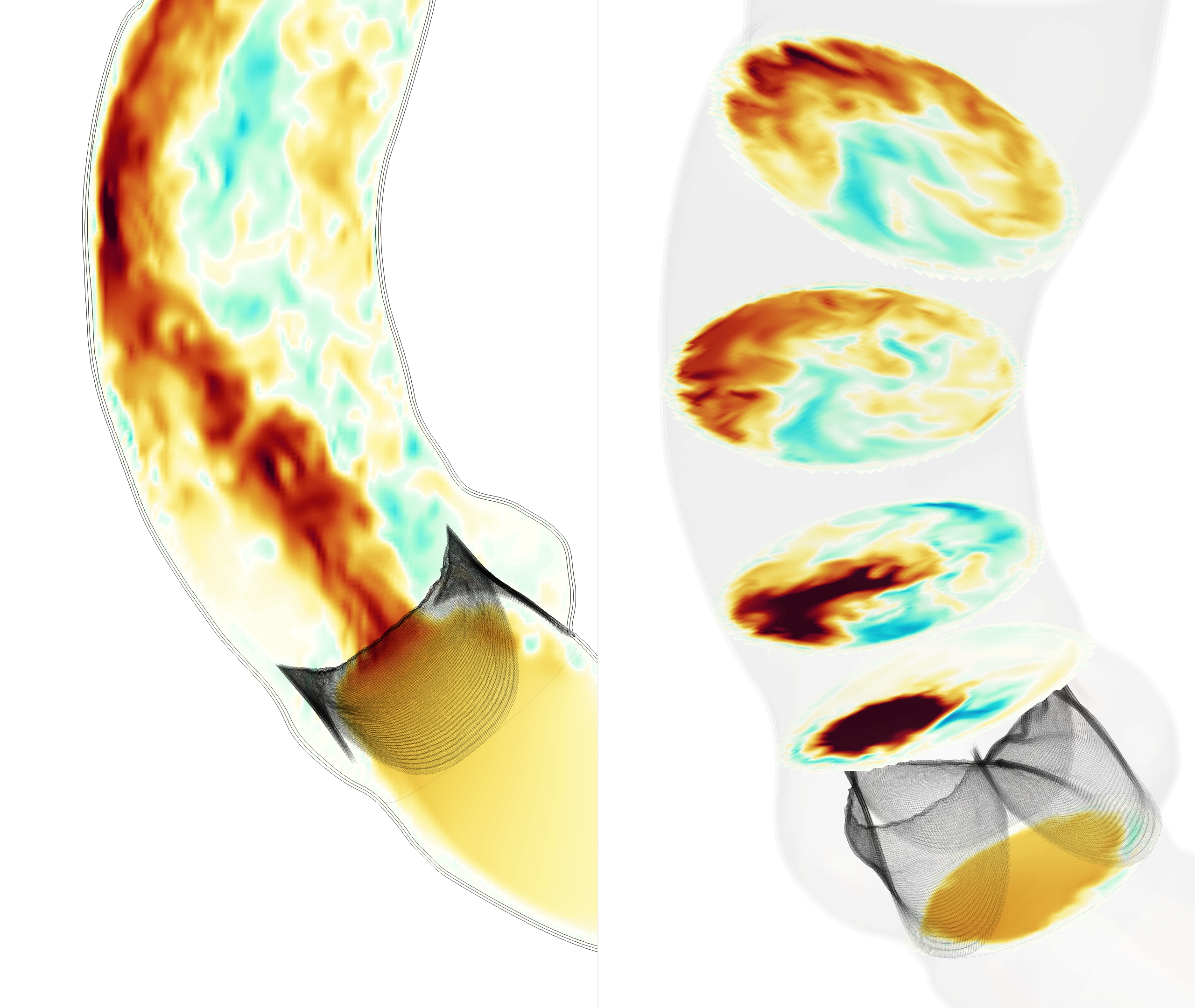} \\ 
        \rotatebox[origin=l]{90}{ $t = 2.375 $ s } &
        \includegraphics[width=.24\textwidth]{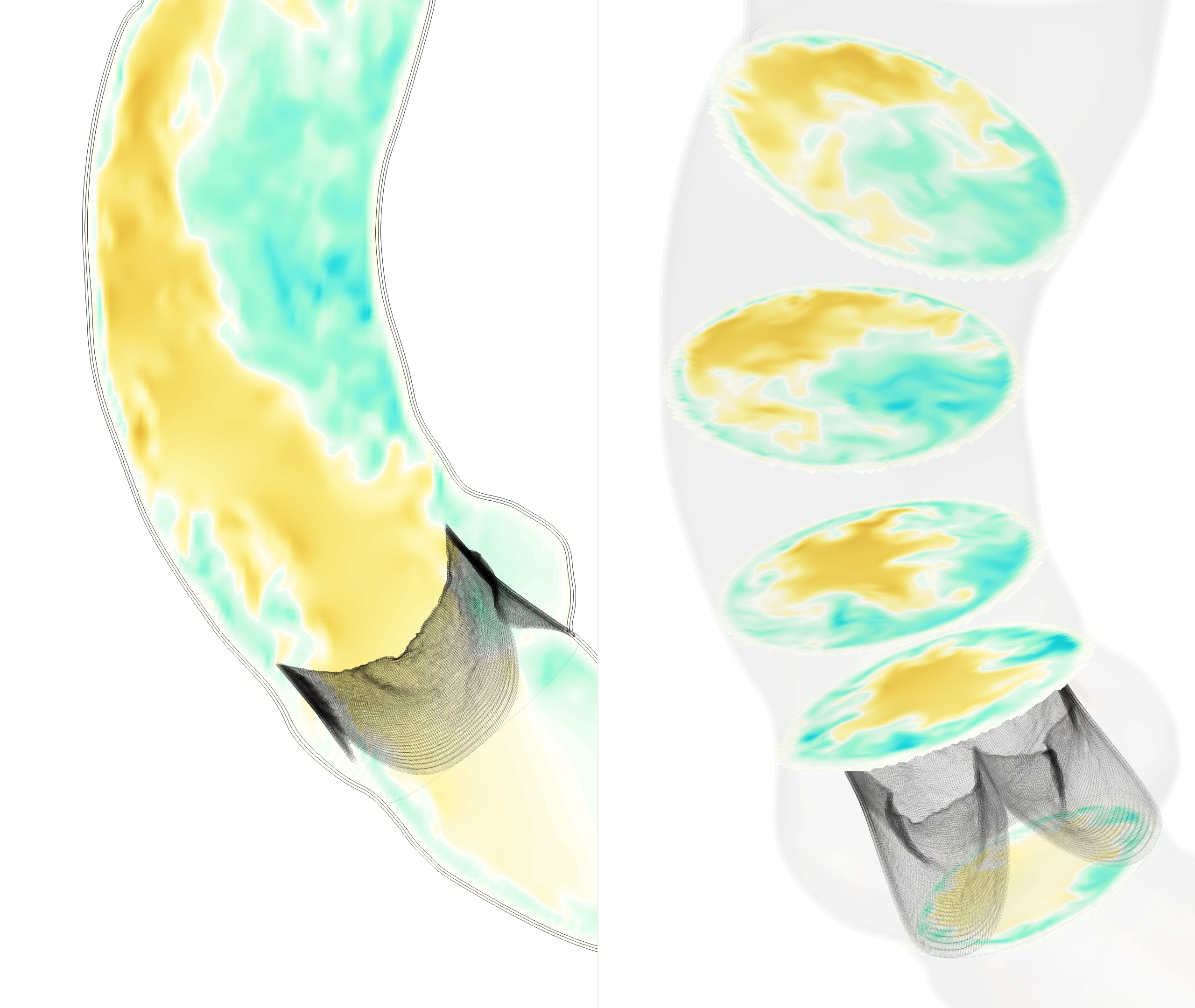}   &
        \includegraphics[width=.24\textwidth]{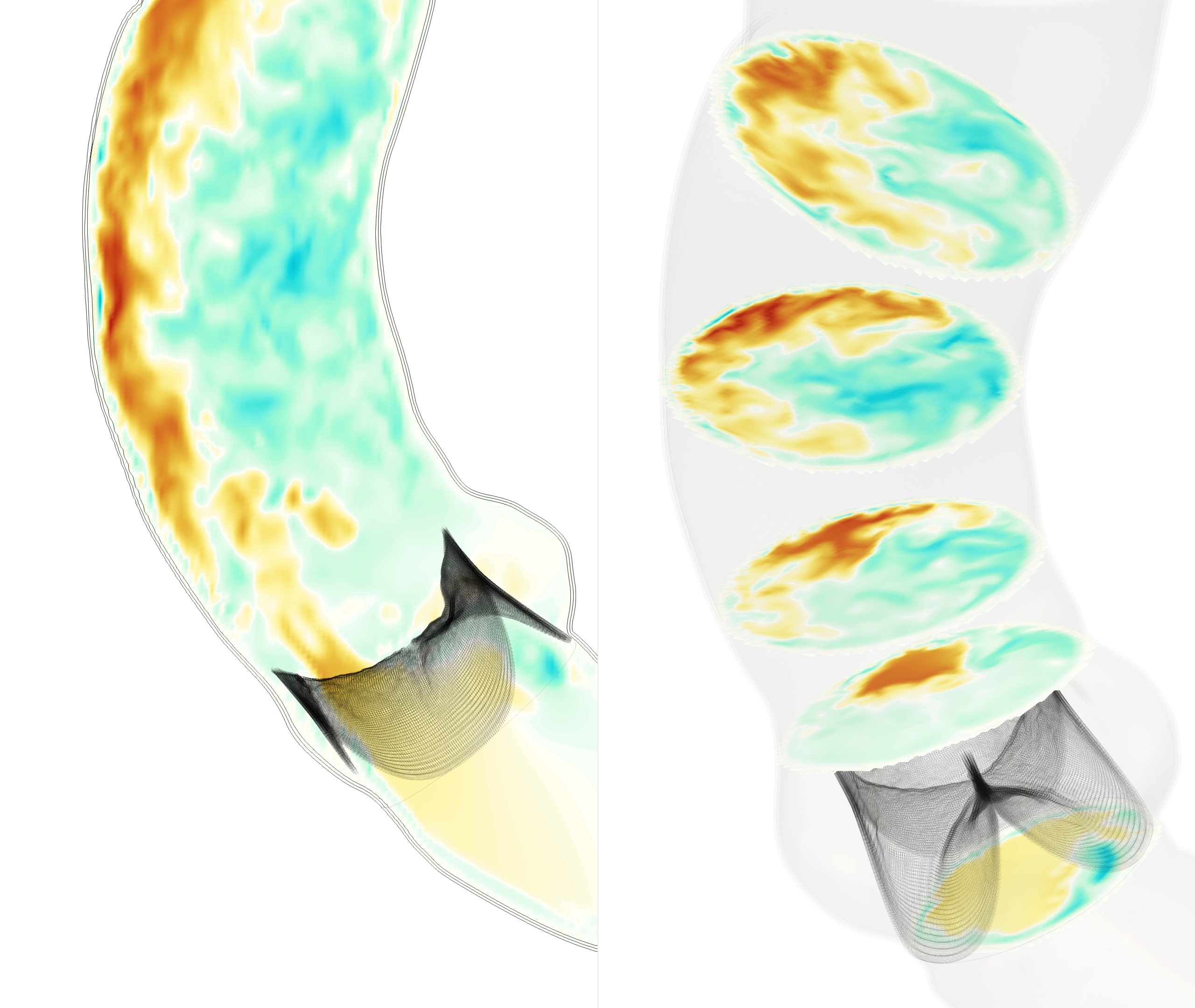}  &
        \includegraphics[width=.24\textwidth]{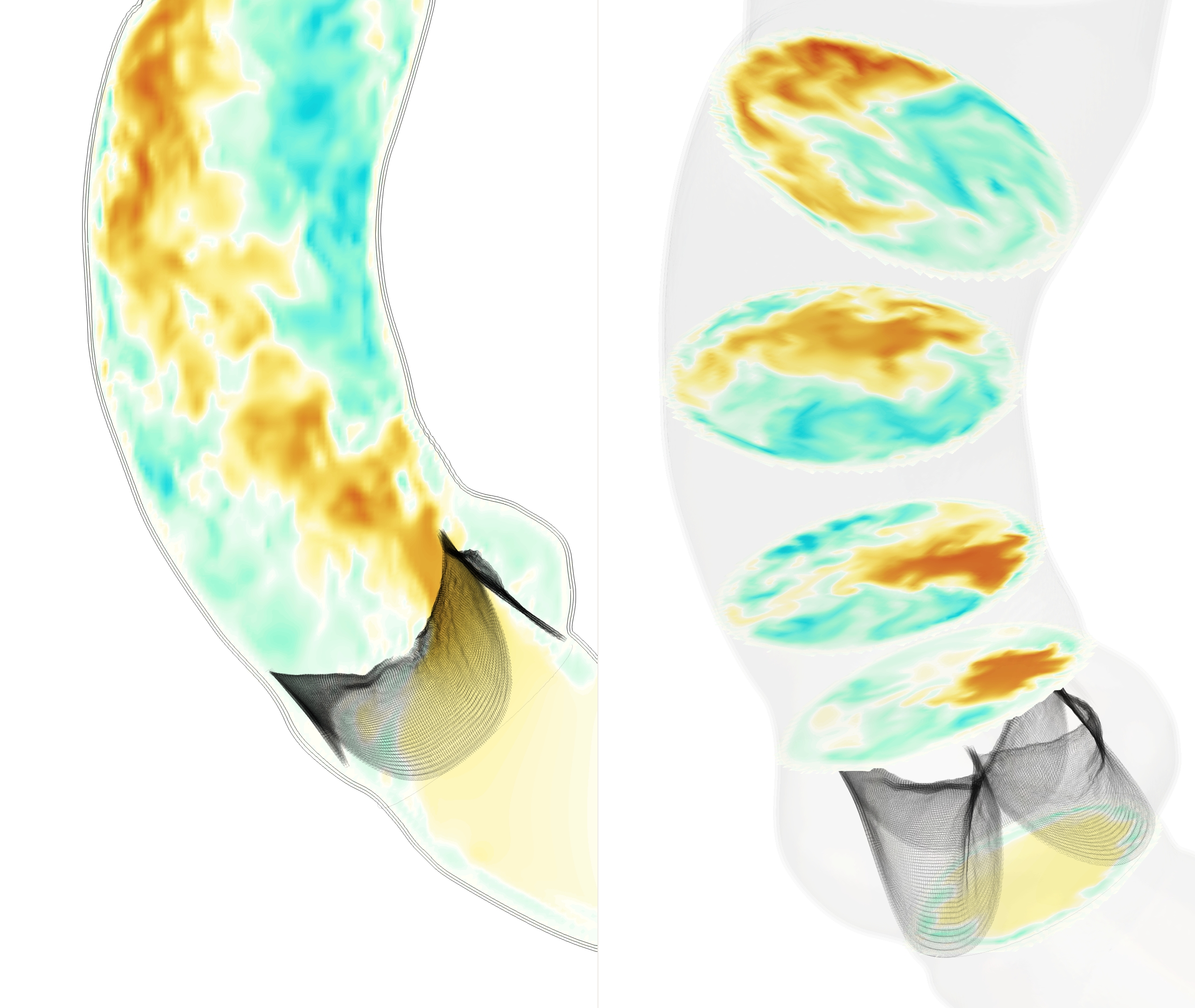}  & 
        \includegraphics[width=.24\textwidth]{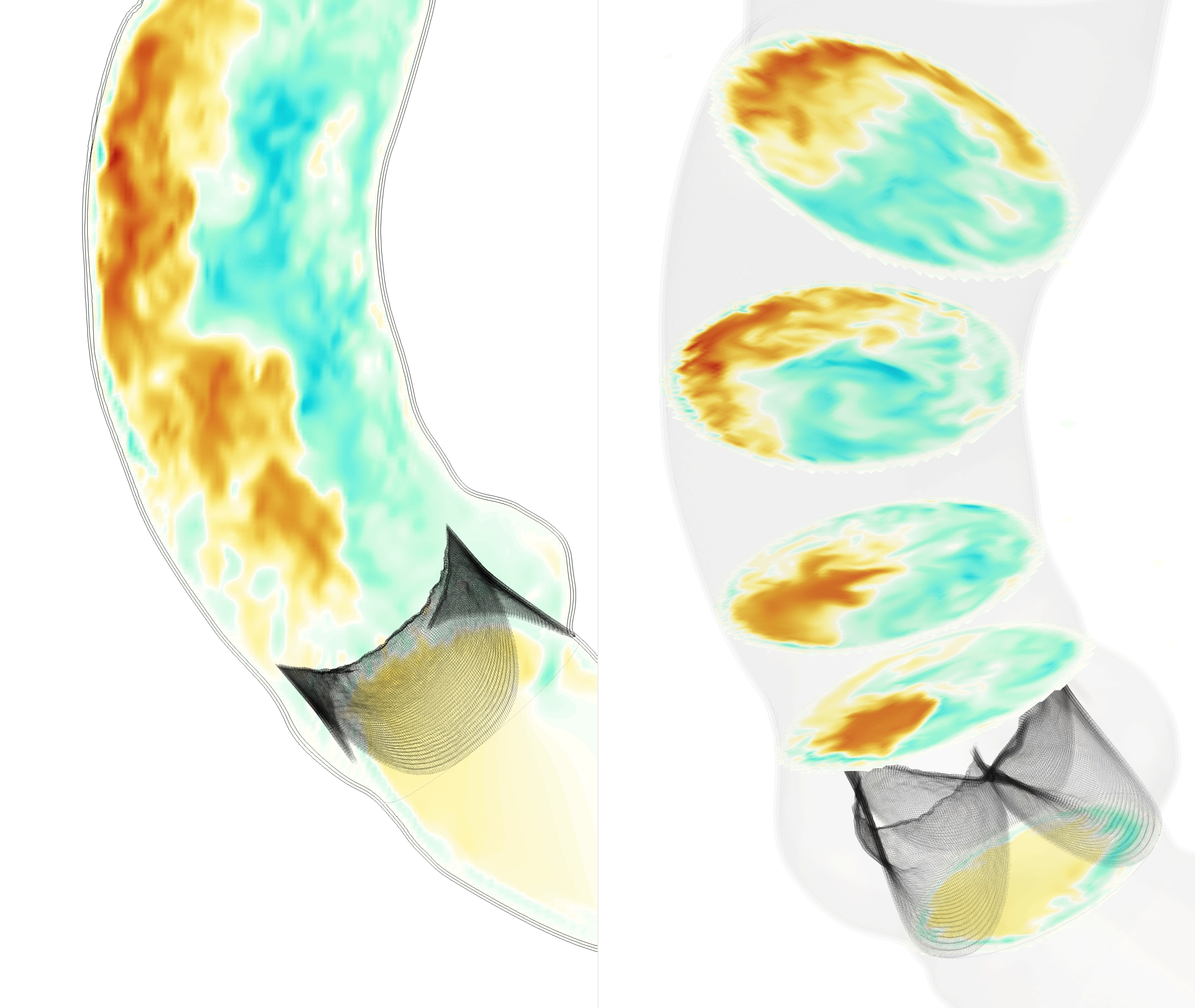} \\ 
        \end{tabular}
	}
	
	\vspace{5pt}
	{\Large B:}
	\vspace{2pt}
	
	{
        \centering
        \setlength{\tabcolsep}{1.18pt}        
        \begin{tabular}{c  c  c  c  c} 
       \rotatebox[origin=l]{90}{ $t = 2.271 $ s } & 
        \includegraphics[width=.24\textwidth]{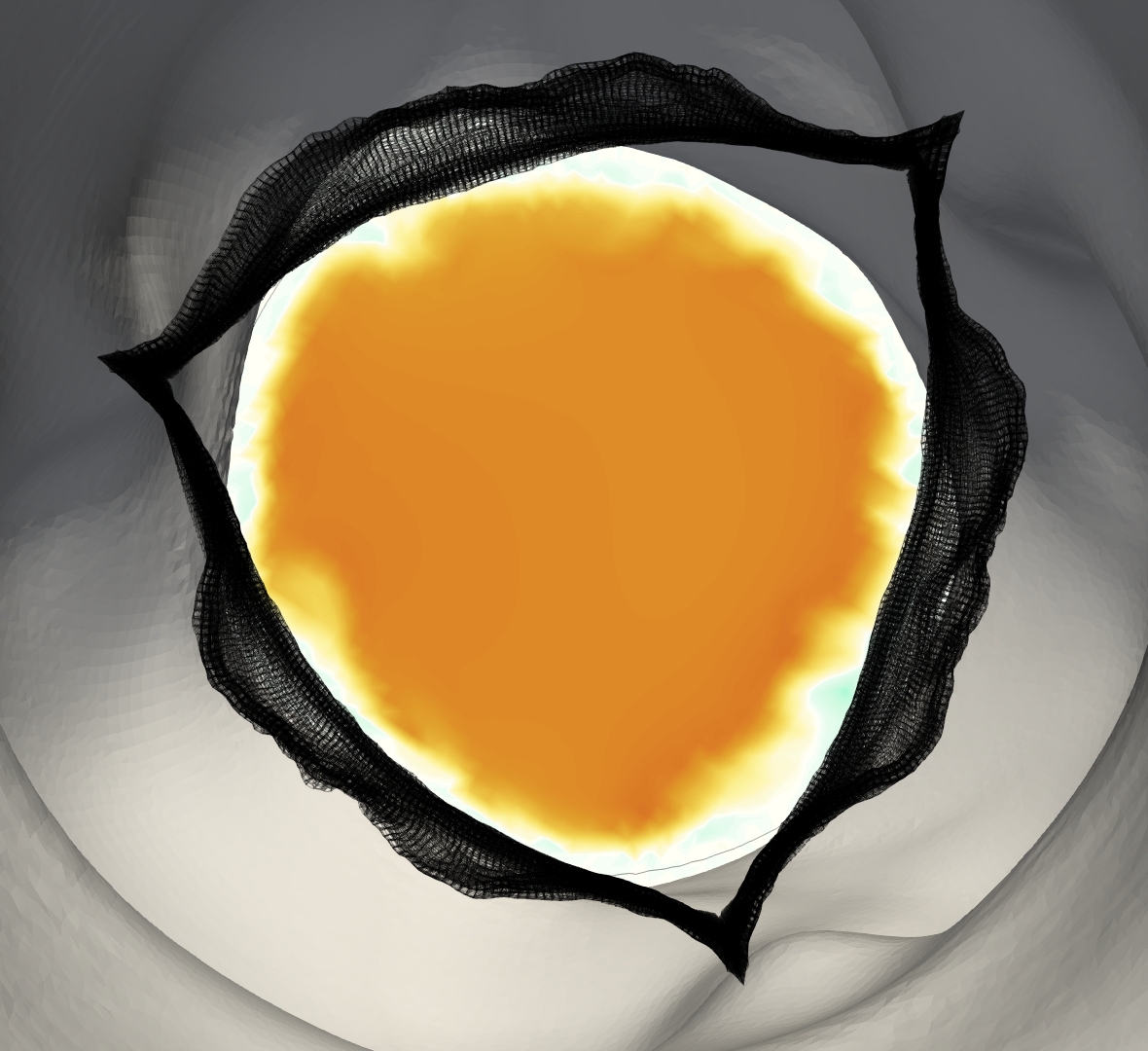}   &
        \includegraphics[width=.24\textwidth]{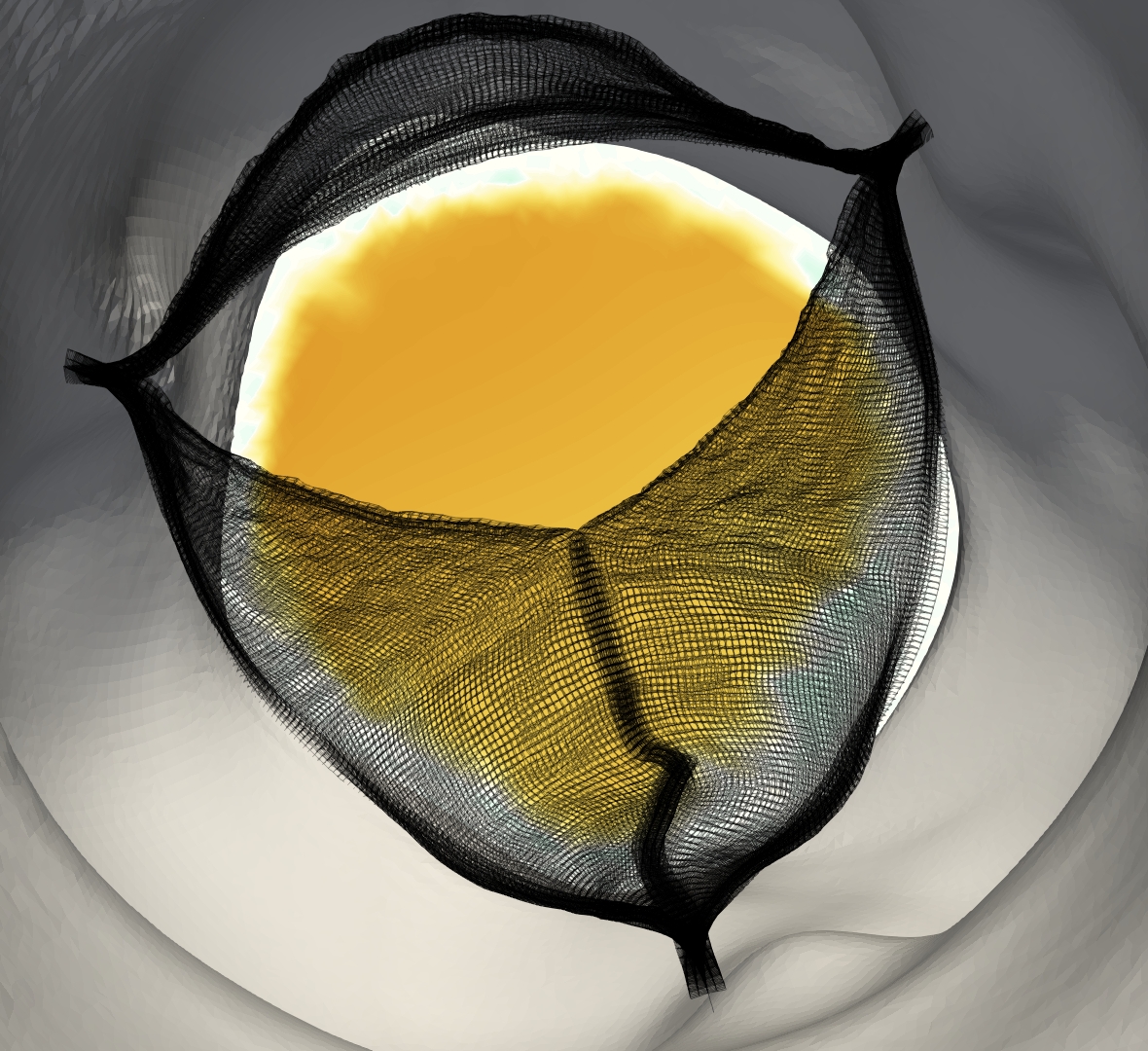}  &
        \includegraphics[width=.24\textwidth]{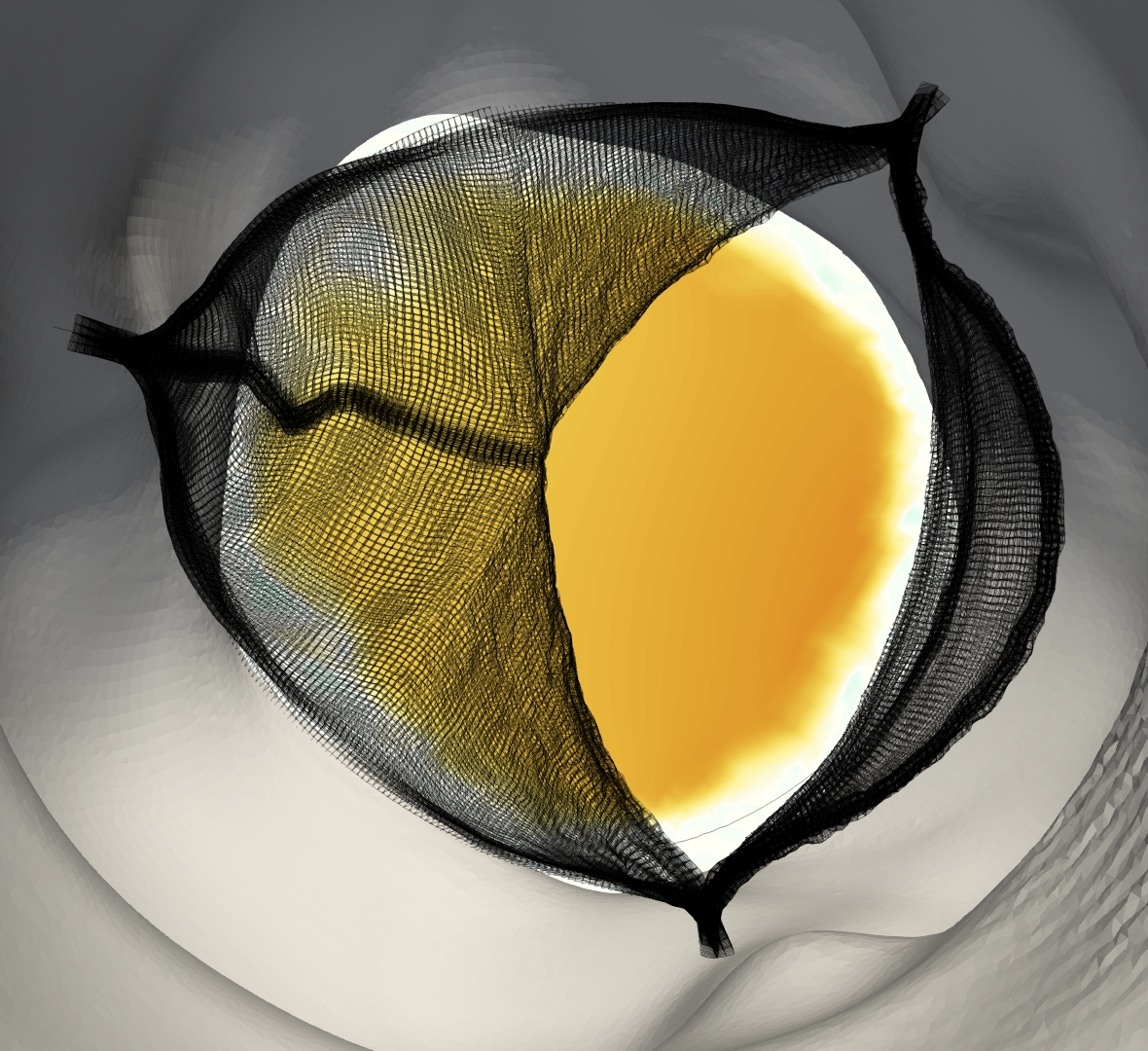}  & 
        \includegraphics[width=.24\textwidth]{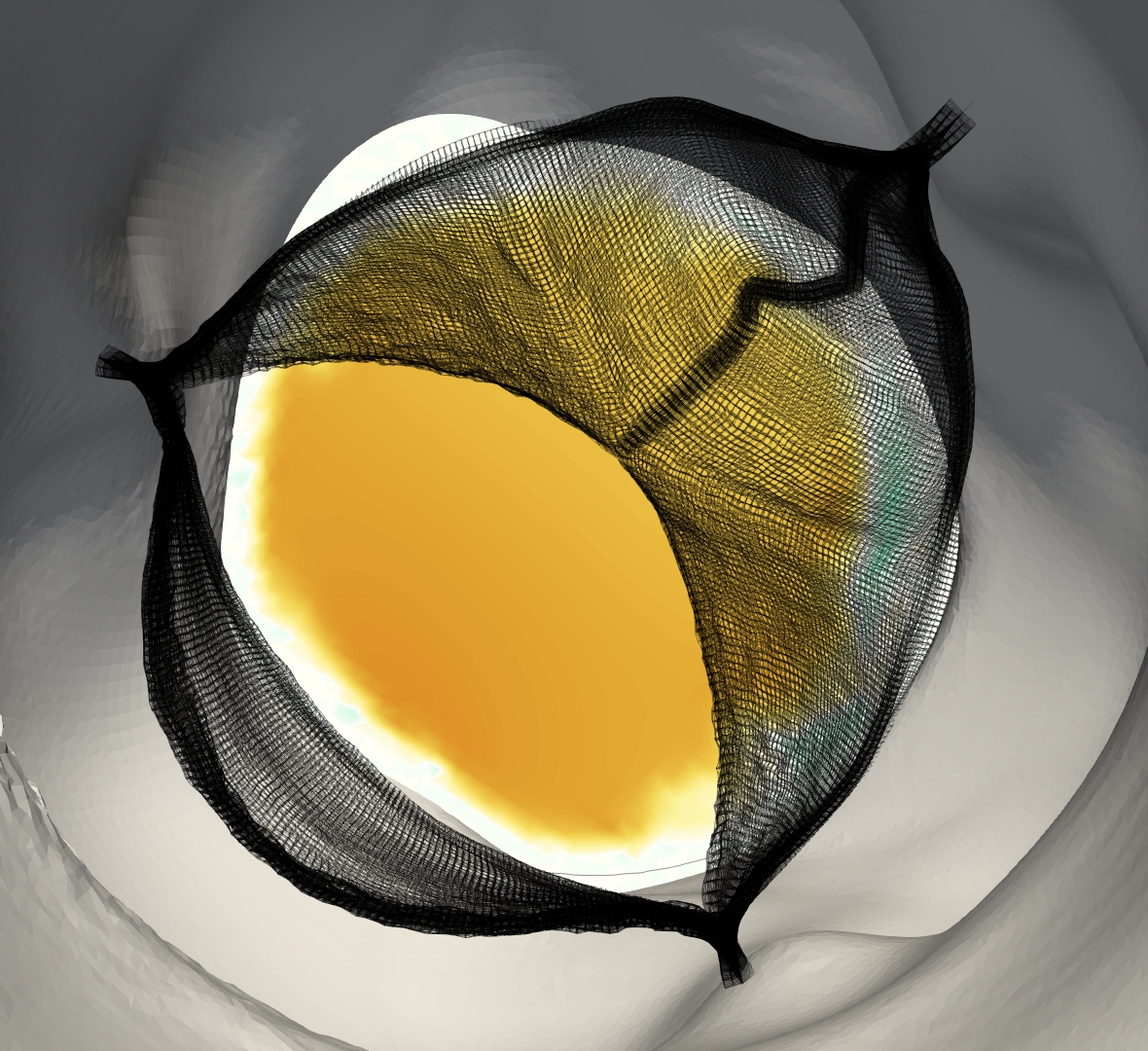} \\ 
        \end{tabular}
	}

\caption{A: Flows through four model aortic valves showing, from left to right: a tricuspid, bicuspid valve with LC/RC fusion, bicuspid valve with RC/NC fusion, and bicuspid valve with NC/LC fusion. 
For each model we show the vertical component of velocity and the velocity normal to five selected slices. 
The frames shown correspond to, from top to bottom, early systole, peak systole, mid systole and early diastole. 
B: Valves viewed from above at mid systole. 
} 
\label{flow_panels}
\end{figure*}

\begin{figure*}[t!]  
{
\centering
\setlength{\tabcolsep}{1.0pt}        
\begin{tabular}{cc}        
tricuspid &
LC/RC fusion  \\
\includegraphics[width=.48\textwidth]{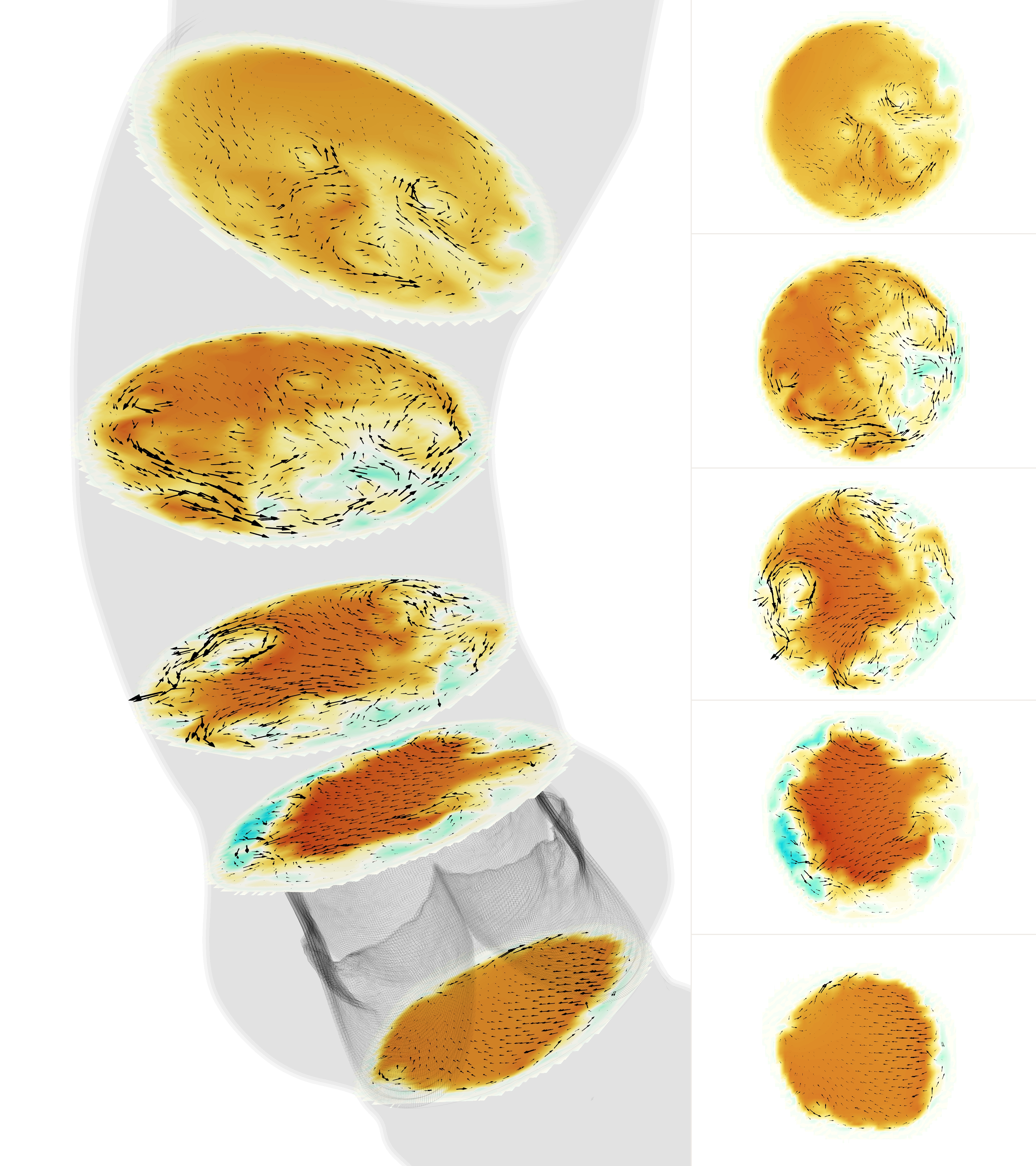} & 
\includegraphics[width=.48\textwidth]{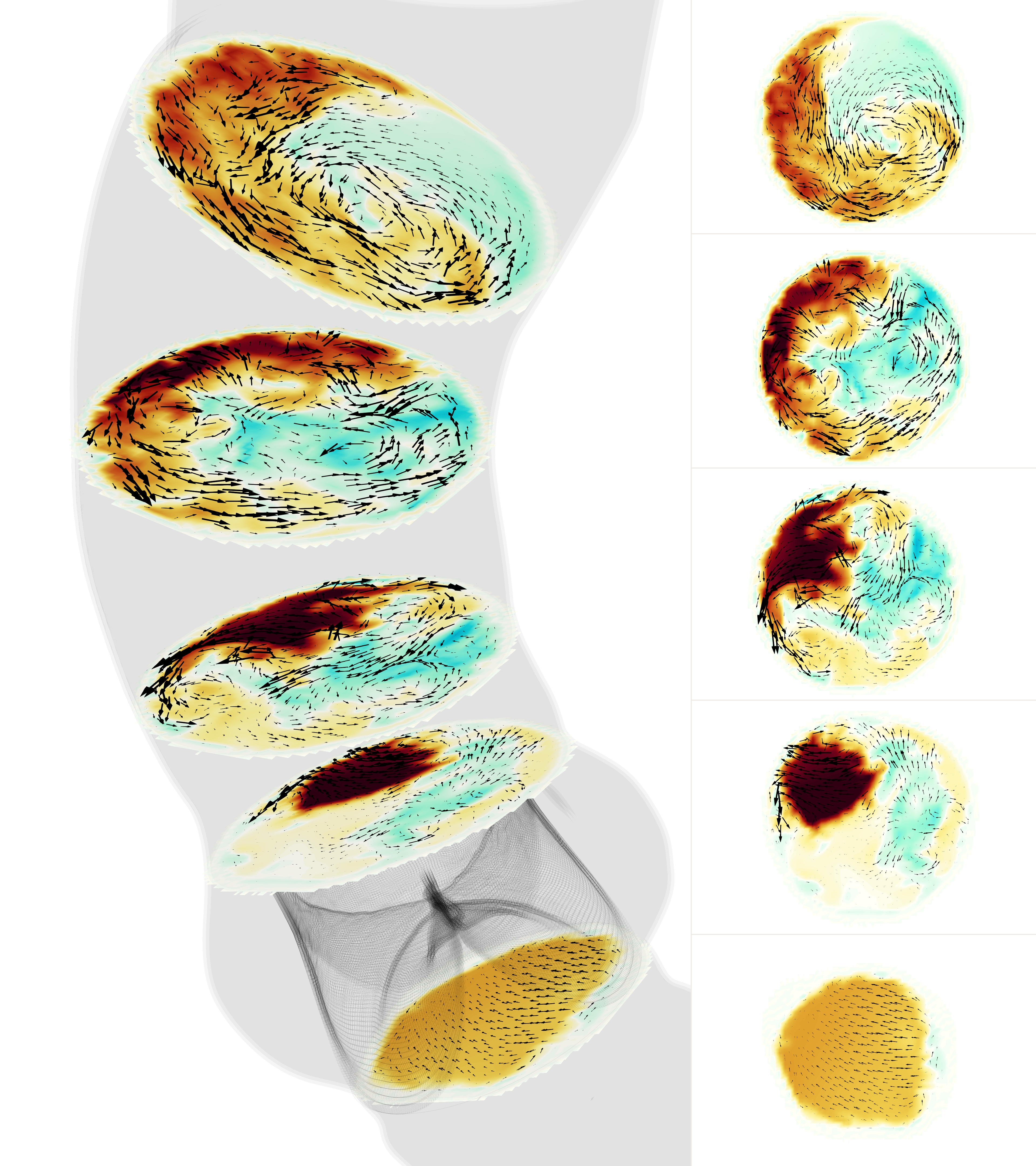} \\
RC/NC fusion & 
NC/LC fusion \\ 
\includegraphics[width=.48\textwidth]{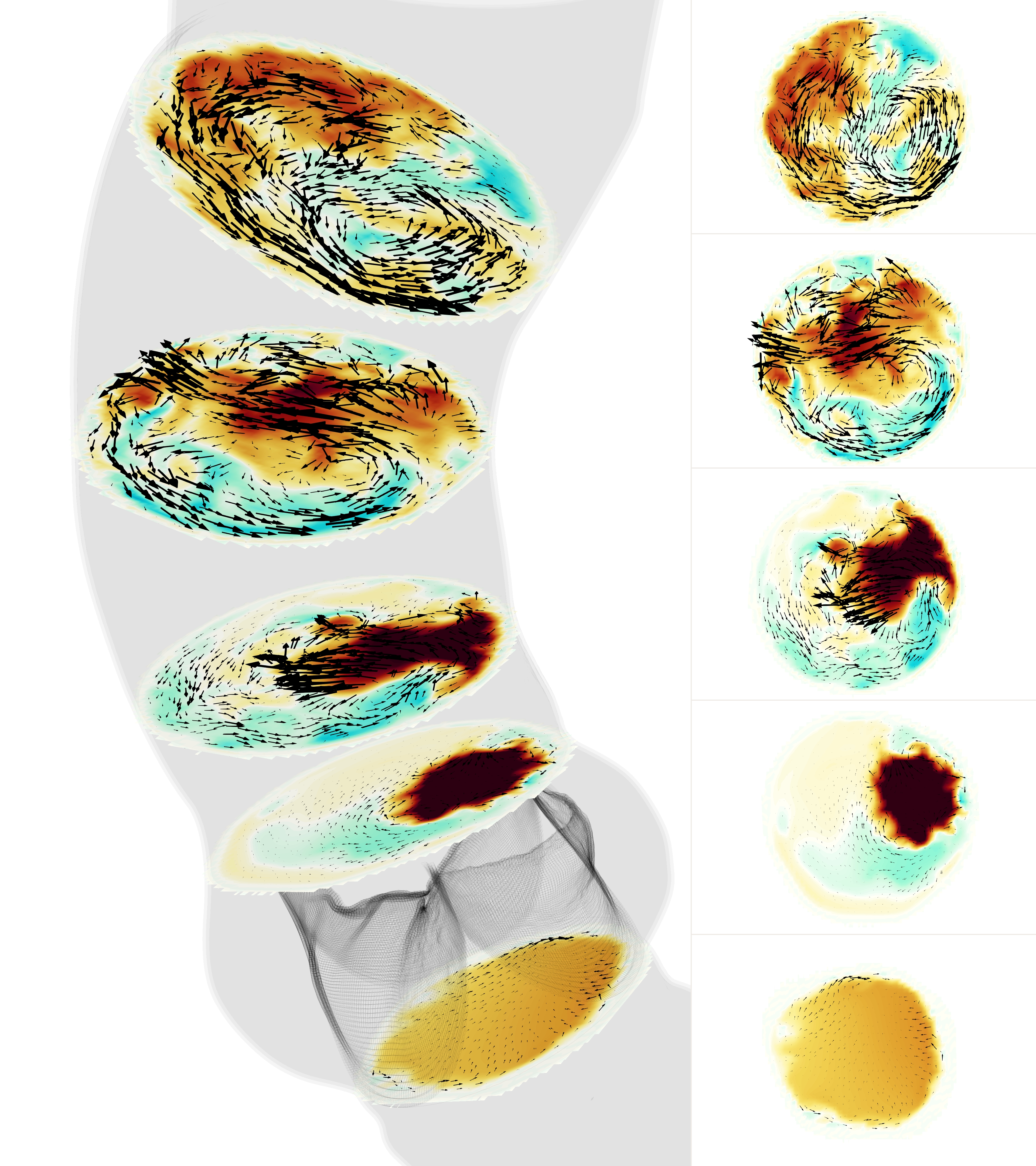} & 
\includegraphics[width=.48\textwidth]{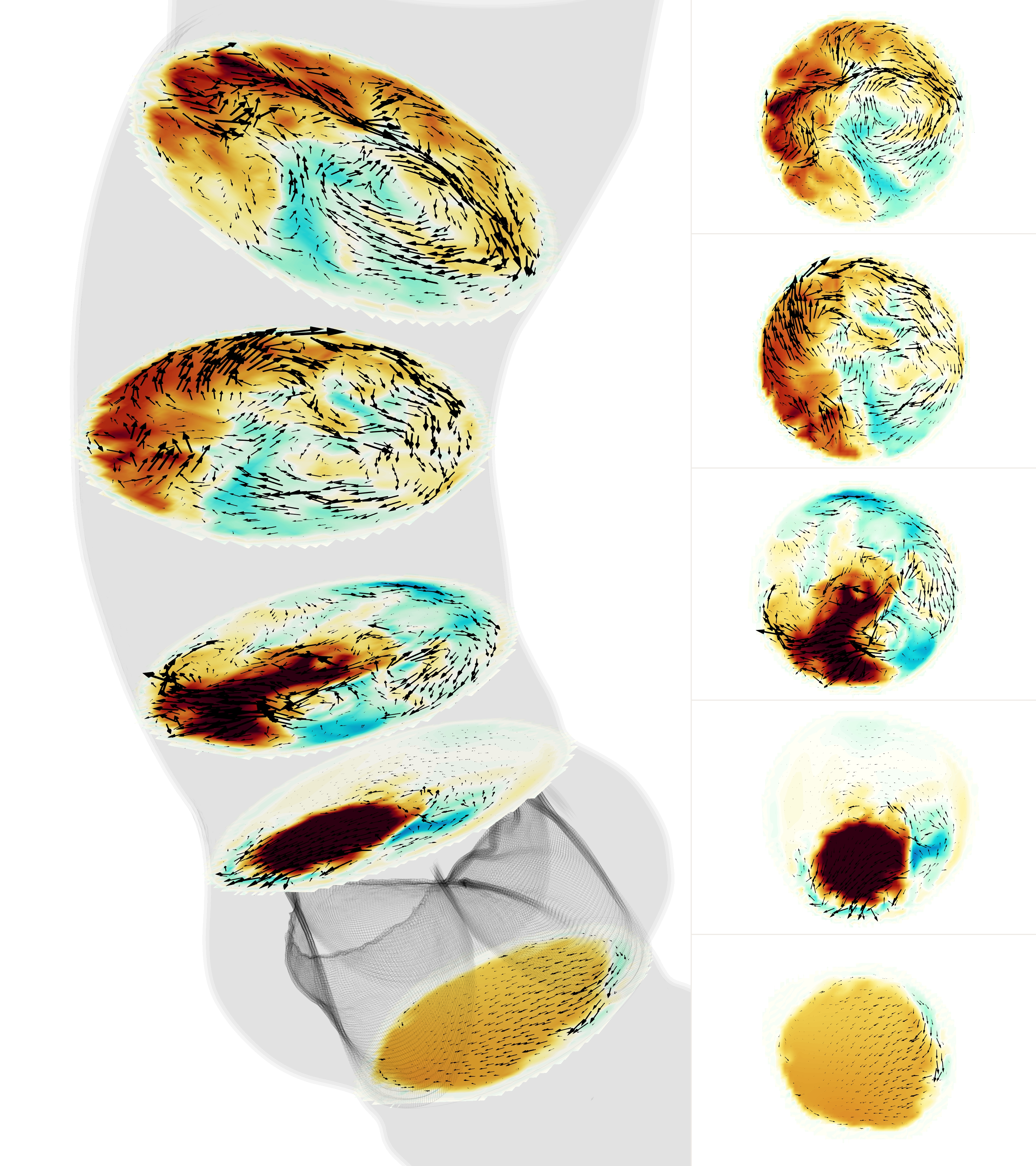} 
\end{tabular}
\caption{Vector plots of the tangential component of velocity at mid systole ($t = 2.271 $ s). 
Normal velocity is shown in the background with the same velocity scale as in Figure \ref{flow_panels}.
Each panel shows slice views as oriented in the aorta and the same slices projected onto a two dimensional plane.}
\label{vector_plots}
}
\end{figure*}

Simulations were run with a tricuspid valve and three bicuspid bicuspid valves: LC/RC fusion, RC/NC fusion and NC/LC fusion.
In all four cases, the simulated valves functioned as expected over multiple cardiac cycles, opening to allow forward flow and closing under back pressure. 
The flow fields and forward jets showed substantial differences between all cases. 
Visualizations of velocity are shown in Figure \ref{flow_panels}, depicting the vertical component of velocity on a slice through the annulus, and the normal component of velocity on fives slices approximately normal to the local axis.
These slices are located the aortic annulus, the sinotubular junction, and three locations in the ascending aorta (numbered $1\dots5$ streamwise). 
Movies of these flows are included in the supplemental information. 
Vector plots of the tangential component of velocity are shown in Figure \ref{vector_plots}, which illustrate the locations and orientation of secondary flows.

In the tricuspid case, the flow appeared plug-like immediately after opening. 
As the flow accelerated during peak systole, a jet developed through the center of the valve orifice, then separated slightly from the inner wall at mid systole. 
At peak and mid systole, there were regions of slight recirculation back towards the valve orifice. 
We refer to the outer side of the aorta with greater radius of curvature, also called the convexity, as the \emph{greater curvature} and the inner side with lesser radius of curvature, also called the concavity, as the \emph{lesser curvature}.
As the valve began to close, the jet velocity decreased and the flow reversed along the lesser curvature first, where the flow was already decelerating due to the jet angle towards the greater curvature.

In each of the bicuspid cases, the valve orifice was far narrower than in the tricuspid case, and a more complex flow pattern with greater recirculation occurred. 
Due to the orifice orientation, the jet developed in close approximation to the vessel wall, in contrast to the tricuspid case where the jet remained approximately centered in the vessel lumen. 
This difference in location and increased jet intensity produced substantially different flow features in all four cases. 

In the bicuspid case with LC/RC fusion, the valve orifice was oriented along the greater curvature.
At peak systole, a concentrated jet developed immediately distal to the valve orifice and remained adherent to the greater curvature. 
As the jet progressed downstream, it spread along the greater curvature normal to the flow direction while maintaining intensity.
There was a region of reverse flow distal to the fused LC and RC leaflets.
The strongest reverse flows were present at peak systole, generating recirculation in the vertical component of velocity (Figure \ref{flow_panels}).
Downstream of the aortic sinus, lower velocity fluid moved away from the center of the jet along the vessel wall towards the lesser curvature.
This generated a counter-rotating vortex pair at peak systole in the middle of the ascending aorta, which began to shift and lose strength as the jet spread along the greater curvature at mid systole (Figure \ref{vector_plots}).
This vortex pair structure (slices 3 and 4) persisted with an additional secondary flow rotating counterclockwise that extended along almost the entire circumference of aorta.
Further downstream (slice 5), the larger counterclockwise rotation had formed into a vortex and the counter-rotating vortex pair did not form.

In the bicuspid case with RC/NC fusion, the valve orifice was oriented along the lesser curvature, and correspondingly at the sinotubular junction, the jet was centered along the lesser curvature.
As the flow propagated, the jet traveled along the wall from the lesser curvature to the greater curvature and began to spread.
Further downstream, (slice 5) there is reverse flow at the lesser curvature of the aorta, directly distal of the valve orifice. 
The forward flow extended into the center of the vessel, especially at mid systole, separating from the wall more than in the LC/RC fusion case.
Strong secondary flows accompanied the movement of the jet.
An asymmetric counter-rotating vortex pair was generated that was eventually subsumed by a counterclockwise rotation around the vessel wall.

With NC/LC fusion, the valve orifice was oriented along the greater curvature.
The streamwise flow behaved similarly to the LC/RC fusion case. 
The jet was concentrated immediately downstream of the valve orifice and then spread as flow progressed through the aorta, with reverse flow distal to the fused leaflets. 
The secondary flows have rotation along the vessel wall that traveled clockwise, particularly in the ascending aorta, and no distinct counter-rotating vortex pair appeared.
Therefore, the secondary flows are substantially different from the two other bicuspid cases.

Thus, in all cases, the jet ultimately traveled towards the greater curvature of the aorta.
The LC/RC and RC/NC fusion cases generated a counterclockwise rotation, whereas the NC/LC case generated a clockwise rotation.
In addition, RC/NC fusion produced more substantial rotating flow, as the jet traveled from the lesser to greater curvature.

\begin{figure*}[t!]
\setlength{\tabcolsep}{2.0pt}
\begin{tabular}{M{.02\textwidth} M{.02\textwidth} M{.46\textwidth} M{.46\textwidth}}        
& & tricuspid & LC/RC fusion  \\
& \rotatebox[origin=c]{90}{\small pressure (mmHg)} \vfill  &  \includegraphics[width=.46\textwidth]{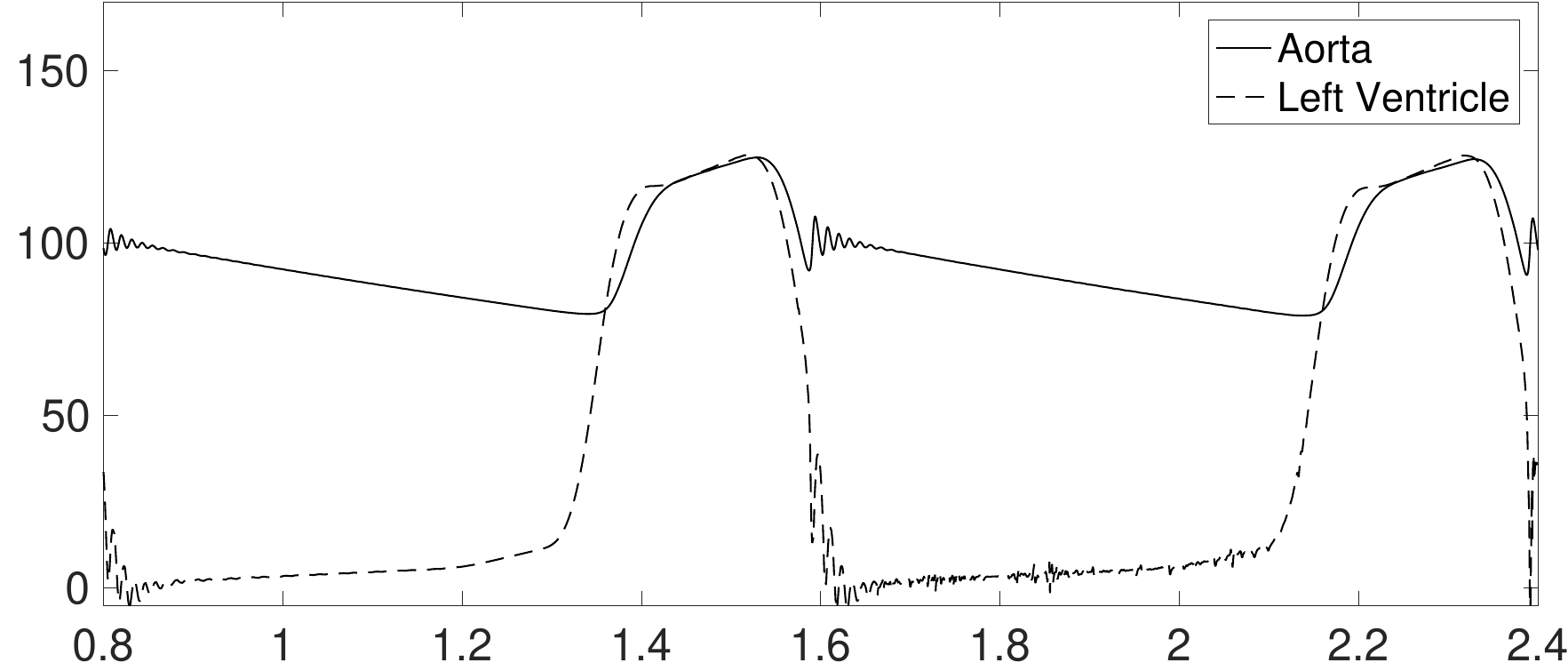} & \includegraphics[width=.46\textwidth]{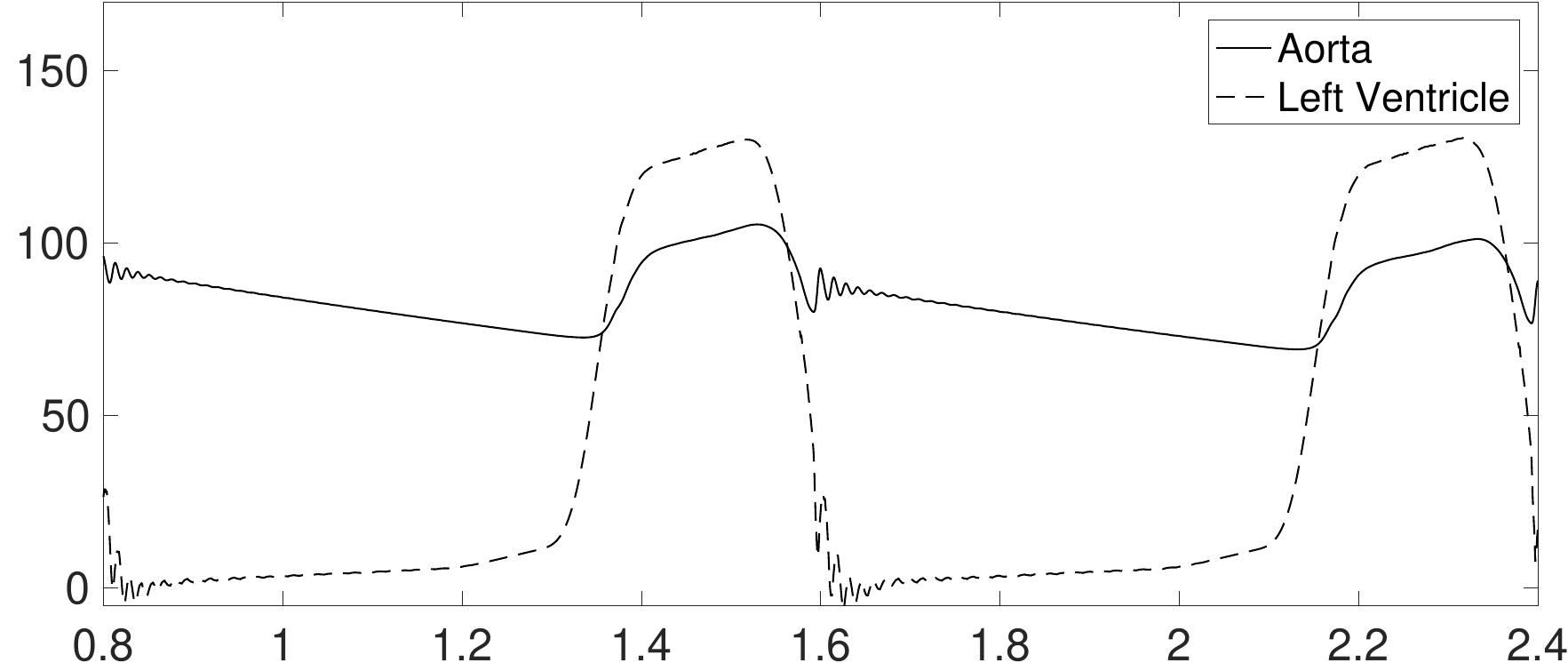}   \\  
 \rotatebox[origin=c]{90}{\small flow (ml/s),} \vfill & \rotatebox[origin=c]{90}{\small cumulative flow (ml)} \vfill &  \includegraphics[width=.46\textwidth]{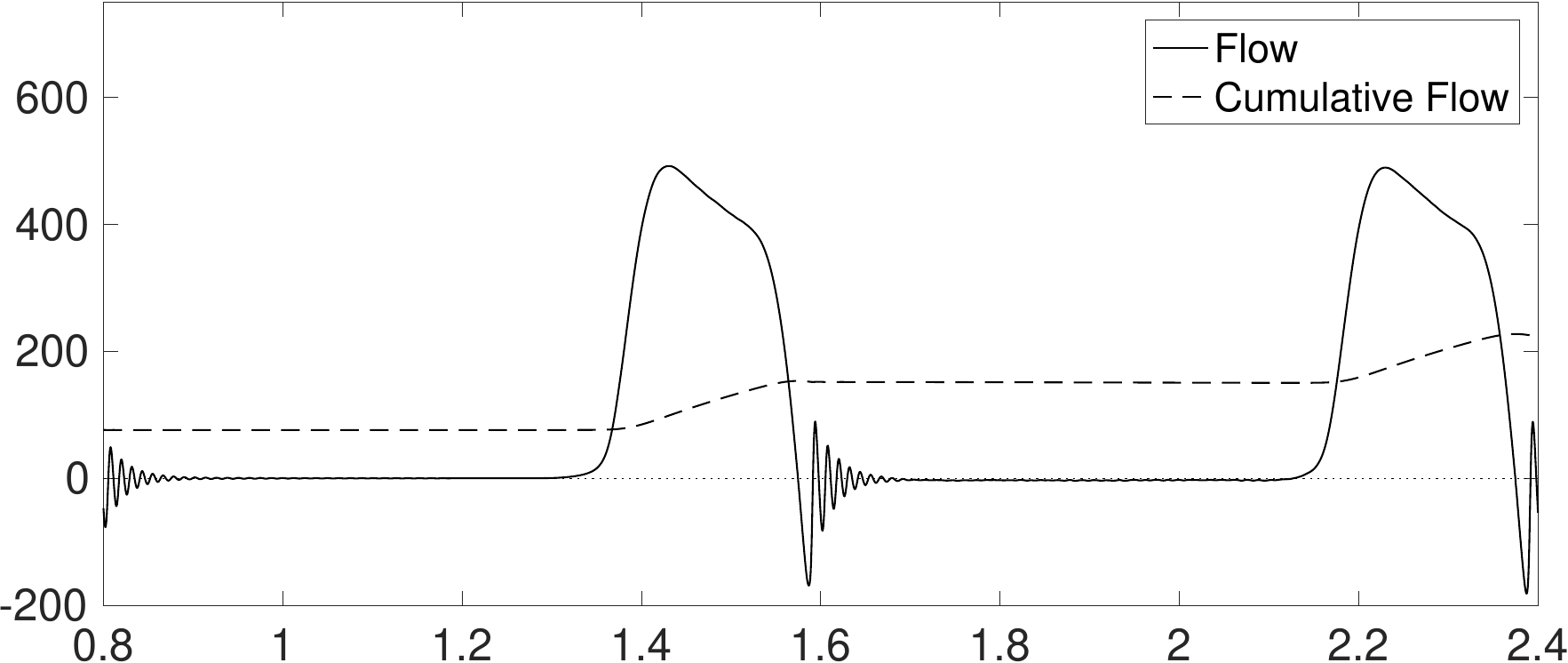}  &  \includegraphics[width=.46\textwidth]{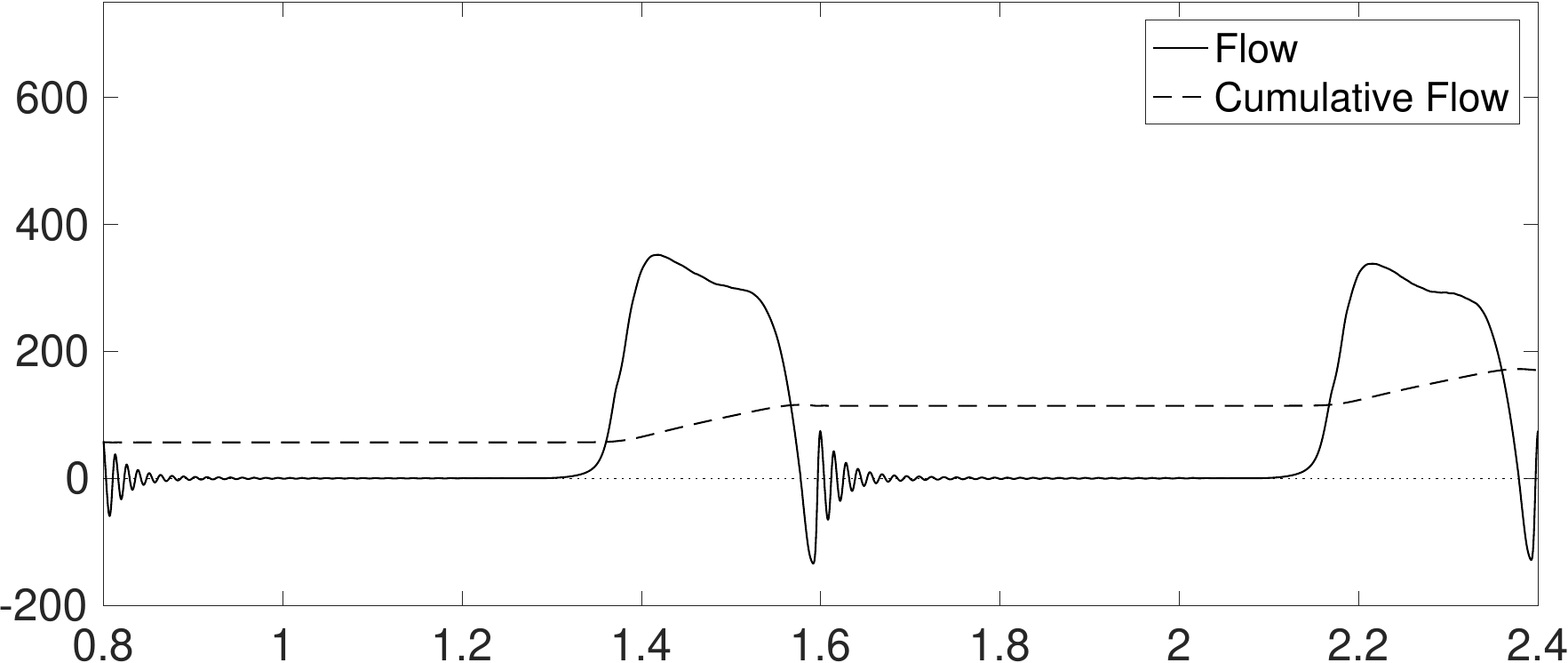} \\
 & & time (s) & time (s)
\end{tabular}   
\caption{Flow rates and pressures through the tricuspid aortic valve (left) and the bicuspid valve with LC/RC fusion (right). 
In the tricuspid case, a negligible pressure gradient occurs across the valve during forward flow. 
In the bicuspid case, a substantial pressure gradient across the valve occurs during systole, indicative of aortic stenosis. 
}
\label{flow_pressure_basic}
\end{figure*}

Pressure and flow waveforms for the tricuspid and LC/RC fusion cases are shown in Figure \ref{flow_pressure_basic}. 
The second and third cardiac cycles are shown; the first cardiac cycle contains initialization effects and is omitted. 
The aortic pressure decreased gradually through diastole, while the ventricular pressure remained low. 
Next, the ventricular pressure increased as systole began. 
A transient forward pressure gradient across the valve occurred, with peak values as high as 16.3 mmHg. 
Then, the pressure gradient decreased, leaving minimal pressure gradient across the valve. 
At times, the pressures even crossed, while forward flow continued due to inertia. 
Thus, the model valve offered minimal resistance to flow.
When the ventricular pressure began to drop, a prominent dicrotic notch occurred in the aortic pressure, followed by an oscillation associated with valve closure. 
This oscillation is similar to that seen in experimental pressure traces \cite{stergiopulos1999use,laniado1982physiologic} and is caused by a sudden pressure loading of the elastic valve. 
The simulation oscillation has possibly higher amplitude and longer duration due to the rigidity in the model aorta than those of experiments, but noise and sampling rates in experimental traces makes this difficult to assess.
An oscillation also occurs in the ventricular pressure immediately following valve closure, again due to sudden pressure loading on the valve (recall that the ventricular pressure reported is the pointwise fluid pressure immediately proximal to the valve). 
The cycle then repeated, with pressure gradually decreasing through diastole. 
On the third cycle, trace regurgitation occurred and a corresponding lack of smoothness is seen in the pressure curve.

The flow waveform showed a flow rate of approximately zero during diastole, as expected. 
At the beginning of systole, the flow rate increased rapidly under the transient forward pressure gradient. 
Next, the ventricular pressure began to fall below the aortic pressure, the flow rate decreased and subsequently became negative. 
An oscillation was seen in the flow rate that then decayed and was followed by a flow rate of approximately zero for the remainder of diastole. 
A similar oscillation occurs in experimental measurements of flow rate, though the simulation oscillation is of possibly longer duration due to rigidity in the model aorta, as in the oscillation in pressure \cite{stergiopulos1999use,laniado1982physiologic}.
The cycle then repeated. 
The cumulative flow had an average value of 75.24 ml/cycle, or 5.64 L/min, corresponding to the nominal cardiac output of 5.6 L/min. 
The peak flow rate was $Q_{max}$ = 522.2 ml/s.  
We estimate the peak Reynolds number as $ \rho (Q_{max}/A) 2r / \mu  \approx 6600$, where $r = 1.25$ cm is the valve radius and $A = \pi r^{2}$ is the annulus area.

With LC/RC fusion, a forward pressure gradient across the domain was sustained through systole, with peak pressure gradient of 30.6 mmHg and sustained pressure gradients over 18 mmHg. 
These pressure gradients represent clinically significant, stage B, progressive, mild-to-moderate aortic stenosis \cite{Nishimura_2014_stenosis_guidelines}.
This model does not include calcification or other material property changes. 
The stenosis, therefore, was exclusively caused by fusion of the free edges.  
The occurrence of stenosis is in stark contrast to the tricuspid case, in which a near zero pressure gradient occurs throughout forward flow, as expected for normal aortic valve morphology. 

The bicuspid flow cases had lower forward flow during systole, with a cumulative flow rate of 56.90 ml/cycle, or 4.27 L/min. 
Thus, without any changes to boundary conditions or modeling any compensatory measures in the circulatory system, the stenosis associated with the bicuspid valves resulted in a loss of about 1.37 L/min (or 24\%) of cardiac output. 
The valve sealed reliably for all three cycles. 
Pressure and flow rates for the RC/NC and NC/LC fusion cases are similar to those of the LC/RC fusion case.

\begin{figure*}[t!]
{
\setlength{\tabcolsep}{2.0pt}        
\begin{tabular}{M{.26\textwidth} M{.0\textwidth} M{.21\textwidth} M{.26\textwidth} M{.0\textwidth} M{.21\textwidth}}
\multicolumn{3}{c}{ tricuspid} & \multicolumn{3}{c}{ LC / RC fusion } \\ 
\includegraphics[width=.26\textwidth]{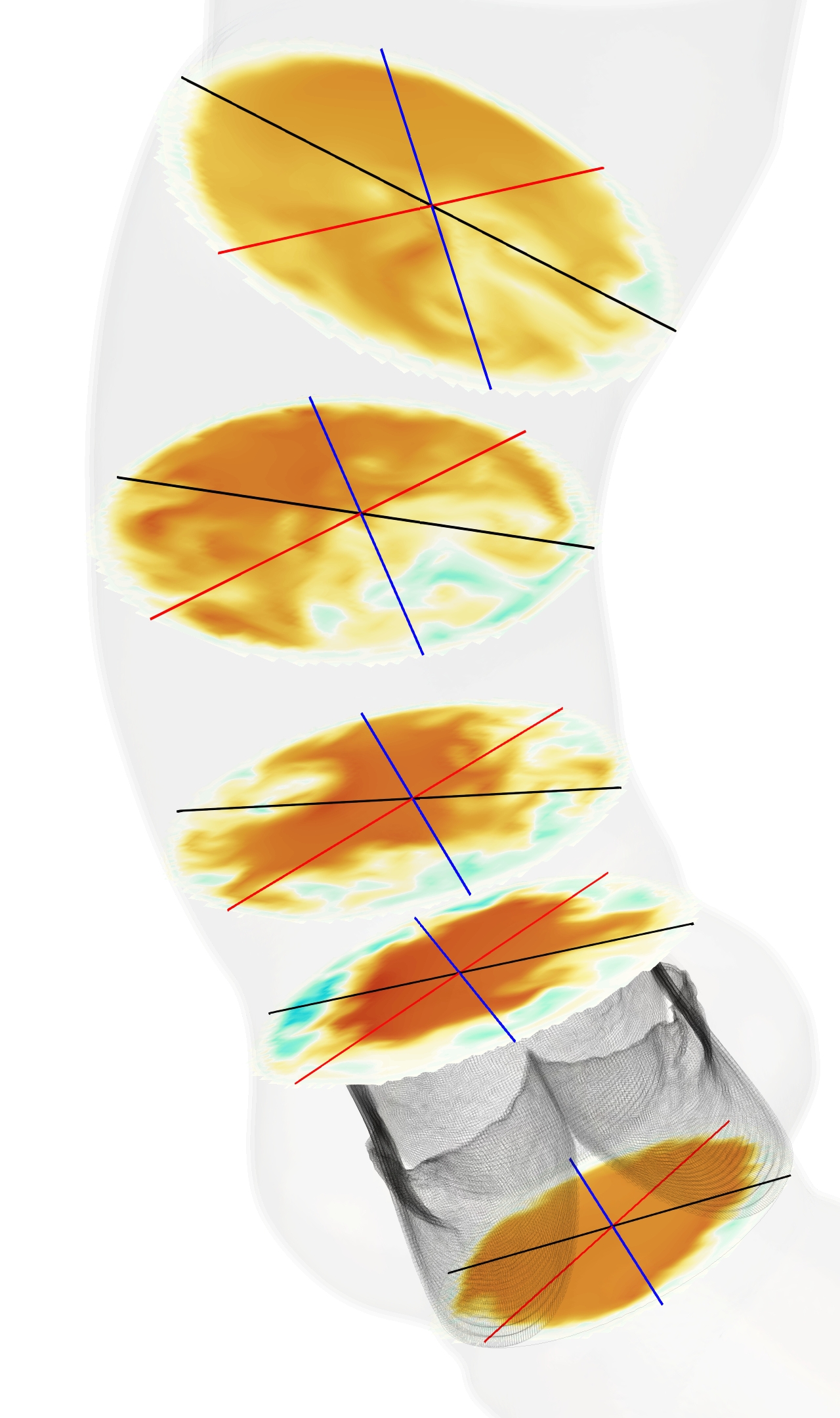} \vfill 
&
\hspace{-13pt} \rotatebox[origin=c]{90}{\small velocity (cm/s)} \vfill
&
\begin{tabular}{c}
\includegraphics[width=.21\textwidth]{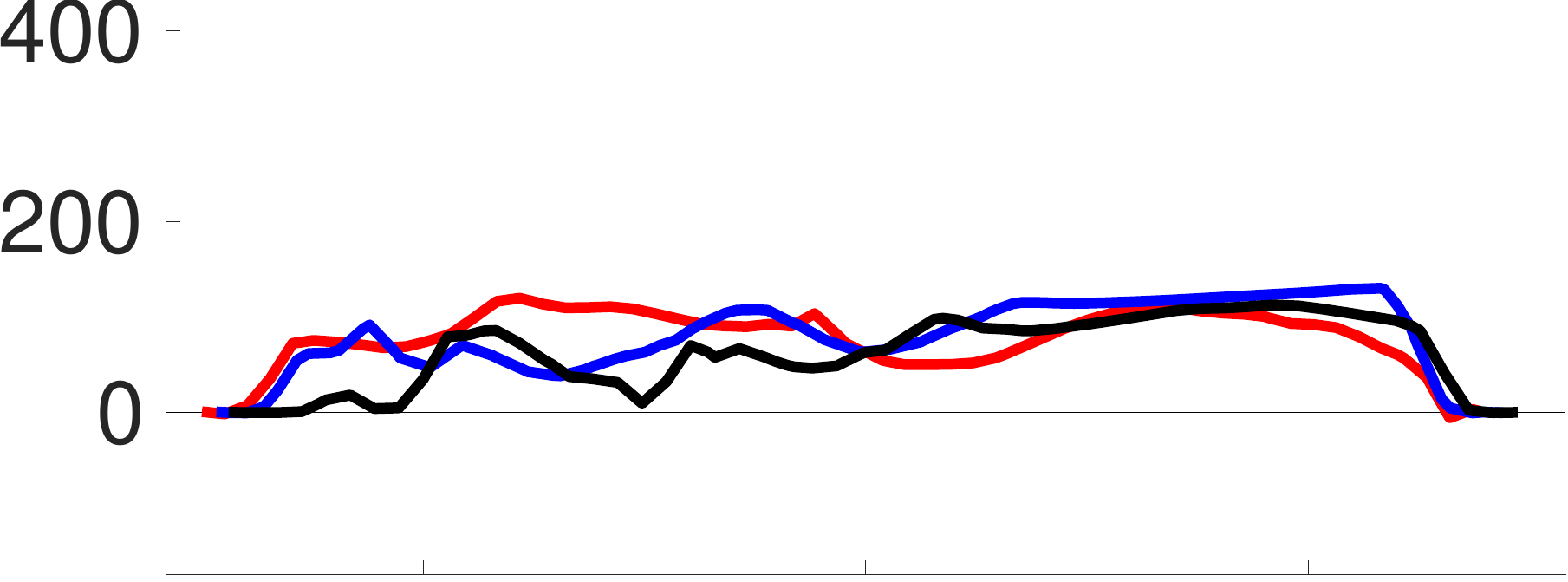} \\
\includegraphics[width=.21\textwidth]{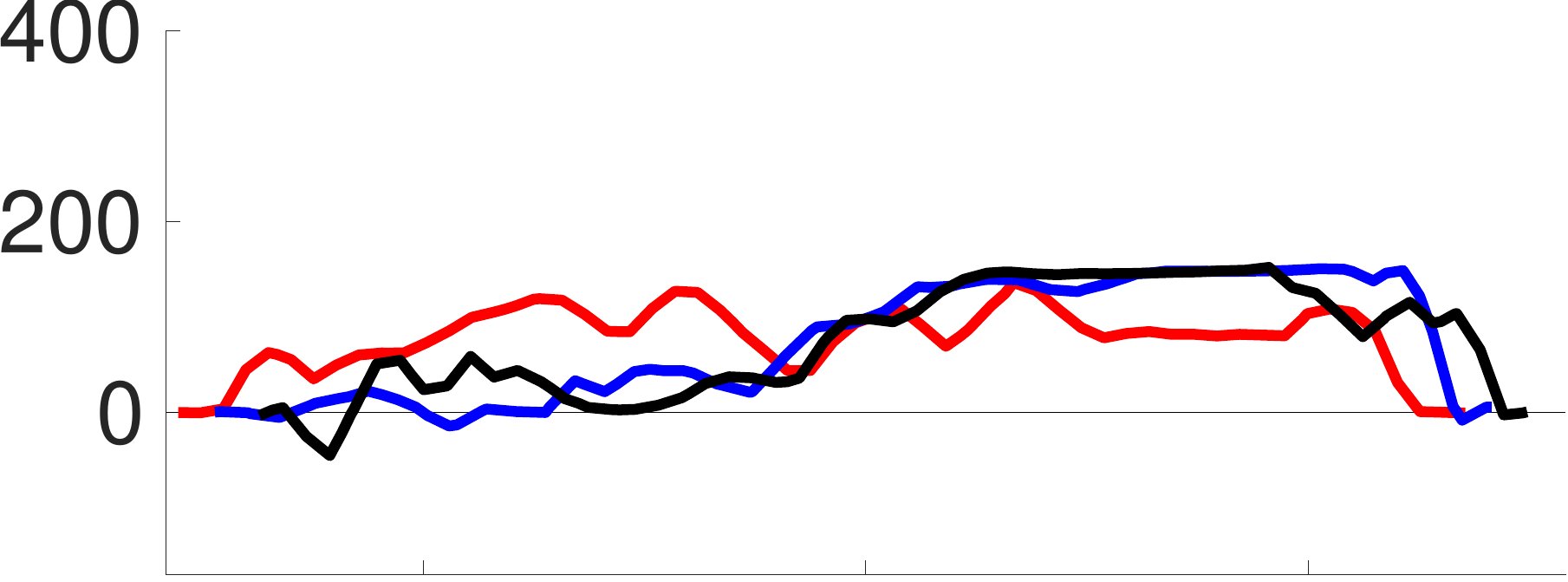} \\
\includegraphics[width=.21\textwidth]{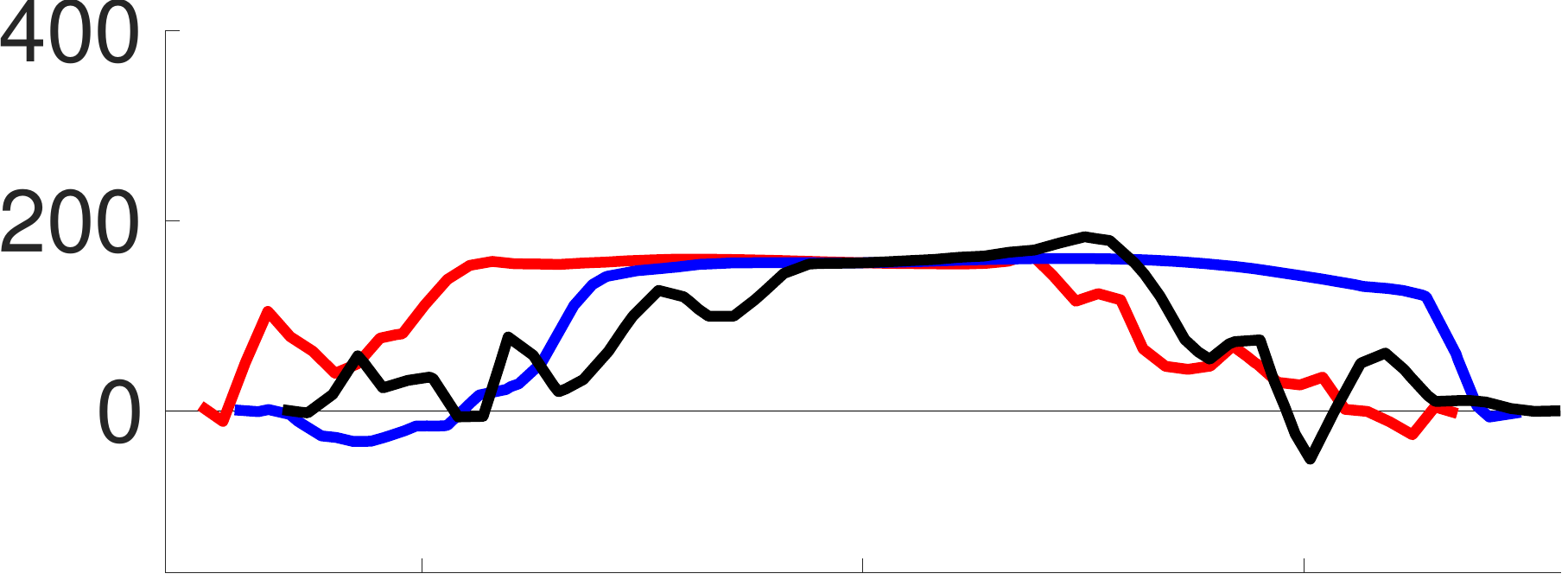} \\
\includegraphics[width=.21\textwidth]{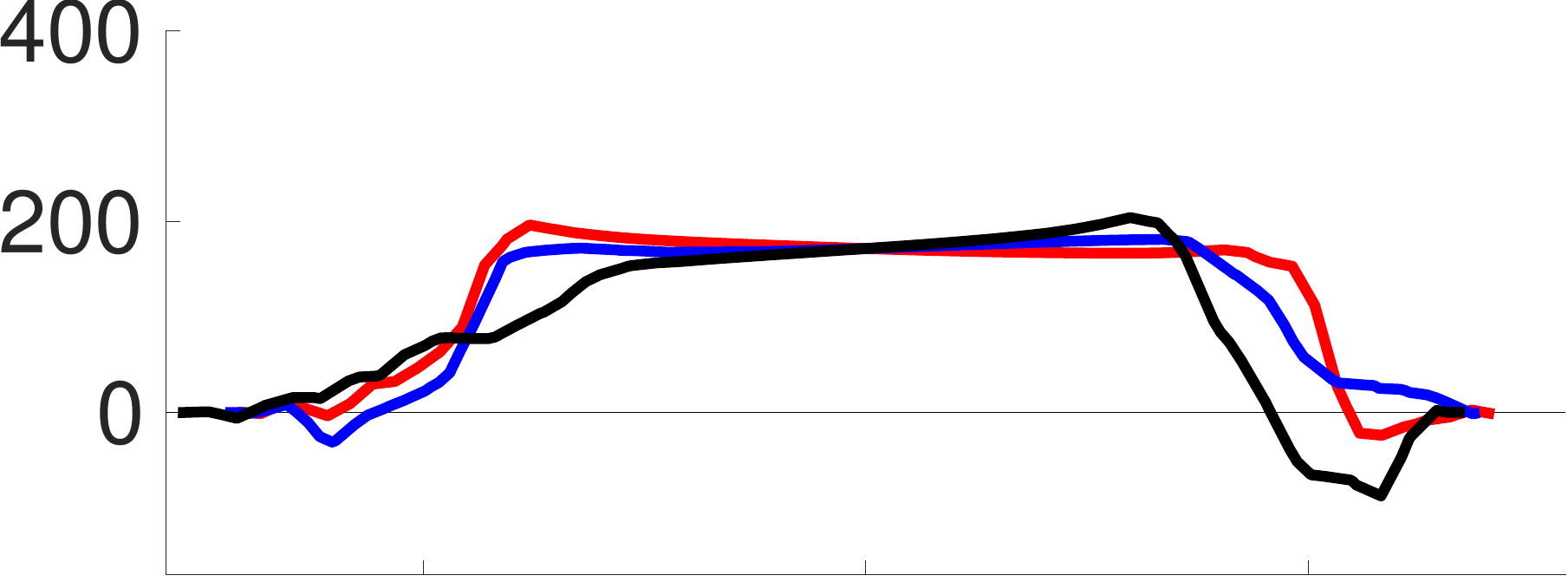} \\ 
\includegraphics[width=.21\textwidth]{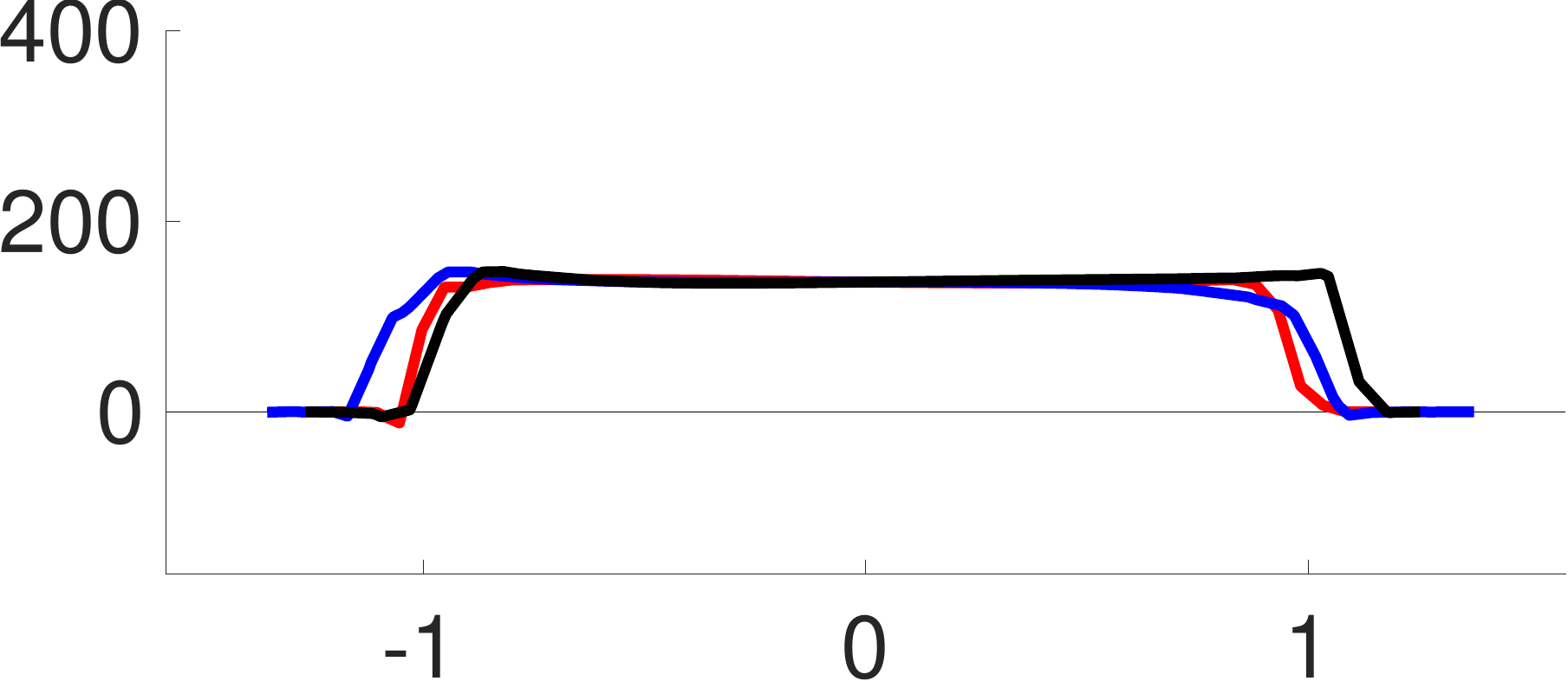} \\
{\small cm from center } 
\end{tabular} \vfill
&
\includegraphics[width=.26\textwidth]{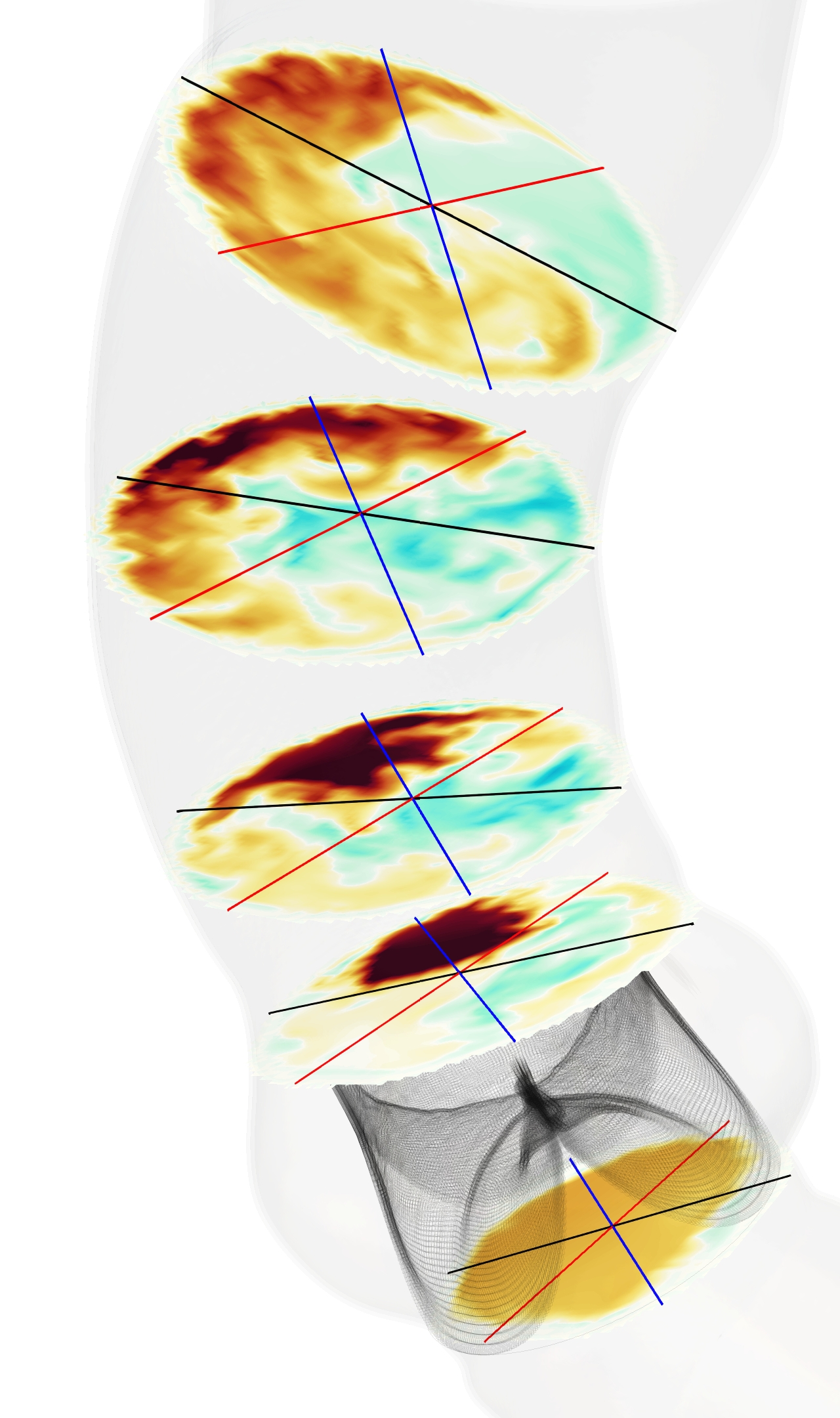} \vfill 
&
\hspace{-13pt} \rotatebox[origin=c]{90}{\small velocity (cm/s)} \vfill 
&
\begin{tabular}{c}
 \includegraphics[width=.21\textwidth]{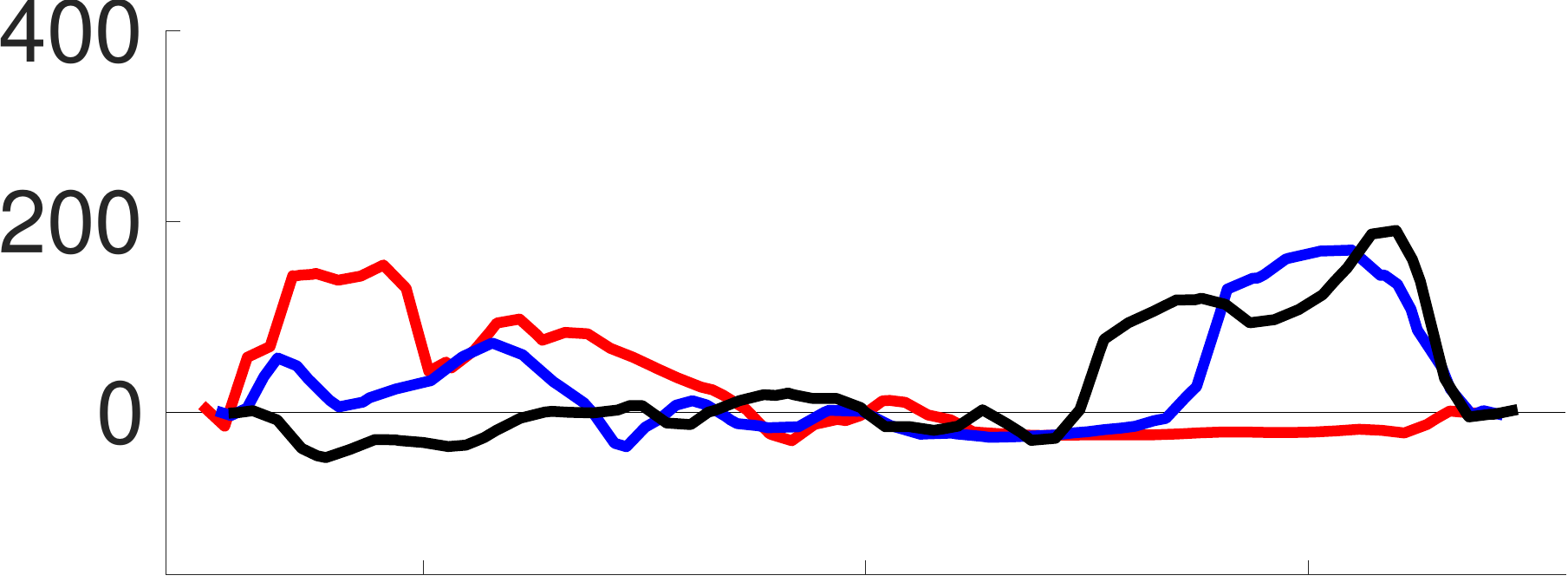} \\ 
 \includegraphics[width=.21\textwidth]{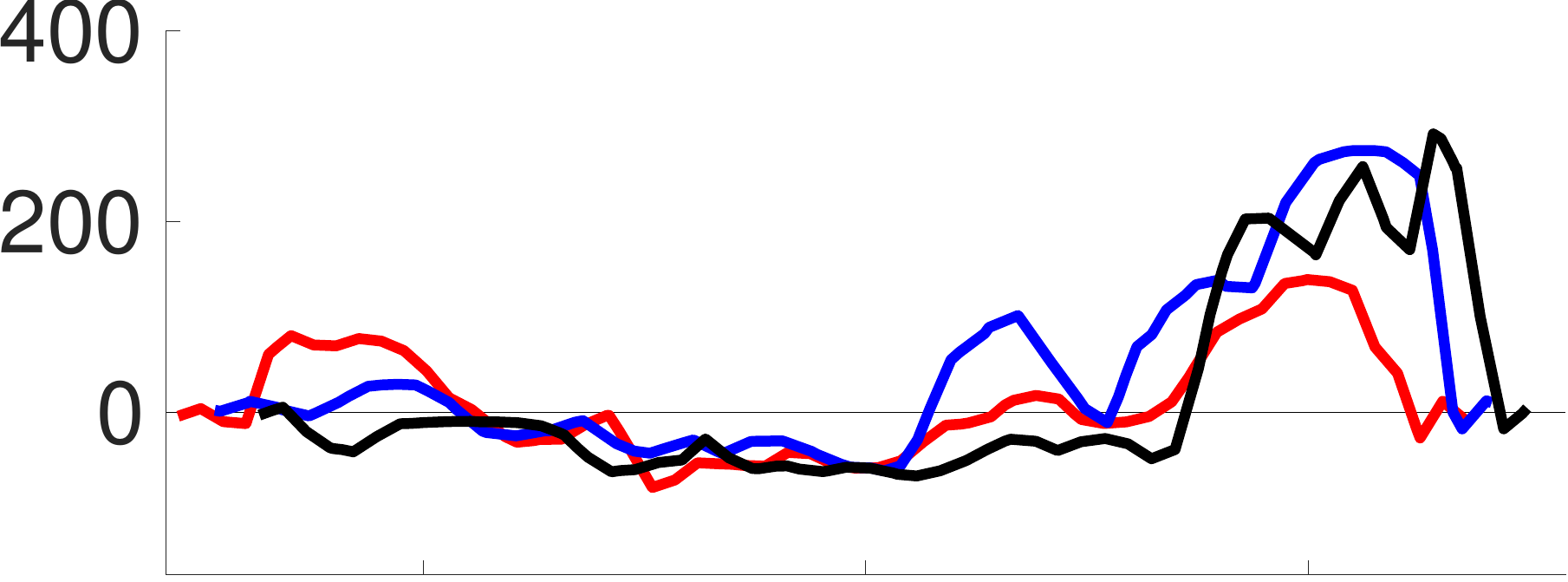} \\ 
 \includegraphics[width=.21\textwidth]{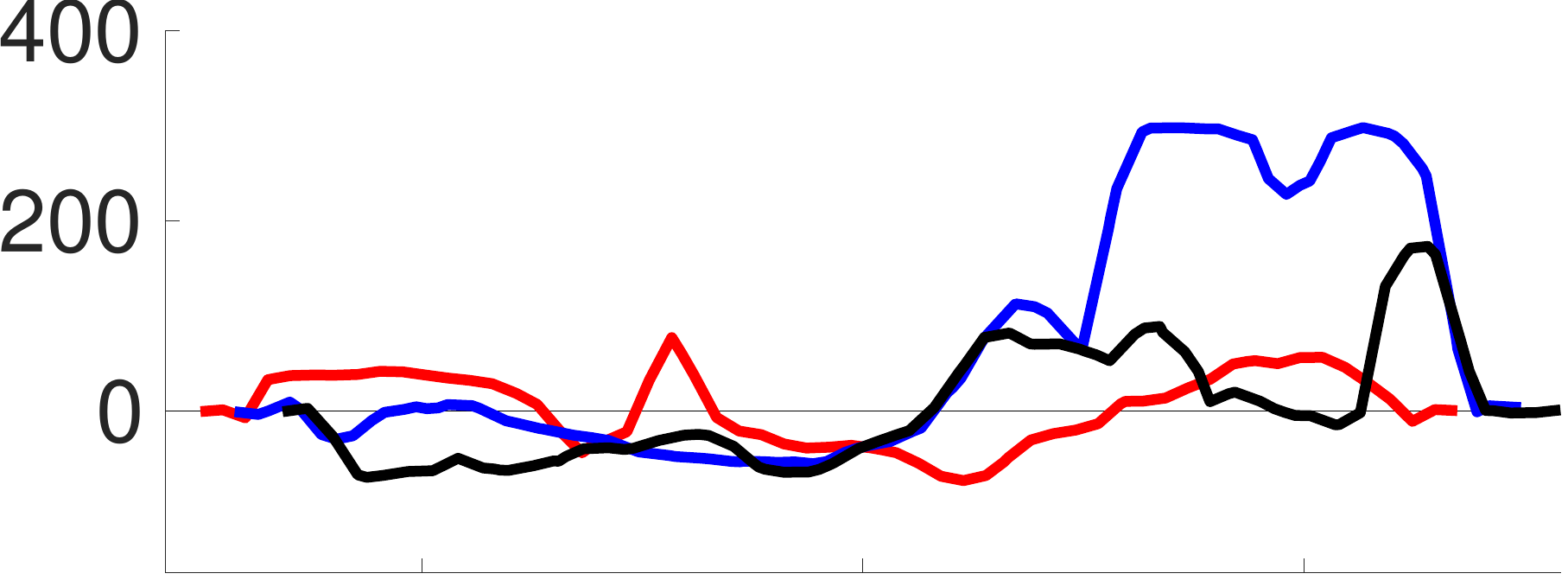} \\ 
 \includegraphics[width=.21\textwidth]{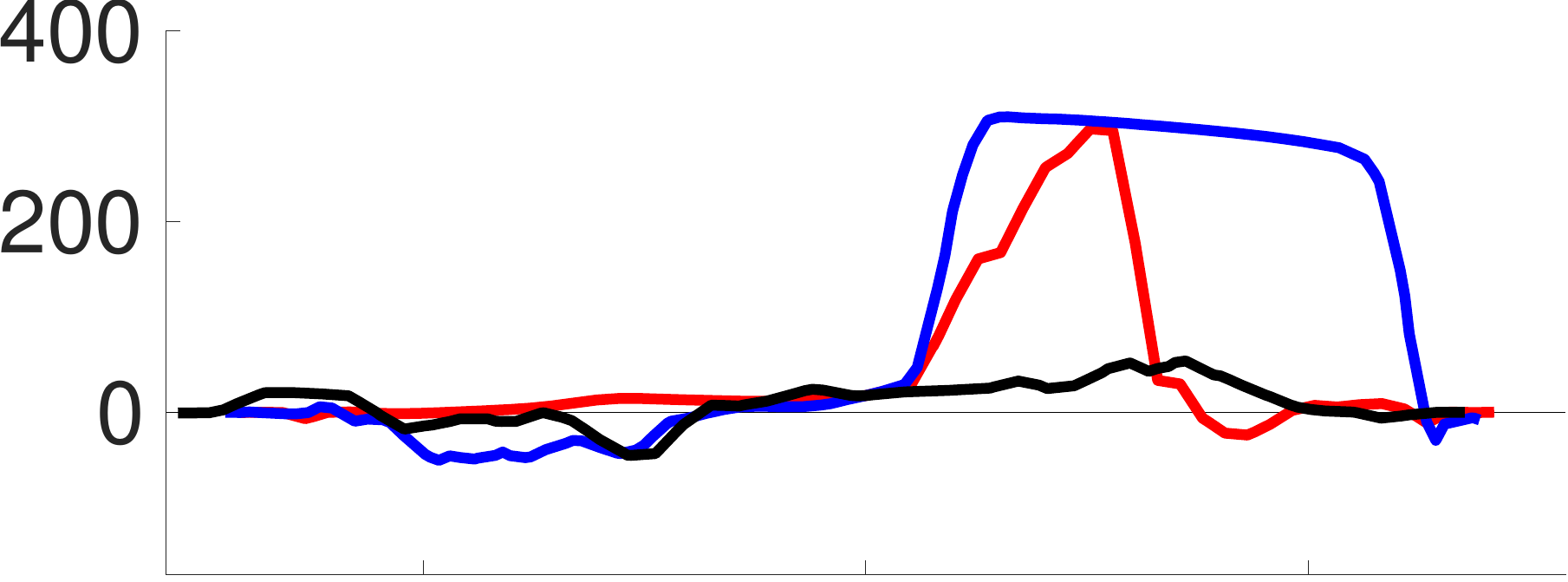} \\ 
 \includegraphics[width=.21\textwidth]{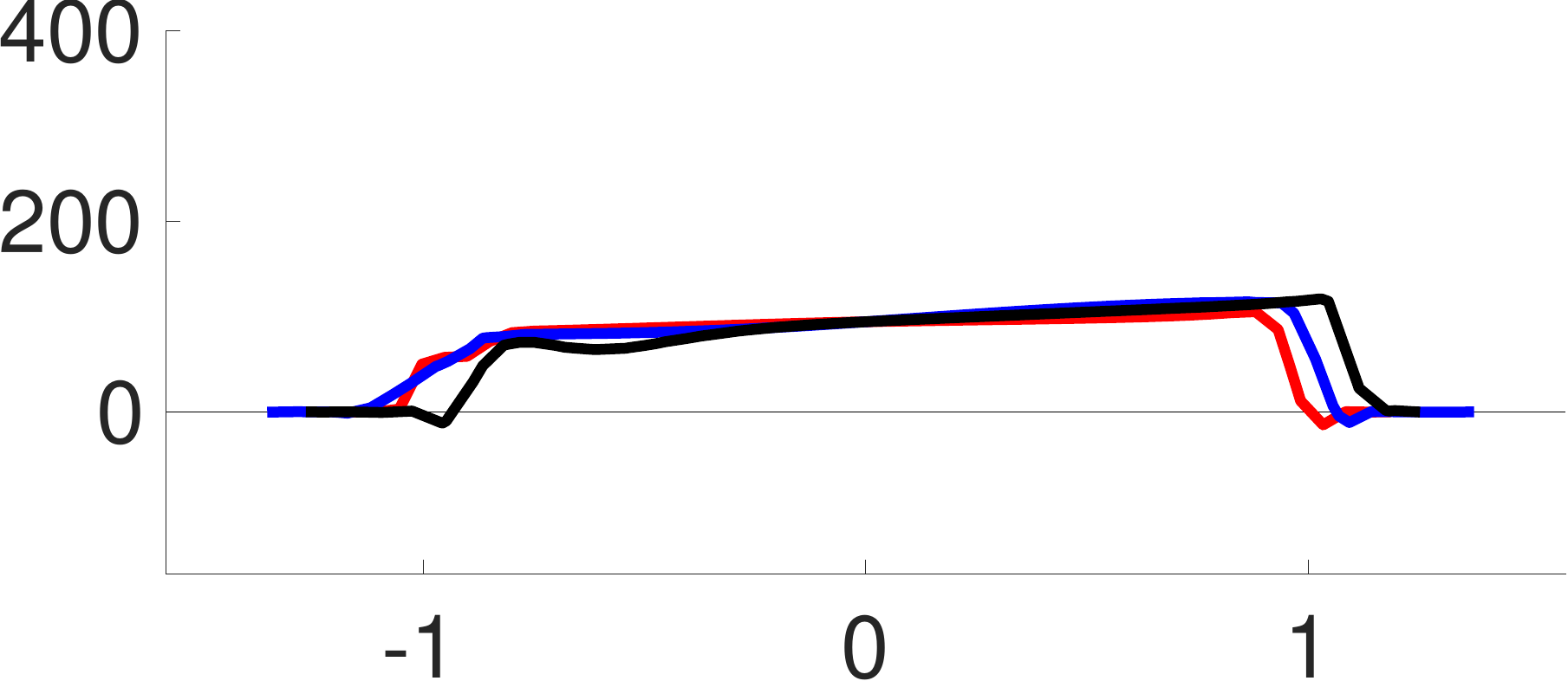} \\ 
{\small cm from center }
\end{tabular} \vfill 
\\
\multicolumn{3}{c}{ RC / NC fusion } & \multicolumn{3}{c}{ NC / LC fusion } \\  
\includegraphics[width=.26\textwidth]{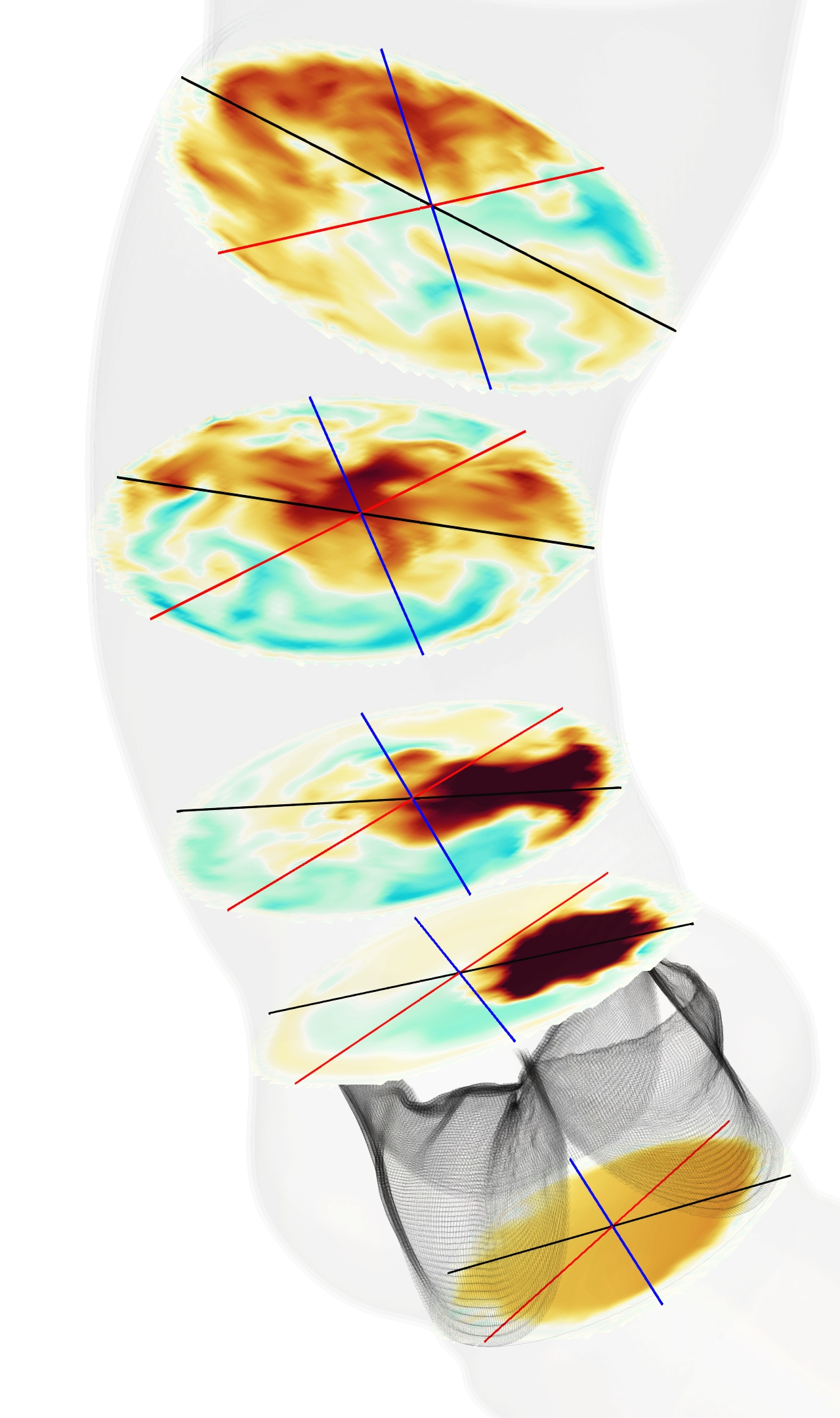} \vfill 
&
\hspace{-13pt} \rotatebox[origin=c]{90}{\small velocity (cm/s)} \vfill  
&
\begin{tabular}{c}
\includegraphics[width=.21\textwidth]{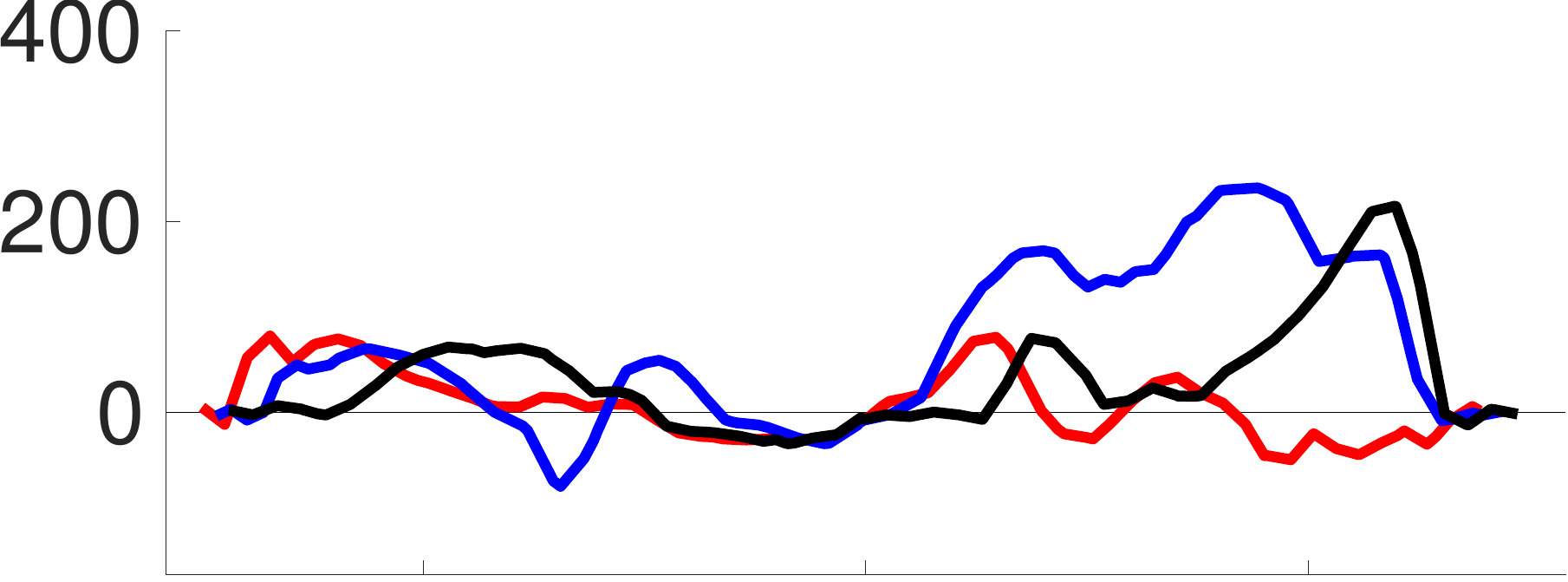} \\
\includegraphics[width=.21\textwidth]{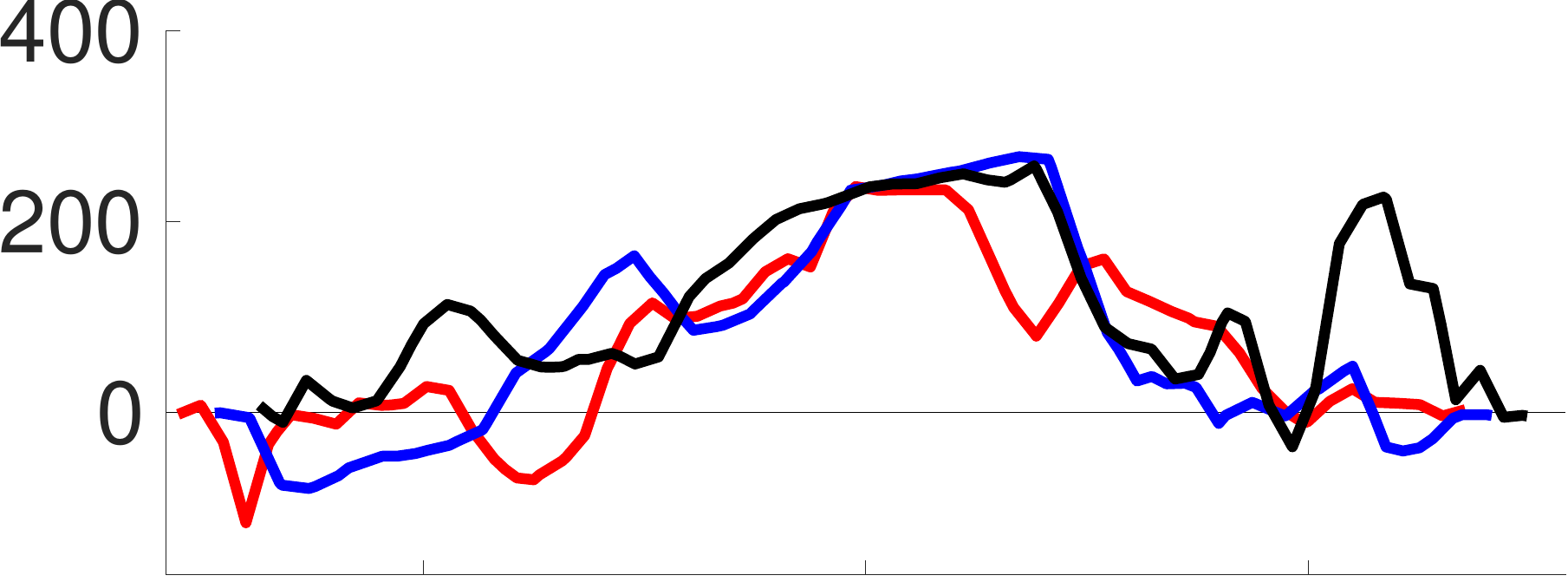} \\
\includegraphics[width=.21\textwidth]{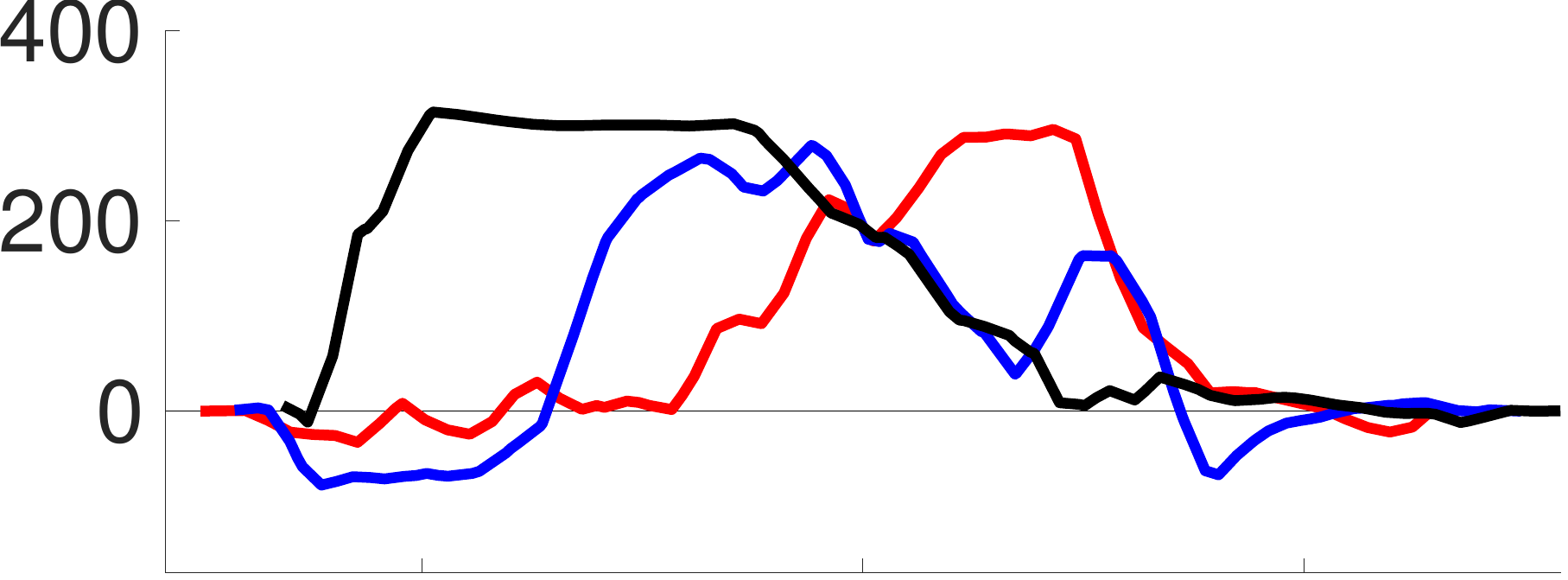} \\
\includegraphics[width=.21\textwidth]{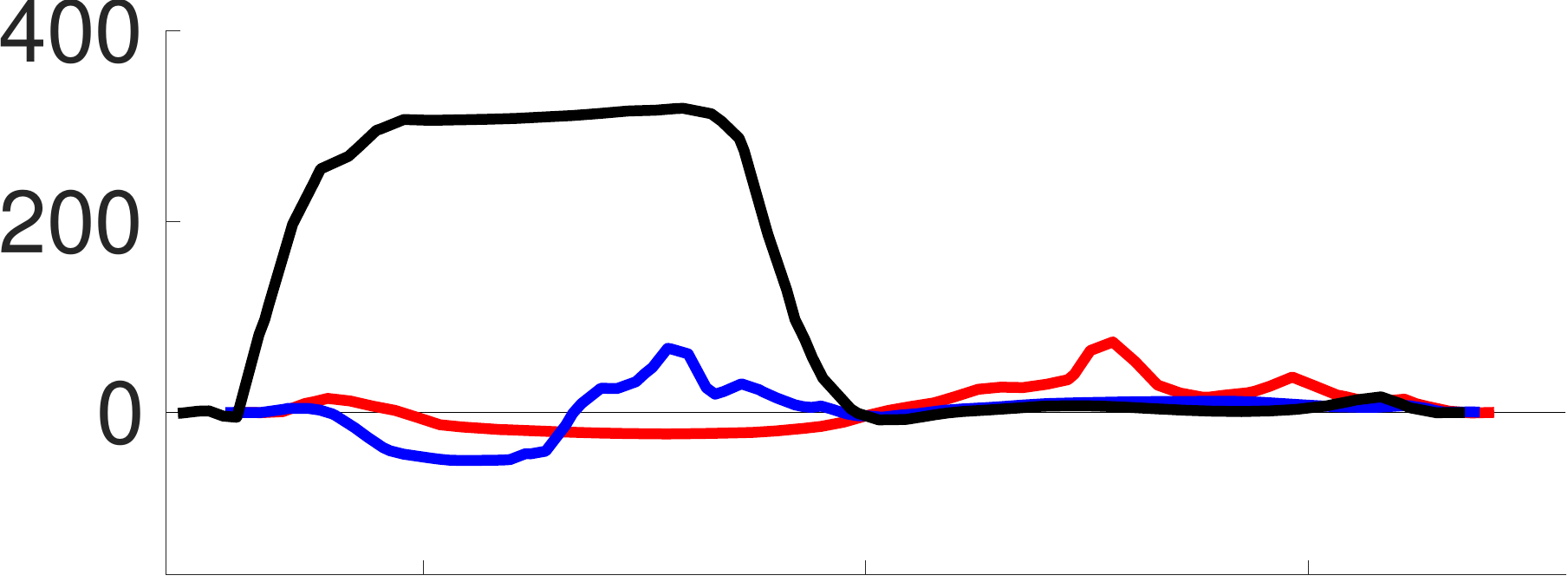} \\
\includegraphics[width=.21\textwidth]{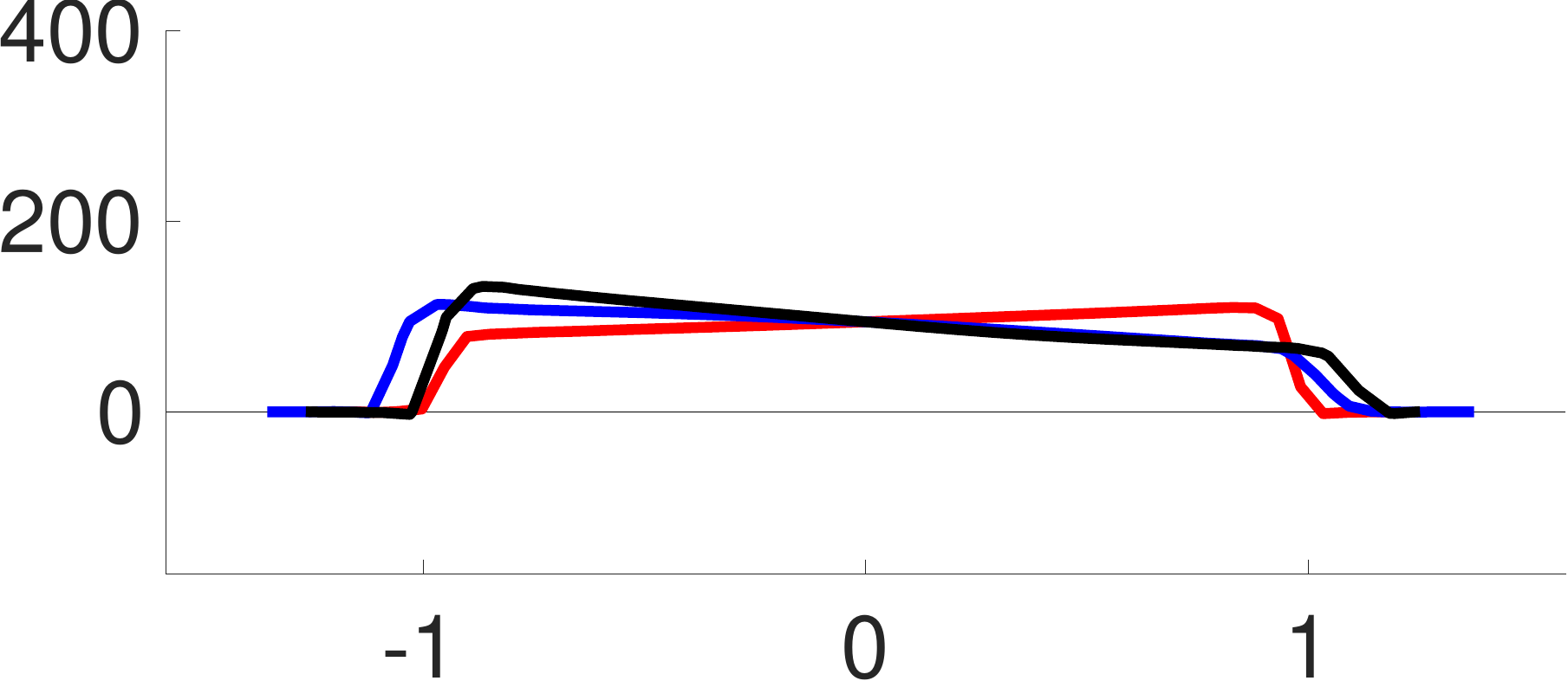} \\
{\small cm from center }
\end{tabular} \vfill 
&
\includegraphics[width=.26\textwidth]{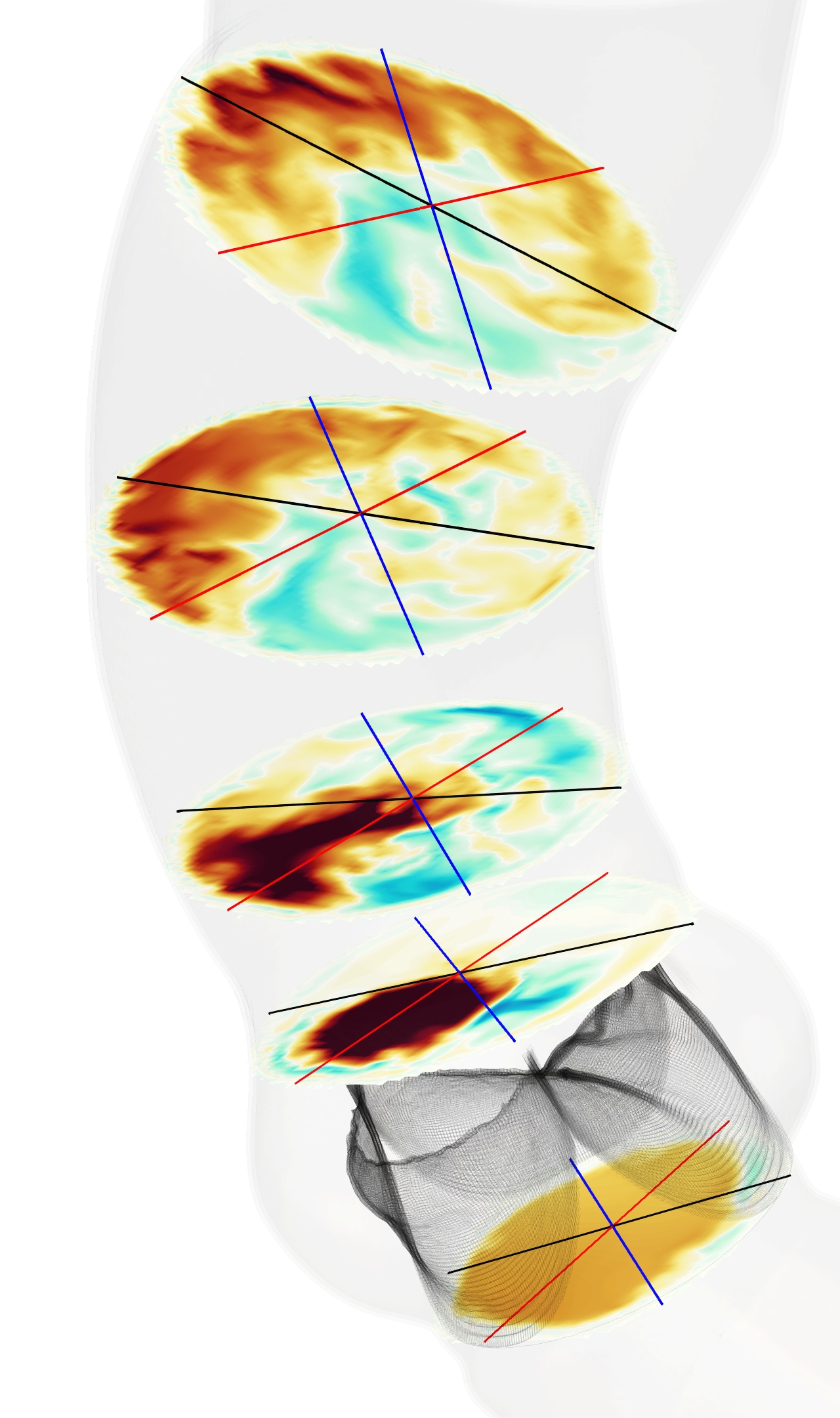} \vfill 
&
\hspace{-13pt} \rotatebox[origin=l]{90}{\small velocity (cm/s)} \vfill 
&
\begin{tabular}{c}
 \includegraphics[width=.21\textwidth]{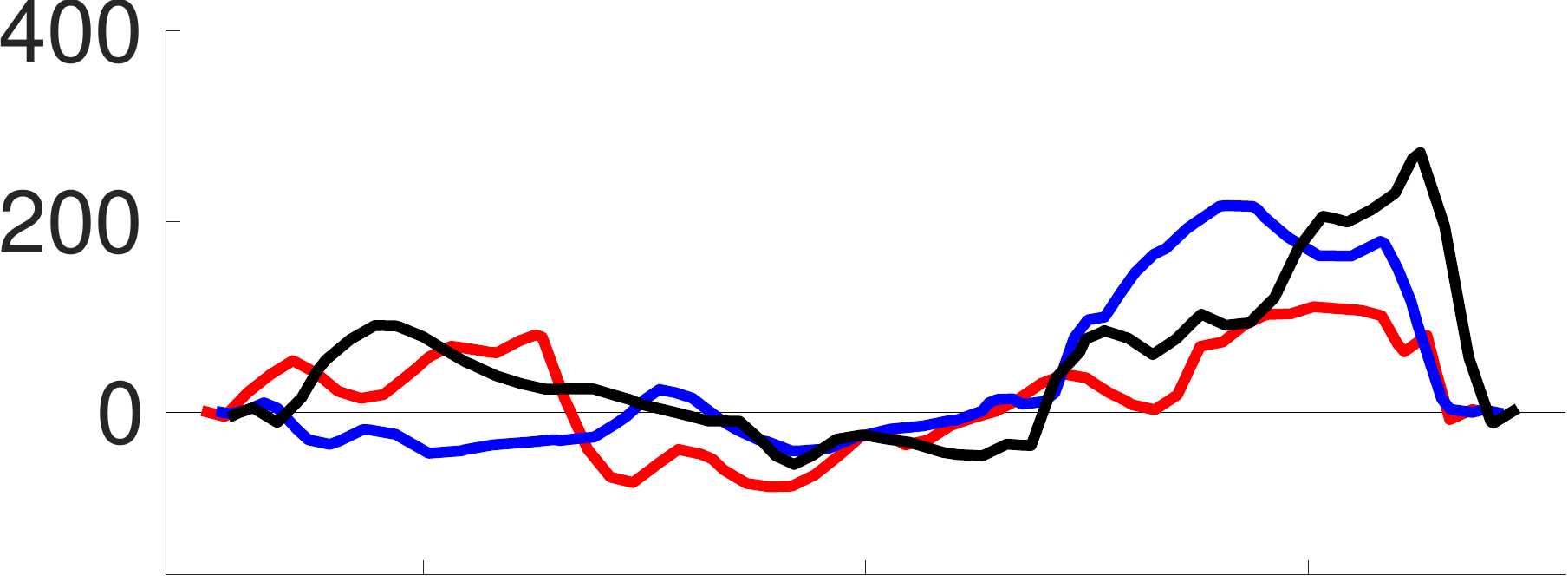} \\ 
 \includegraphics[width=.21\textwidth]{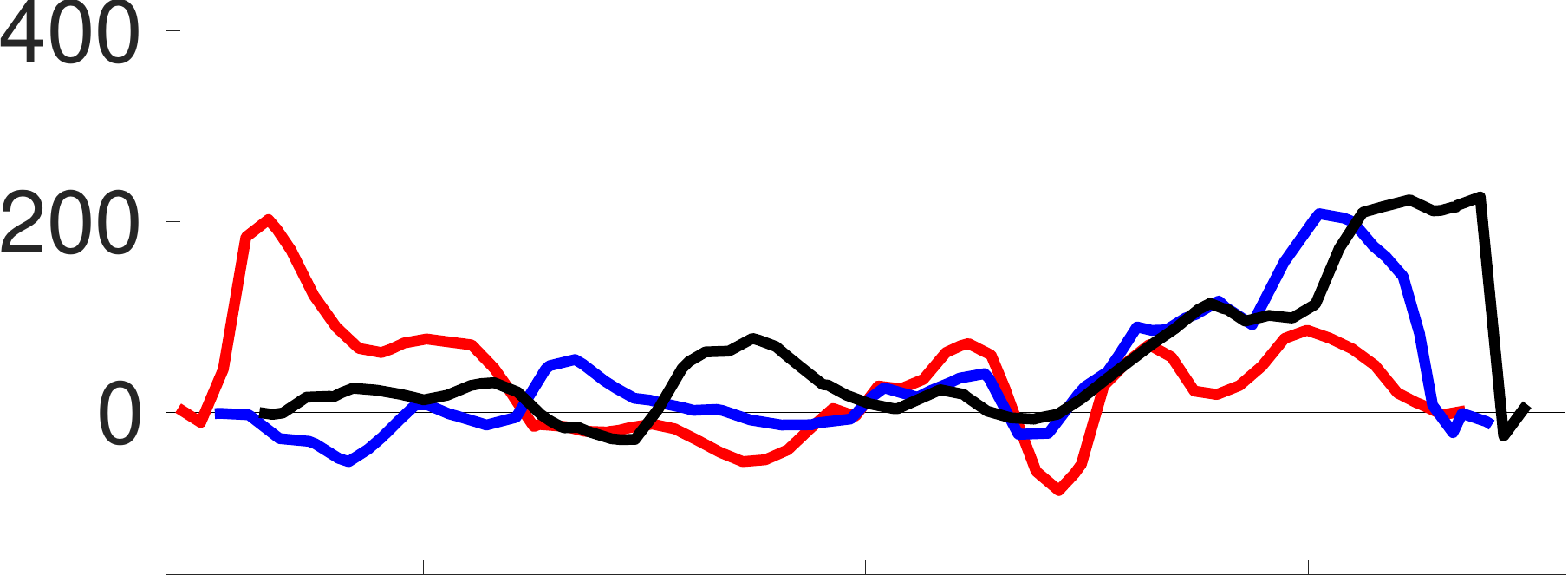} \\ 
 \includegraphics[width=.21\textwidth]{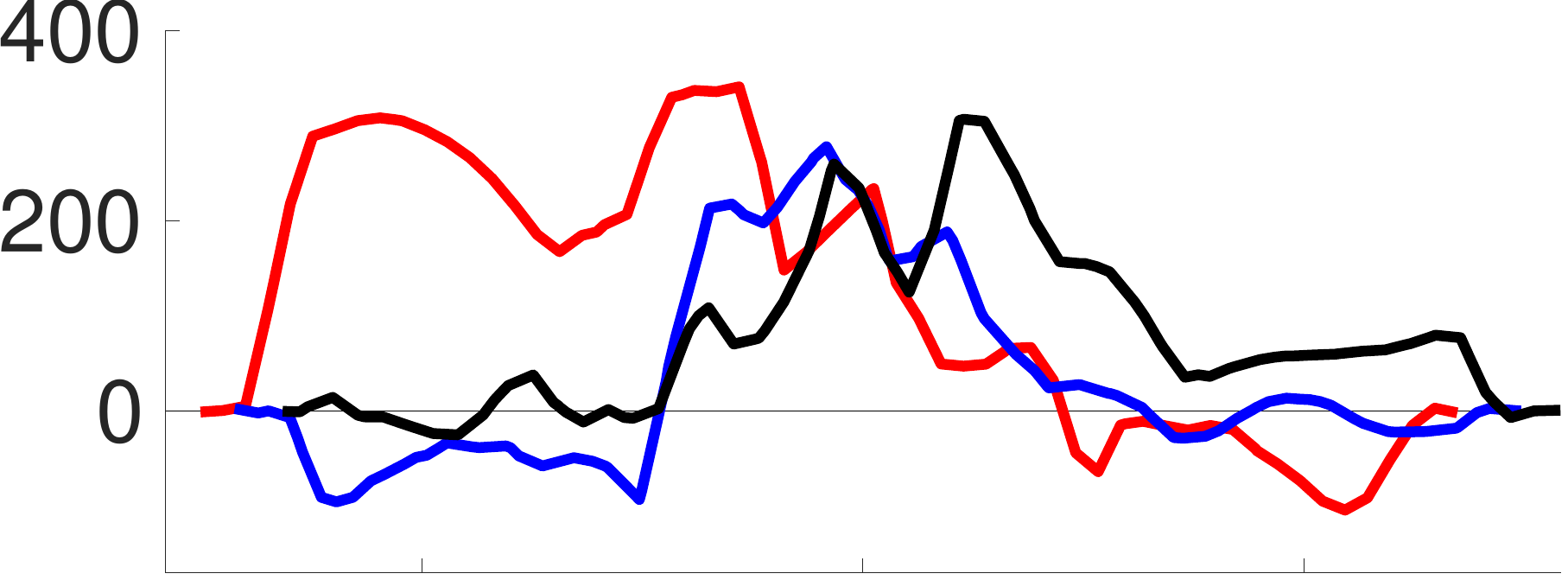} \\ 
 \includegraphics[width=.21\textwidth]{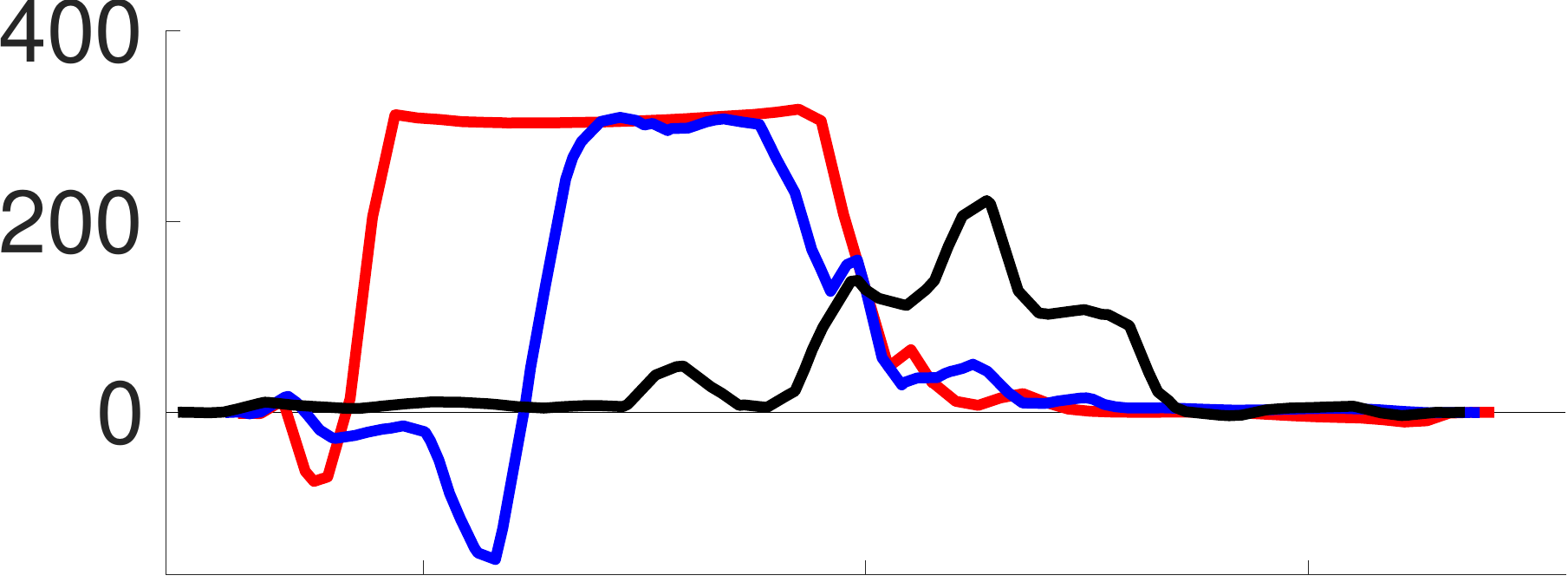} \\ 
 \includegraphics[width=.21\textwidth]{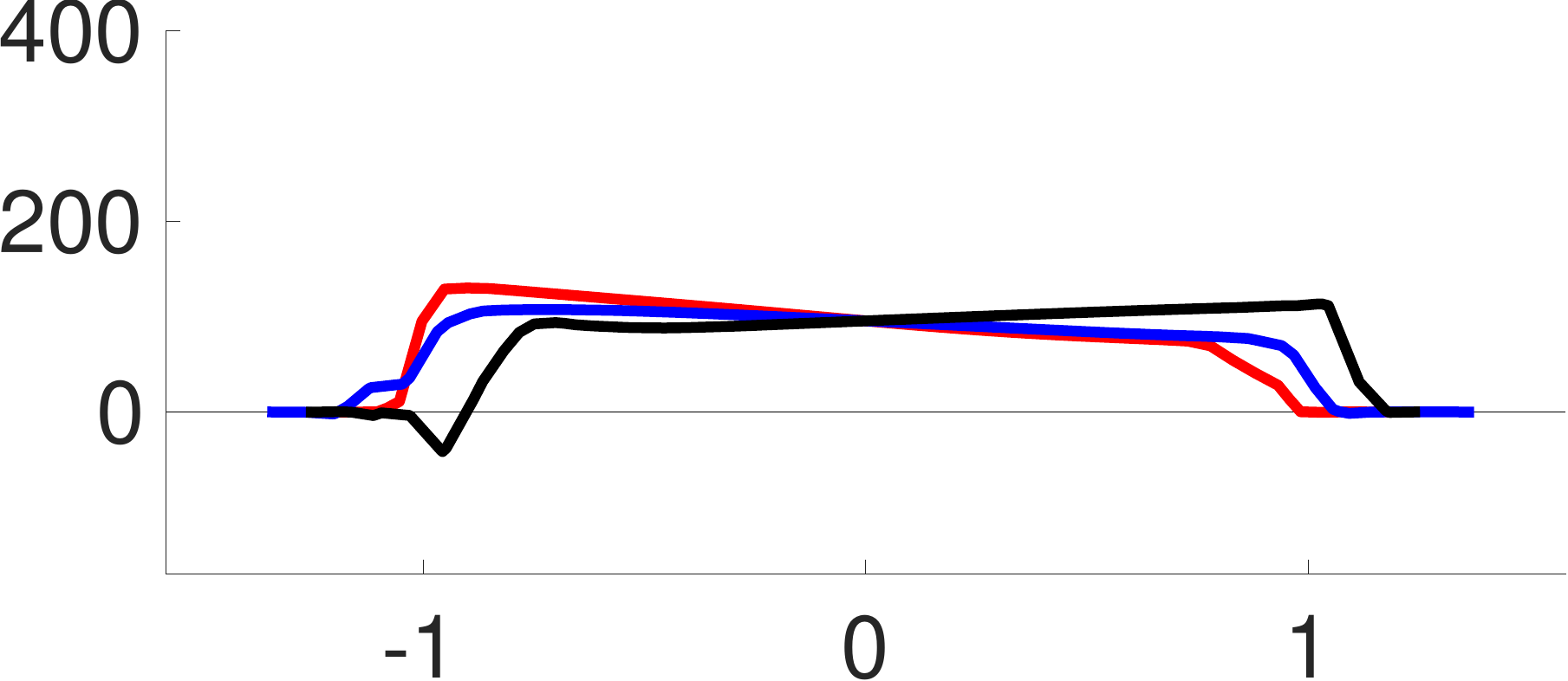} \\
{\small cm from center }
\end{tabular} \vfill 
\end{tabular}
\caption{Line plots of the normal component of velocity at mid systole ($t = 2.271 $ s). 
The blue, black and red lines are aligned at the annulus with center of the non-coronary, left-coronary, right-coronary leaflets, respectively. 
Normal velocity is shown with the same scale as in Figure \ref{flow_panels}. }
\label{line_plots}
}
\end{figure*}

Line plots of velocity for all cases are shown in Figure \ref{line_plots}, depicting the velocity normal to the slices along three lines through the center of each slice, enabling direct comparison. 
In the tricuspid case, at the annulus the flow was plug-like. 
Immediately downstream of the valve, the center of the jet had nearly constant velocity. 
Some recirculation occurred along the edge of the jet, particularly along the greater curvature at the sinotubular junction (slice 2).
Further downstream, trace local reverse flow appeared, especially on the lesser curvature.
The normal velocity later became somewhat uniform in the ascending aorta, as the wide jet spread throughout the vessel.

For all three bicuspid valves, flow proximal to the valve was less symmetric. 
Distal to the valve, this asymmetry became pronounced.
There was a localized plug-like jet immediately distal to the valve, corresponding to the unrestricted leaflet in each case.
The line plots for downstream slices capture how the jet dispersed and spread along the outer wall in each case, creating secondary and reverse flows.

\subsection{Integral metrics}

\begin{figure*}[t!] 
\hfill 
\includegraphics[width=.12\textwidth]{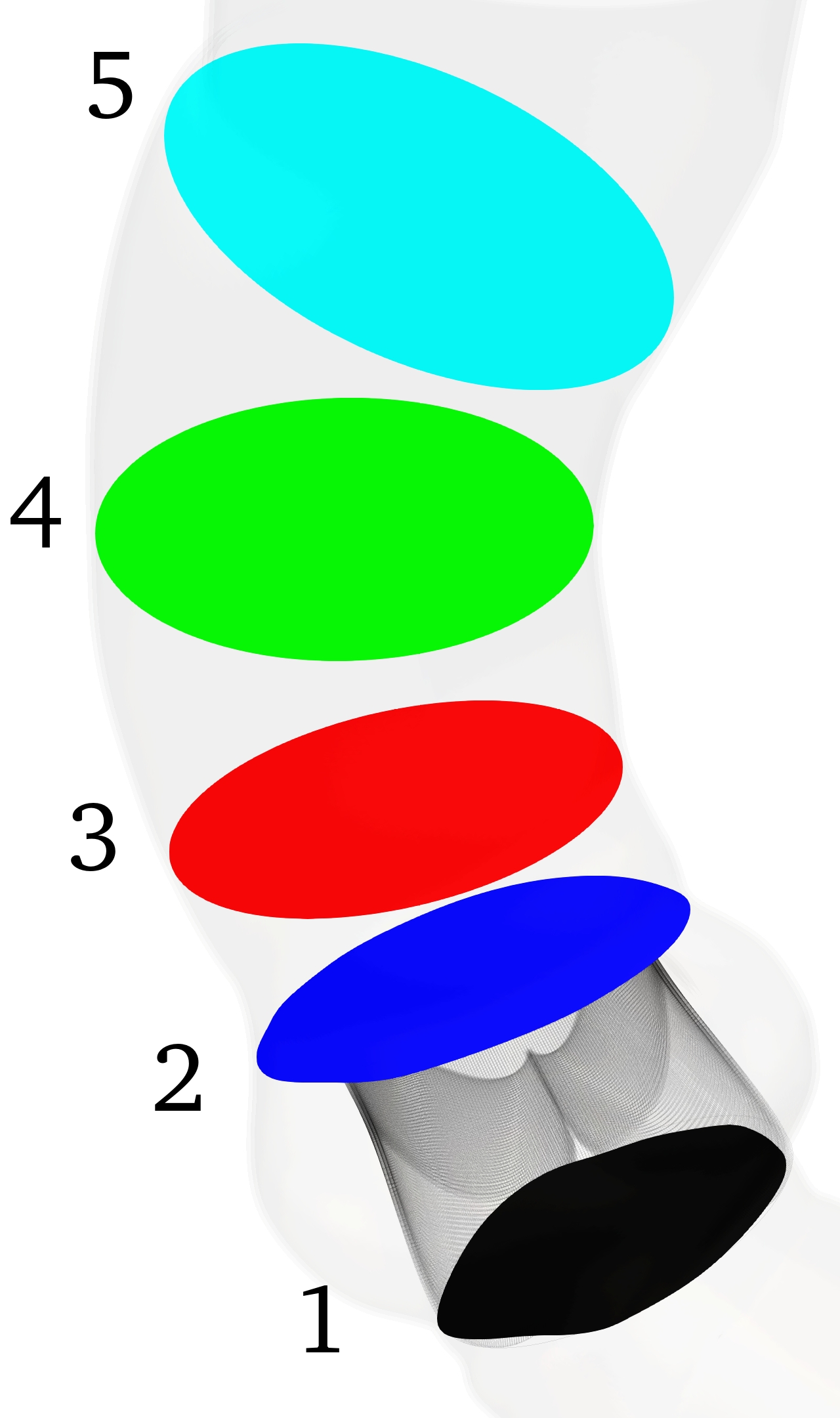} 
\raisebox{20pt}{\includegraphics[width=.15\textwidth]{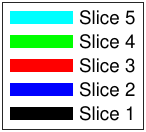}}  \\ 
\setlength{\tabcolsep}{0.0pt}        
\begin{tabular}{ M{.03\textwidth} M{.24\textwidth} M{.24\textwidth} M{.24\textwidth} M{.24\textwidth}}        
&
 tricuspid &
LC/RC fusion & 
RC/NC fusion & 
NC/LC fusion \\ 
 $I_{1}$ & 
\includegraphics[width=.23\textwidth]{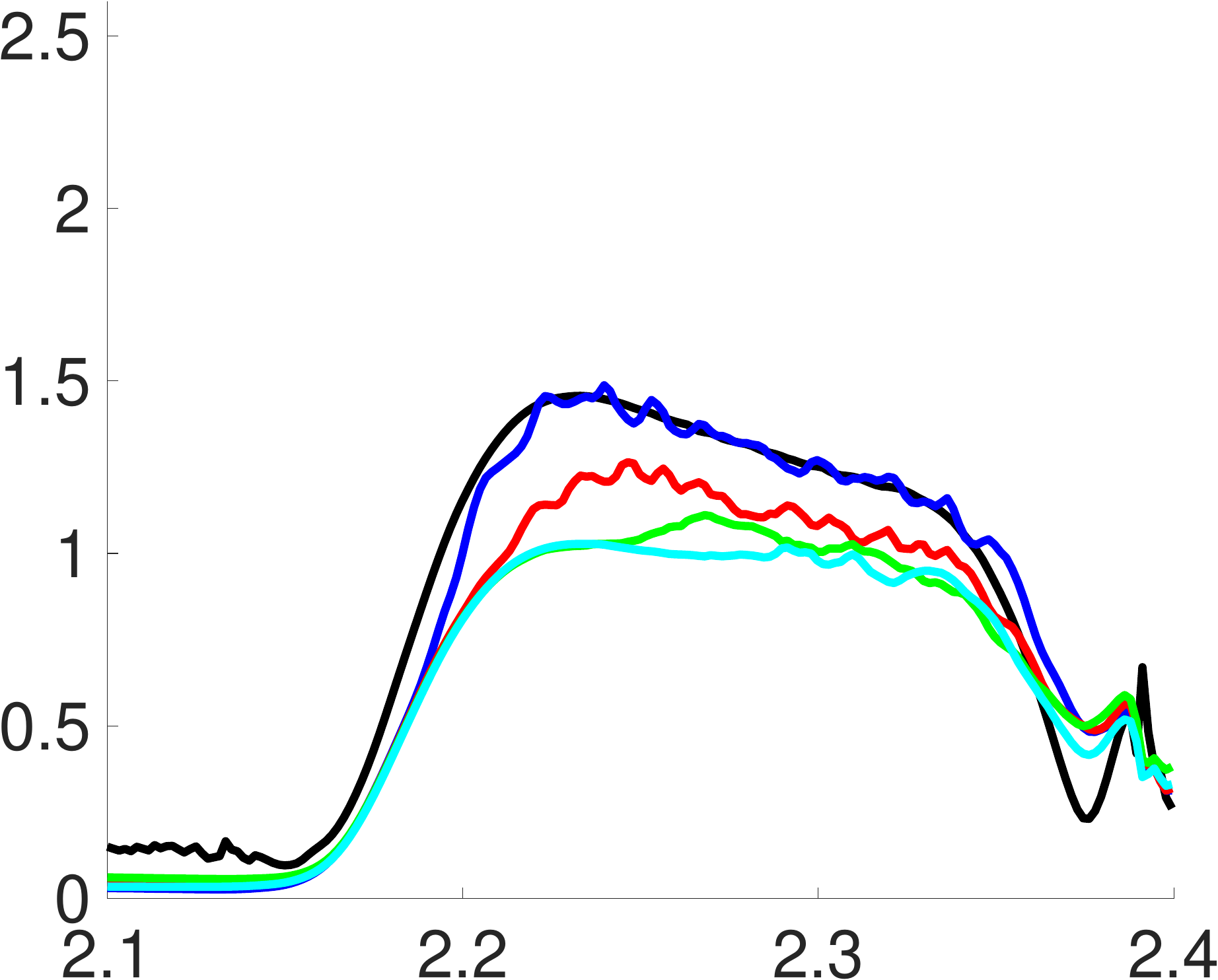} & 
\includegraphics[width=.23\textwidth]{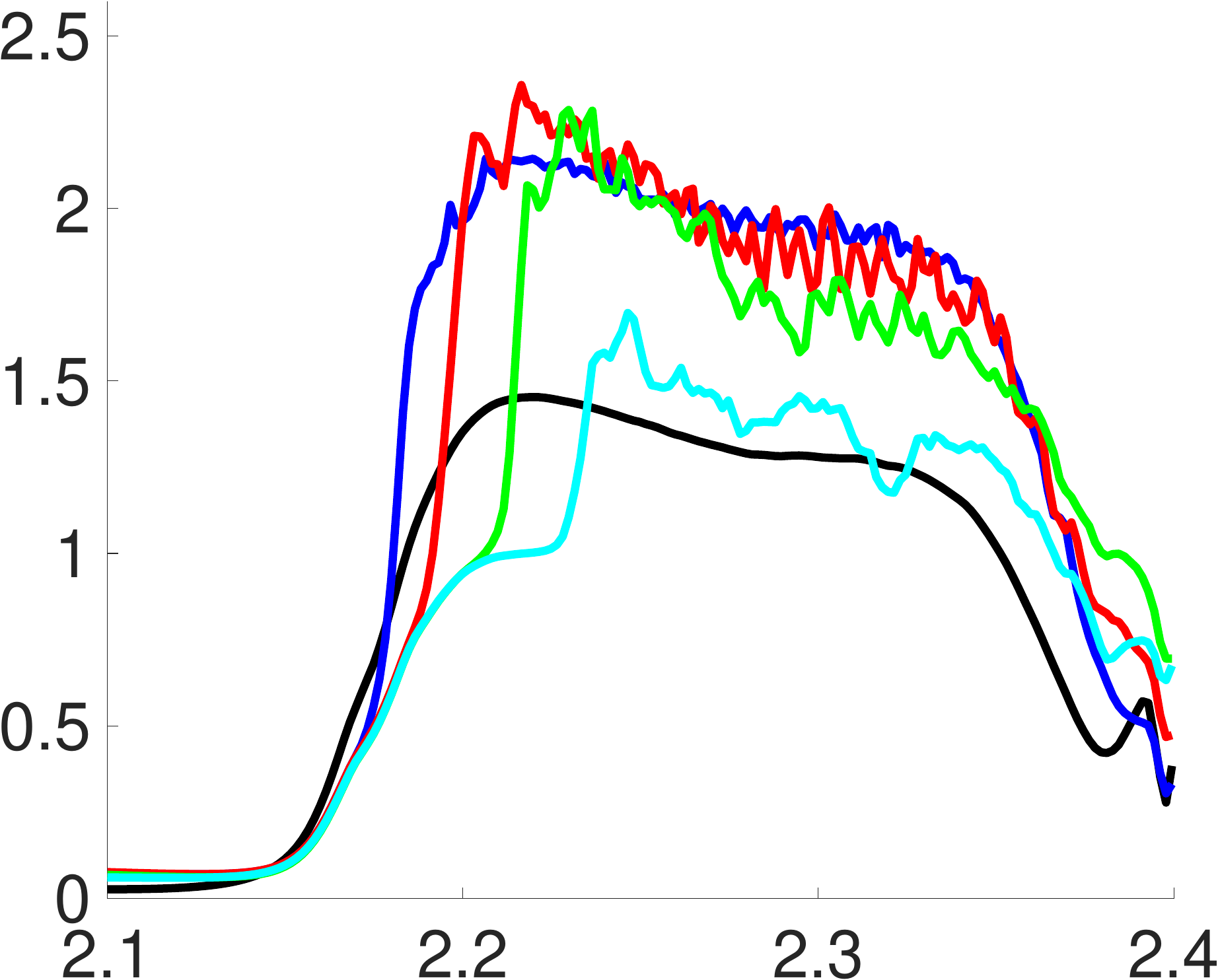} &
\includegraphics[width=.23\textwidth]{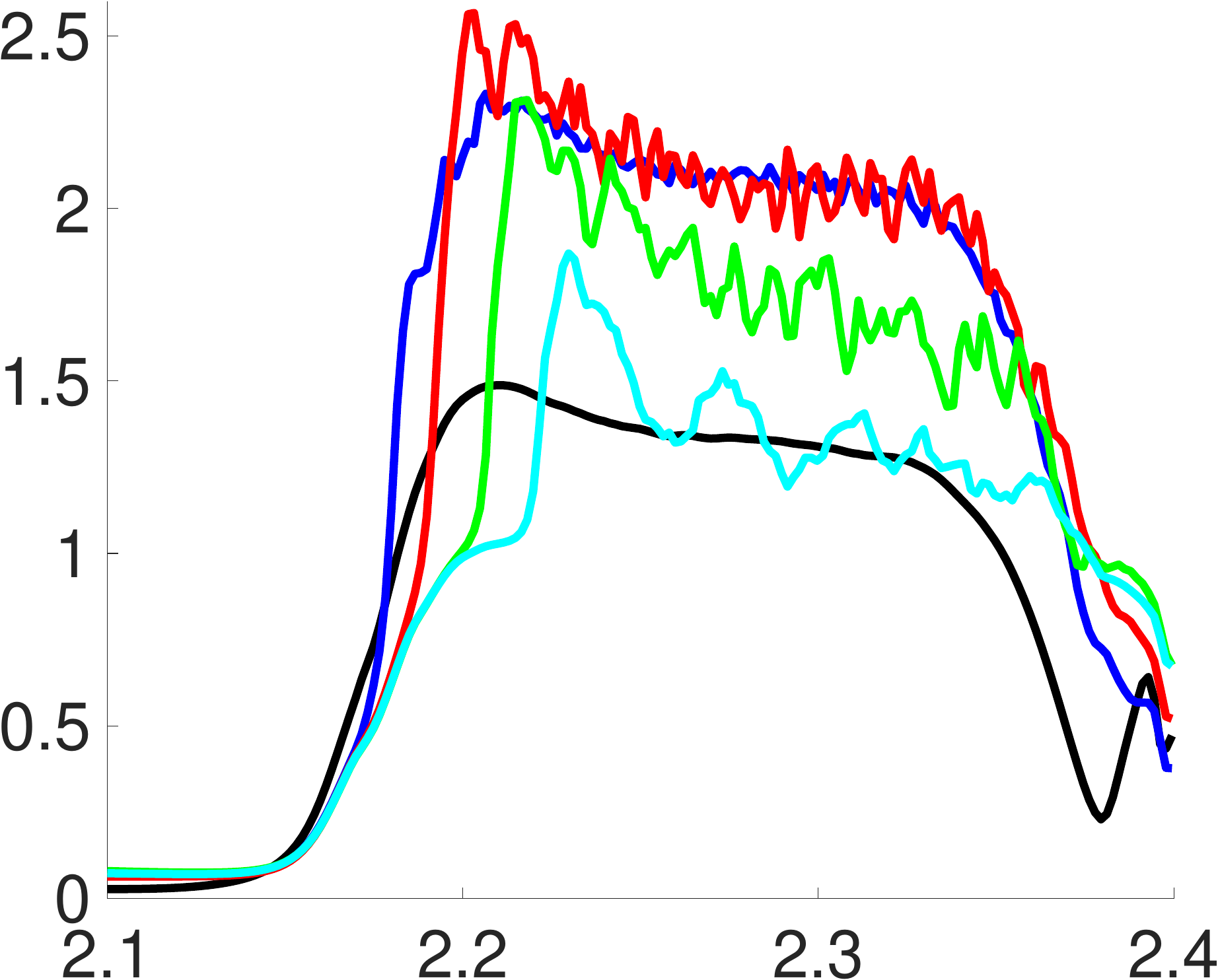} & 
\includegraphics[width=.23\textwidth]{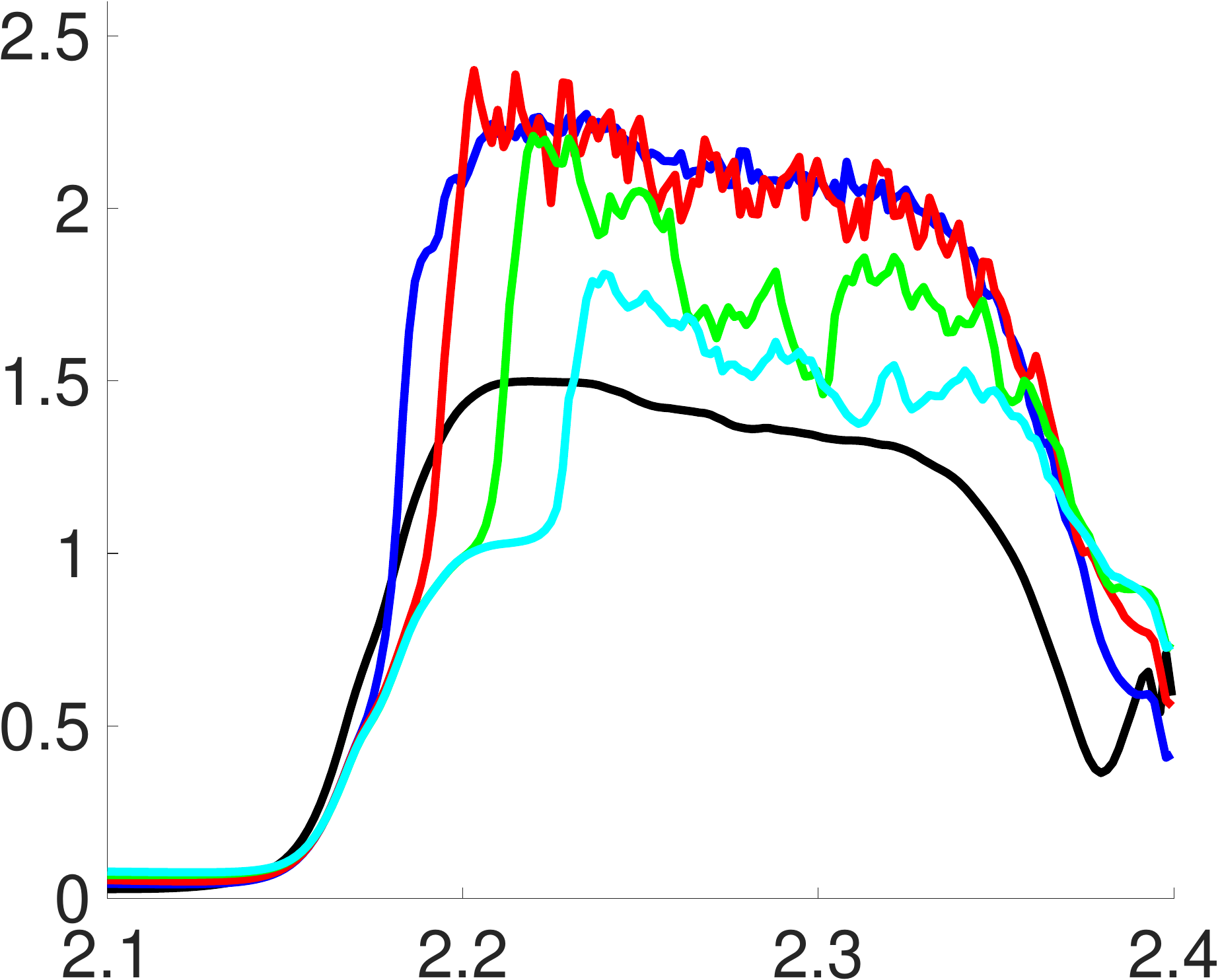} \\ \addlinespace[4.0pt]
$I_{2}$ & 
\includegraphics[width=.23\textwidth]{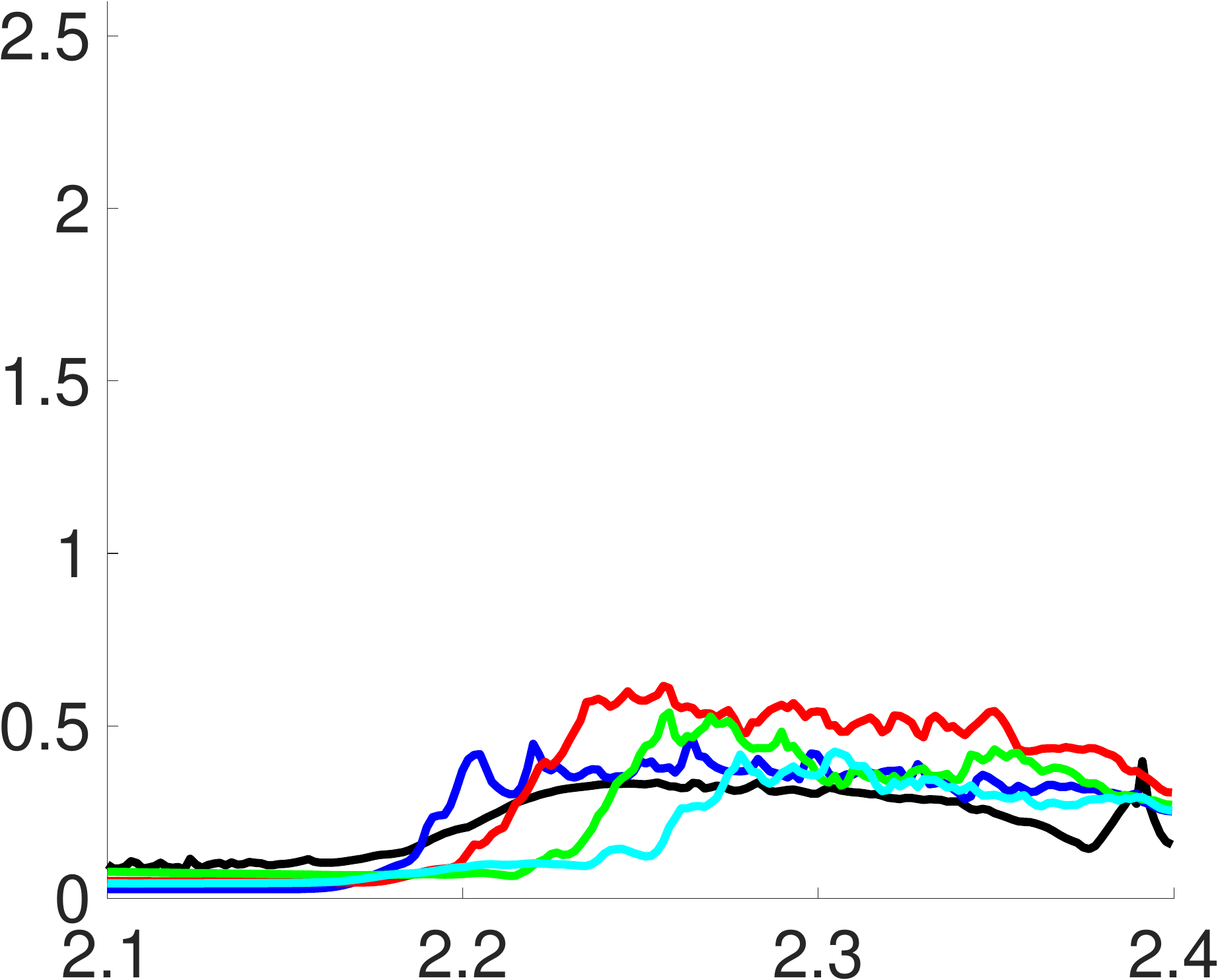} & 
\includegraphics[width=.23\textwidth]{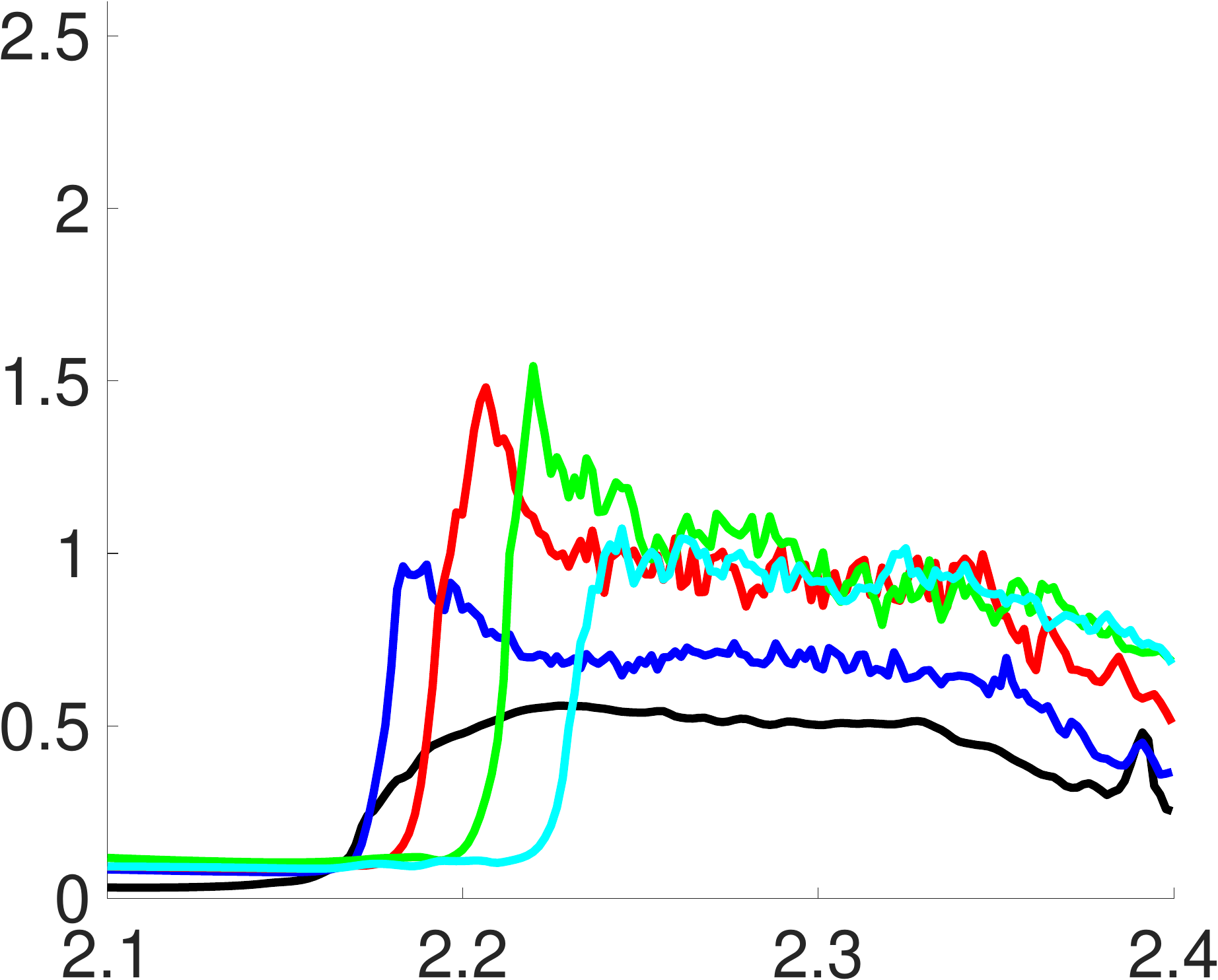} &
\includegraphics[width=.23\textwidth]{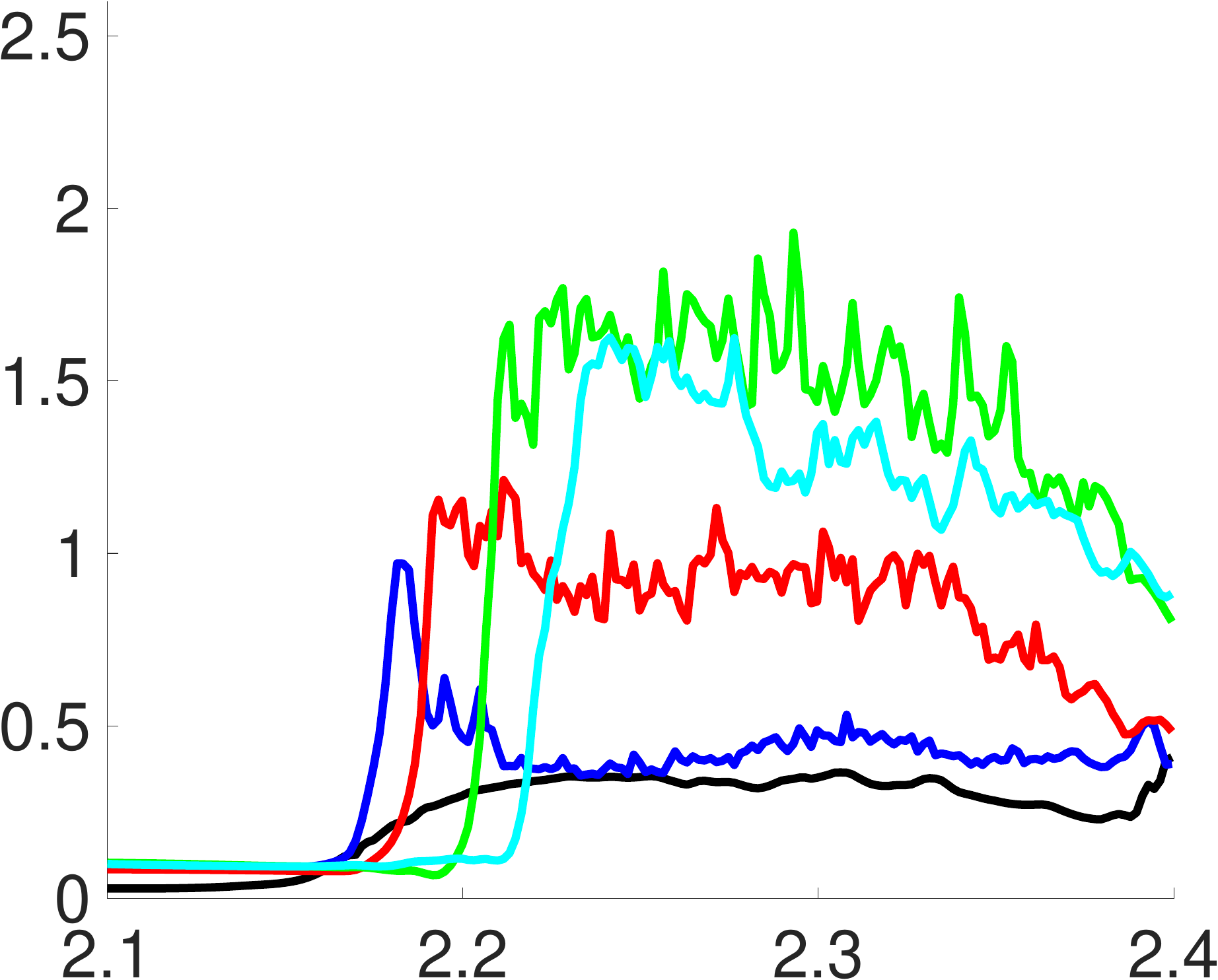} &
\includegraphics[width=.23\textwidth]{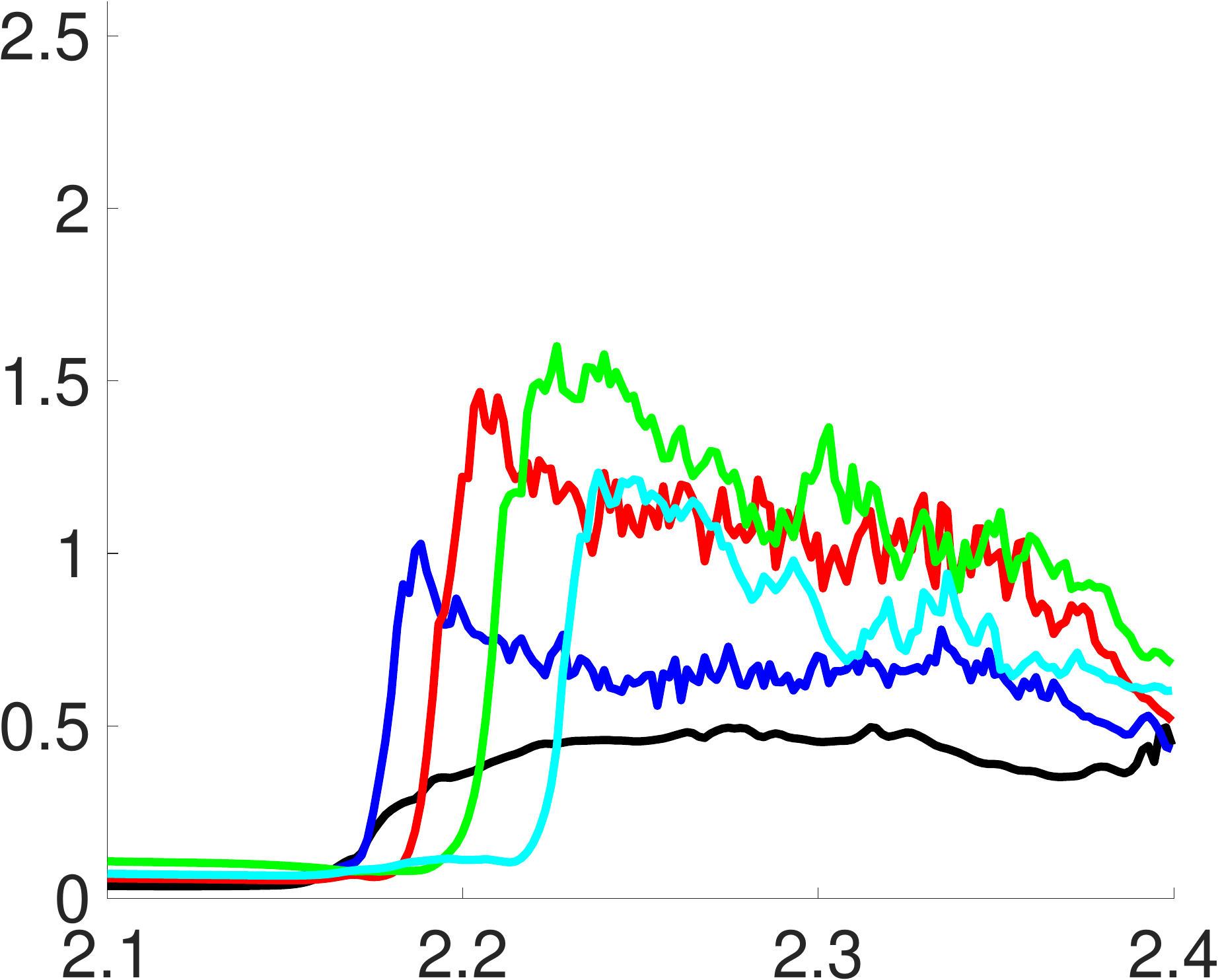} \\ \addlinespace[4.0pt]
$I_{R}$ & 
\includegraphics[width=.23\textwidth]{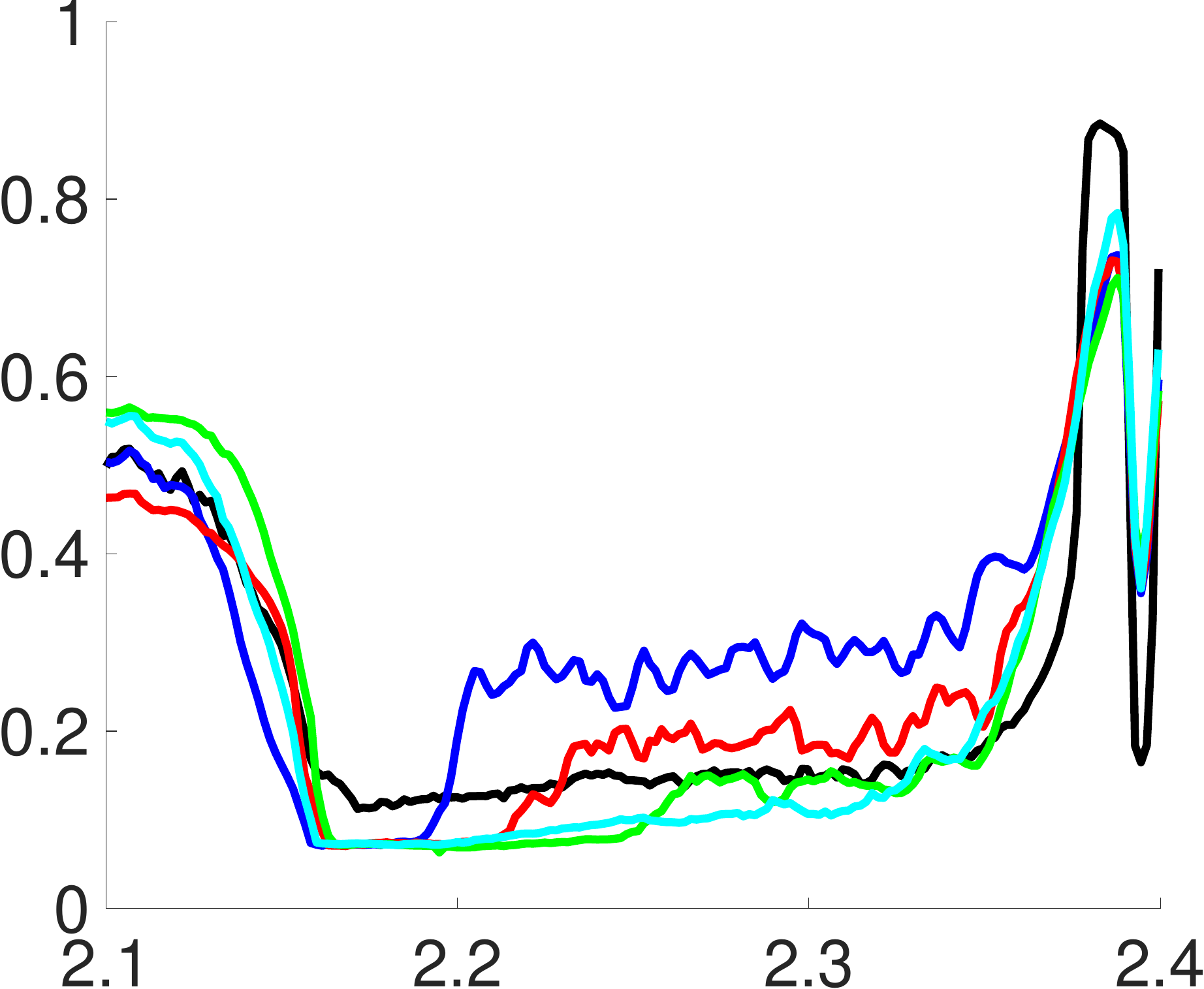} & 
\includegraphics[width=.23\textwidth]{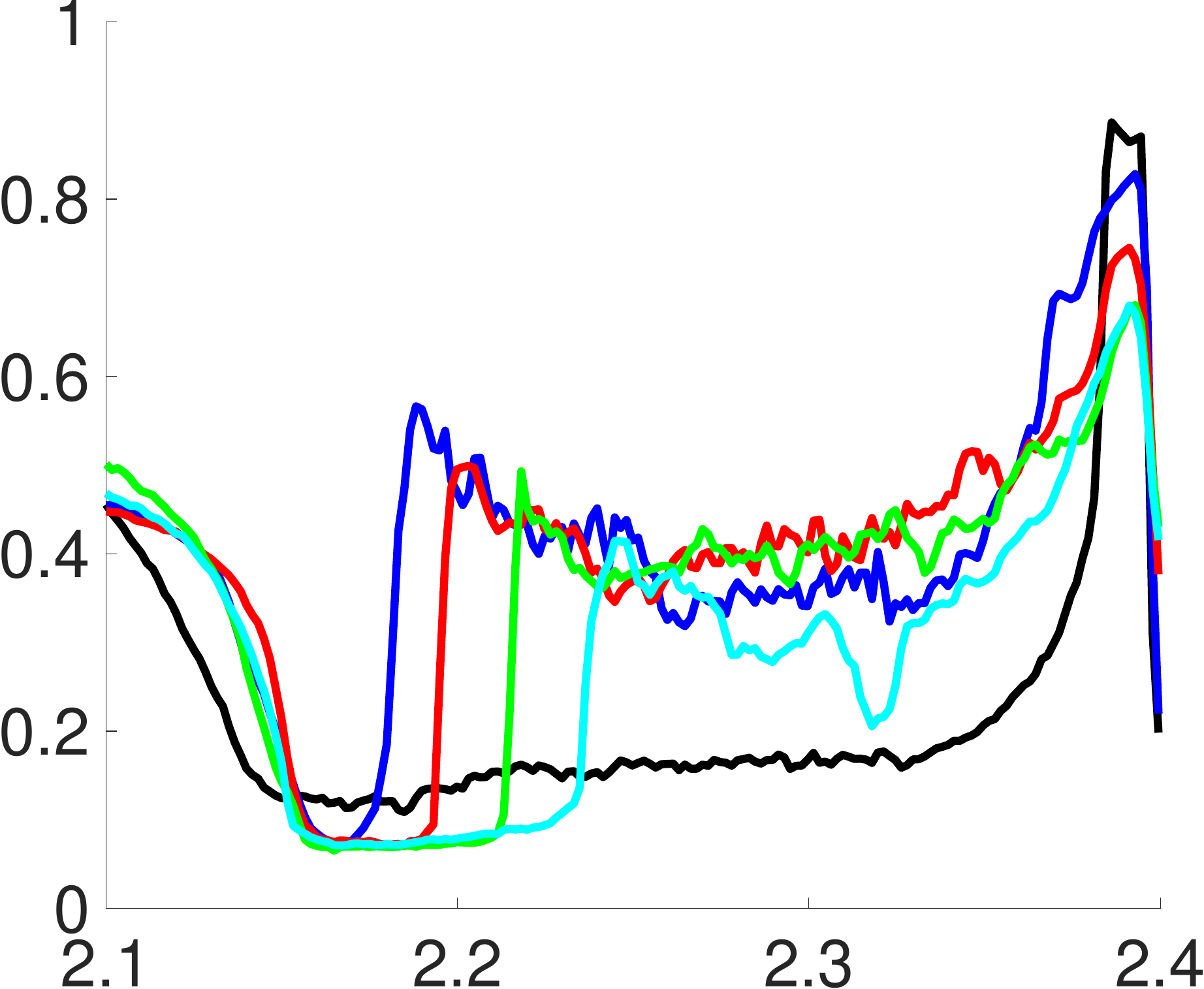} &
\includegraphics[width=.23\textwidth]{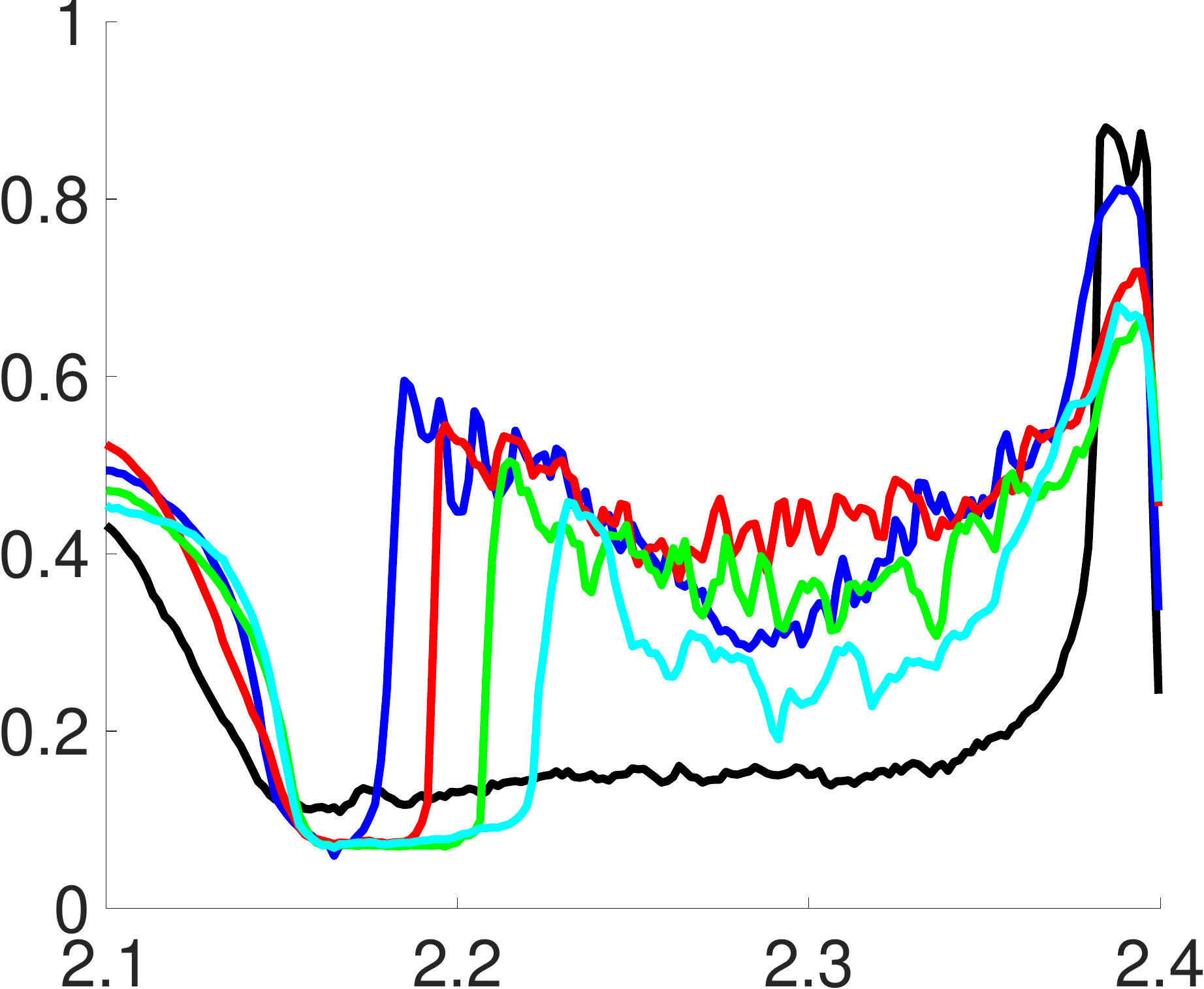} & 
\includegraphics[width=.23\textwidth]{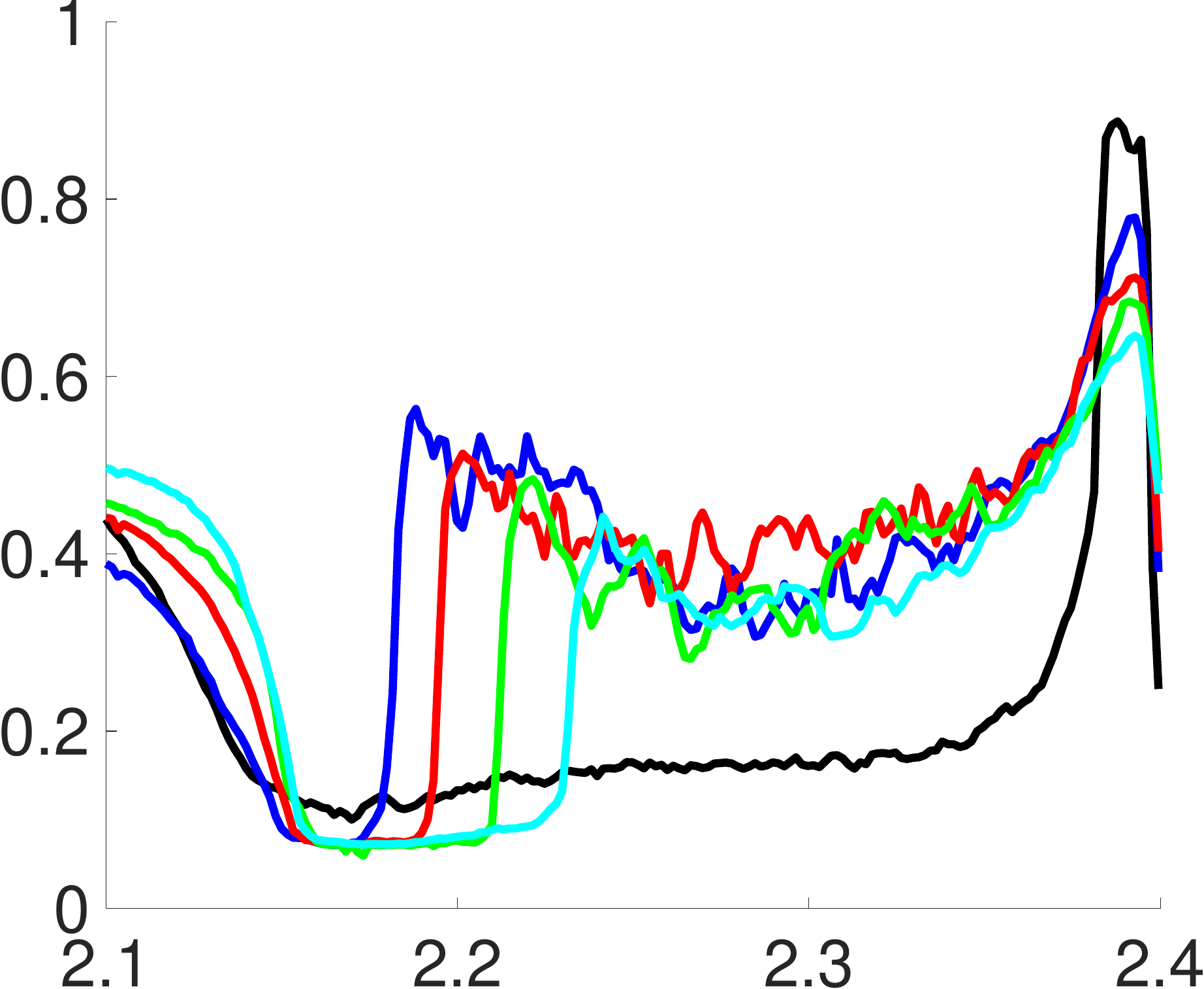} \\
& {\small time (s)} & {\small time (s) } & {\small time (s) } & {\small time (s) }   
\end{tabular}
\caption{Integral metrics $I_{1}$, non-dimensional excess streamwise momentum (top row), $I_{2}$, secondary flow strength (middle row) and $I_{R}$, reverse flow fraction (bottom row).
Numbering of the slice views is shown in the legend.
The times $t = 2.1, 2.2 $ and $ 2.4$ s approximately correspond to end diastole, peak systole, and early diastole in the next cycle, respectively.
}
\label{integral_metric_fig}
\end{figure*}

We further analyzed the flow with three non-dimensional integral metrics: 
$I_{1}$, representing the excess streamwise momentum, equation \eqref{I1_def}, 
$I_{2}$, representing secondary flow strength, equation \eqref{I2_def}, 
and $I_{R}$, the fraction of reverse flow, equation \eqref{IR_def}.

Figure \ref{integral_metric_fig} shows the integral metrics in the final 0.4 s of simulation, including all of systole in the final cycle.
For the tricuspid case, the slices at the annulus and immediately downstream of the valve had $I_{1}$ values greater than 1, indicating higher momentum than uniform flow. 
As the flow moved downstream, the streamwise momentum decreased, with the slices in the ascending aorta having $I_{1}$ values of approximately 1, indicating a return to uniform flow.

For the bicuspid cases, the streamwise momentum at slice 1, the annulus slice, is similar to the tricuspid case throughout the cardiac cycle. 
At slices 2 and 3, the streamwise momentum drastically increased, with values of $I_{1}$ reaching nearly 2.5, thereafter decreasing at slice 4 and more still at slice 5. 
The exception was in the LC/RC fusion case, where $I_{1}$ did not substantially decline until slice 5, as opposed to slice 4 in the other two bicuspid cases.
Thus, higher momentum flow persisted further downstream with LC/RC fusion.

The metric $I_{2}$ represents the non-dimensional secondary flow strength and quantifies the vortices and rotational motion seen in Figure \ref{vector_plots}.
For the tricuspid case, aortic flow exhibited low $I_{2}$ throughout the cardiac cycle.
For the bicuspid cases at slice 1, though $I_{2}$ remained low, its values are still greater than those of the tricuspid case. 
Immediately downstream of the valve and through the ascending aorta, $I_{2}$ in the bicuspid cases is substantially higher than in the tricuspid case.
For all three bicuspid cases, $I_{2}$ was higher at slice 3 than slice 2, as vortex structures developed downstream. 
With LC/RC fusion and NC/LC fusion, $I_{2}$ increased again at slice 4, indicating strong rotating flow along the vessel wall. 
In the LC/RC case, the counter-rotating vortex pairs also contributed to $I_{2}$.
The value of $I_{2}$ decreased at slice 5 as the jet and secondary flow structures broke up. 
In RC/NC fusion, $I_{2}$ is substantially higher at slices 4 and 5, as the jet moved from the lesser to greater curvature.
Thus, differences in leaflet morphology cause higher secondary flow strength further downstream of the valve.

At a given aortic cross section, $I_{R}$ defines the fraction of area in which flow moves in the reverse direction. 
During systole, the tricuspid case generally had low amounts of reverse flow. 
Notably, there is the most reverse flow at the sinotubular junction (slice 2). 
As the flow decelerated into early diastole, the value of $I_{R}$ increased to nearly one.

For all bicuspid cases, $I_{R} > 0.5$ was seen in slice 2 at early systole.
Since the jet was localized to the area above the freely moving cusp in each case, there was more area for reverse flow to develop and flow back towards the fused cusps.
This reverse flow generated large areas of recirculation in the aorta (Figure \ref{flow_panels}).
The value of $I_{R}$ remained above 0.3 for almost the entirely of systole in all three bicuspid cases.
The time at which each slice first experienced a high percentage of reverse flow increased monotonically downstream, illustrating that reverse flow was not generated until the jet reached that slice.

\section{Discussion and conclusions}
\label{Discussion}

Using newly developed models of the aortic valve, we simulated flows through four aortic valves: tricuspid and bicuspid with LC/RC, RC/NC and NC/LC fusion. 
By using a single patient-specific aortic geometry, we isolated the effects of valve morphology on hemodynamics for systematic comparison. 
Flows and pressures showed substantial differences between the tricuspid and bicuspid cases. 
The flows in the bicuspid cases were less uniform than in the tricuspid case, with narrower valve orifices producing an eccentric, localized jet. 
Aortic stenosis occurred in all bicuspid cases due to fusion of the free edges, without any changes in tissue material model. 
The bicuspid cases have more streamwise momentum ($I_{1}$), secondary flow strength ($I_{2}$), and greater reverse flow ($I_{R}$) compared to the tricuspid case. 
In all the bicuspid cases, qualitative inspection of the flow field and large values of $I_{1}$ suggest locally elevated axial wall shear stress at the locations where the jets impact the walls, though we cannot accurately compute wall shear stress directly using our current methods. 
Similarly, large values of $I_{2}$ suggest elevated circumferential wall shear stress.

Differences in hemodynamics were also prominent among the bicuspid cases themselves. 
The LC/RC and NC/LC cases showed concentrated high velocities along the greater curvature of the entire ascending aorta. 
In the RC/NC case, the jet moved from the lesser curvature to greater curvature of the aorta. 
The LC/RC and RC/NC cases had counterclockwise rotating flow, whereas the NC/LC case rotated clockwise. 
The highest values of $I_{2}$ occurred in the distal ascending aorta in the RC/NC fusion case.

Vascular smooth muscle cells present in the tunica media are a key structural component of the aorta \cite{Verma_nejm_bicuspid_review,6ikonomidis2012aortic}. 
In a variety of aortopathies, including those associated with bicuspid aortic valve disease, researchers have described the fragmentation and disruption of the medial extracellular matrix, and detachment of the aortic smooth muscle cells from the surrounding matrix components \cite{3losenno2012bicuspid,6ikonomidis2012aortic,7chen2017loss,8pedroza2020single}. 
These microstructural defects are present uniformly in the aorta, fundamentally weakening the aortic wall, allowing for dilation and growth into an aneurysmal state \cite{9fedak2003vascular}.
Studies found that 95-100\% of dilation associated with bicuspid aortic valve was asymmetric and localized to the greater curvature and that the lesser curvature had normal morphology \cite{bauer2006configuration,cotrufo2009association}. 
In analysis of diseased aortic tissue, spatial asymmetry from the greater to lesser curvature was found in smooth muscle cell apoptosis \cite{DELLACORTE20088} and matrix protein expression \cite{della2006spatial}. 
Assuming an identical genetic predisposition to aneurysmal growth among bicuspid variants, the presence of localized flow features may explain the asymmetric presentation of aortic dilation between the different bicuspid valve morphologies. 
The precise mechanisms by which smooth muscle cells respond to these hemodynamic features, however, remain poorly understood. 
Reverse and secondary flow may also play a role, but by what mechanism is unclear.

Of clinical interest, in all the bicuspid cases, the jet of forward flow is adherent to the greater curvature of the distal ascending aorta, suggesting that this region is exposed to higher local flow and shear stress. 
Further, in LC/RC fusion, the jet is adherent to the greater curvature immediately above the valve orifice near the aortic root, and remains adherent throughout the ascending aorta. 
In RC/NC fusion, the jet is largely not adherent to any particular side near the root, and settles on the greater curvature only after some distance. 
Estimates based on axisymmetric flow profiles suggest that to maintain a homeostatic level of shear stress at higher flow rates, vessels remodel to increase radius and thickness \cite{Humphrey_Remodeling}. 
From clinical registries it is known that dilation along the greater curvature of the ascending aorta is associated with LC/RC fusion \cite{Verma_nejm_bicuspid_review}. 
Studies have shown that patients with LC/RC fusion have significantly more dilation at the aortic root than those with RC/NC fusion, but have not shown significant differences in dilation in the ascending aorta \cite{SCHAEFER2007686,RUSSO2008937}. 
Thus, we observe that the flow jet remained attached to the wall at the same area in which localized, asymmetric dilation typically occurs in patients. 
These correspondences support the hypothesis that chronic exposure to high velocity regions causes higher local shear leading to local aortic dilation.

This study does have limitations, in particular the current IB method does not allow for accurate computation of shear stress on the valve leaflets or aortic wall. 
The IB method allows for simulation of rapidly deforming elastic structures without re-meshing, is insensitive to changes in fluid-domain topology, and eliminates the need for contact forces. 
In a comparison of numerical methods for heart valve simulations, alternatives such as the Arbitrary Lagrangian-Eulerian method that would provide shear stress were not effective at simulating the bulk flow fields, required contact forces and failed to complete a full cardiac cycle \cite{bavo2016fluid}.
Thus, the lack of shear measurements are a worthwhile tradeoff, given the robustness of the IB method.
In future work, we seek to include mechanistic models of growth and remodeling \cite{valentin2009complementary}, which likely requires a methodology that allows for computation of shear stress. 
Computational studies predicting a growth and remodeling response in the aorta would be of substantial future value in exploring mechanisms of dilation.   
Also, the material properties of the model aorta make the walls nearly rigid, which was chosen for simplicity and to further isolate the effects of changes to valve morphology, but the effects of wall material models could be investigated further. 
Additionally, we use a single patient-specific model aortic geometry. 
We expect that general trends would persist in similar healthy model geometries, though flow fields and quantitative values may change somewhat. 
Further, a case study with an anatomical patient-specific model better illustrates the hemodynamic differences caused by valve morphology than a study with a tube-like geometry that is not anatomical.

Long term, one could study the impact of isolated bicuspid valve disease in animal models without a background genetic disorder. 
Such a study would involve surgically inducing bicuspid valve morphology by suturing together two adjacent leaflets in chronic piglet or calf models. 
This experiment would show if the underlying cellular phenotype and transcriptome are similar to other aortopathies.
Most importantly, this experiment would also reveal whether isolated bicuspid valve disease can independently cause dilation and aneurysm formation.

\section{Acknowledgements}

ADK was supported in part by a grant from the National Heart, Lung and Blood Institute (Grant \linebreak \#1T32HL098049), Training Program in Mechanisms and Innovation in Vascular Disease. 
ADK and ALM were supported in part by the National Science Foundation SSI (Grant \#1663671). 
ADK and NS were supported in part by American Heart Association Transformational Project Award (Grant \#19TPA34910000).
RS was supported in part by the American Heart Association Postdoctoral Fellowship Award (Grant \#834986). 
NS was supported in part by the Stanford Bio-X Bowes Fellowship.
Computing for this project was performed on the Stanford University's Sherlock cluster with assistance from the Stanford Research Computing Center. 
Simulations were performed using the open-source solver package IBAMR, \url{https://ibamr.github.io}.

\section*{Appendix} 

We conducted a convergence study to evaluate the sensitivity of the velocity fields, flow and pressure waveforms, and integral metrics to changes in simulation resolution. 
We selected the bicuspid case with left/right coronary cusp fusion for the study, as we wanted a bicuspid case to evaluate the amount of stenosis caused by geometric changes to the valve as compared with resolution. 
We ran otherwise identical simulations with fluid resolutions of $\Delta x = 0.1, 0.075, 0.05$ and $0.0375$ cm, where $\Delta x = 0.05$ is the fluid resolution used throughout the study.

Precise convergence is difficult to achieve in these flows for a variety of numerical and physical reasons. 
First, the IB method effectively thickens the structure due to $\delta$-function based, diffuse-interface coupling. 
This coupling makes the effective orifice area and thus resistance to forward flow resolution-dependent. 
Further, the boundary conditions at the outlet are determined in a coupled manner with flow through the valve. 
Resistance due to the flow-averaging force is also dependent on flow rate. 
Additionally, the Reynolds number of such flows is in an inertial, potentially transitional, and thus physically unstable regime, which makes precise pointwise agreement in velocity fields impossible.

\begin{figure*}[!p]

       {\Large A:}      
	\vspace{2pt}

        {
        \centering
        \setlength{\tabcolsep}{1.0pt}        
        \begin{tabular}{c | c | c | c | c} 
         &       
         $\Delta x $ = 0.1 cm & 
         $\Delta x $ = 0.075 cm &
         $\Delta x $ = 0.05 cm & 
         $\Delta x $ = 0.0375 cm  
         \\ 
        \rotatebox[origin=l]{90}{ $t = 2.271 $ s } &
        \includegraphics[width=.24\textwidth]{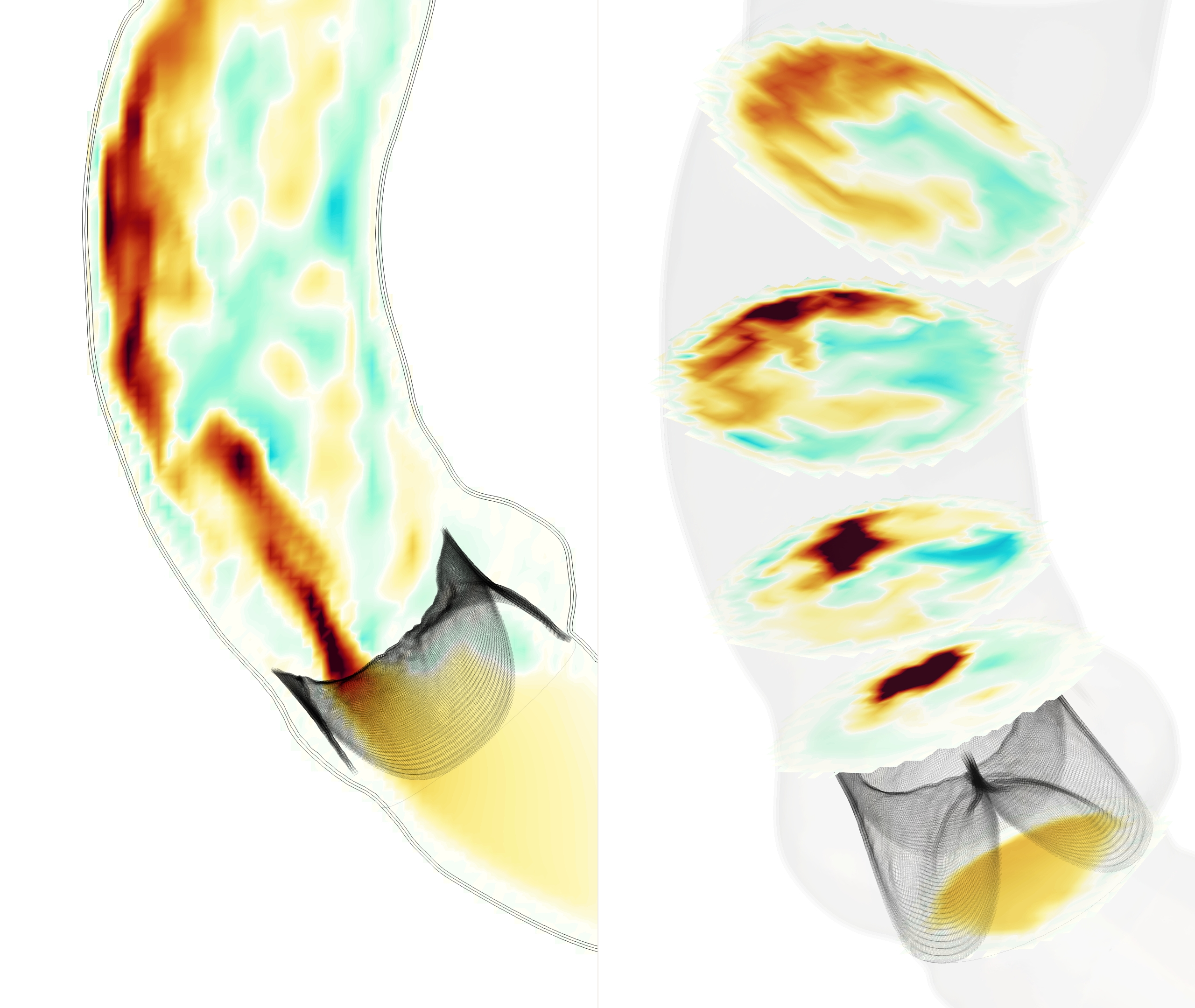}   &
        \includegraphics[width=.24\textwidth]{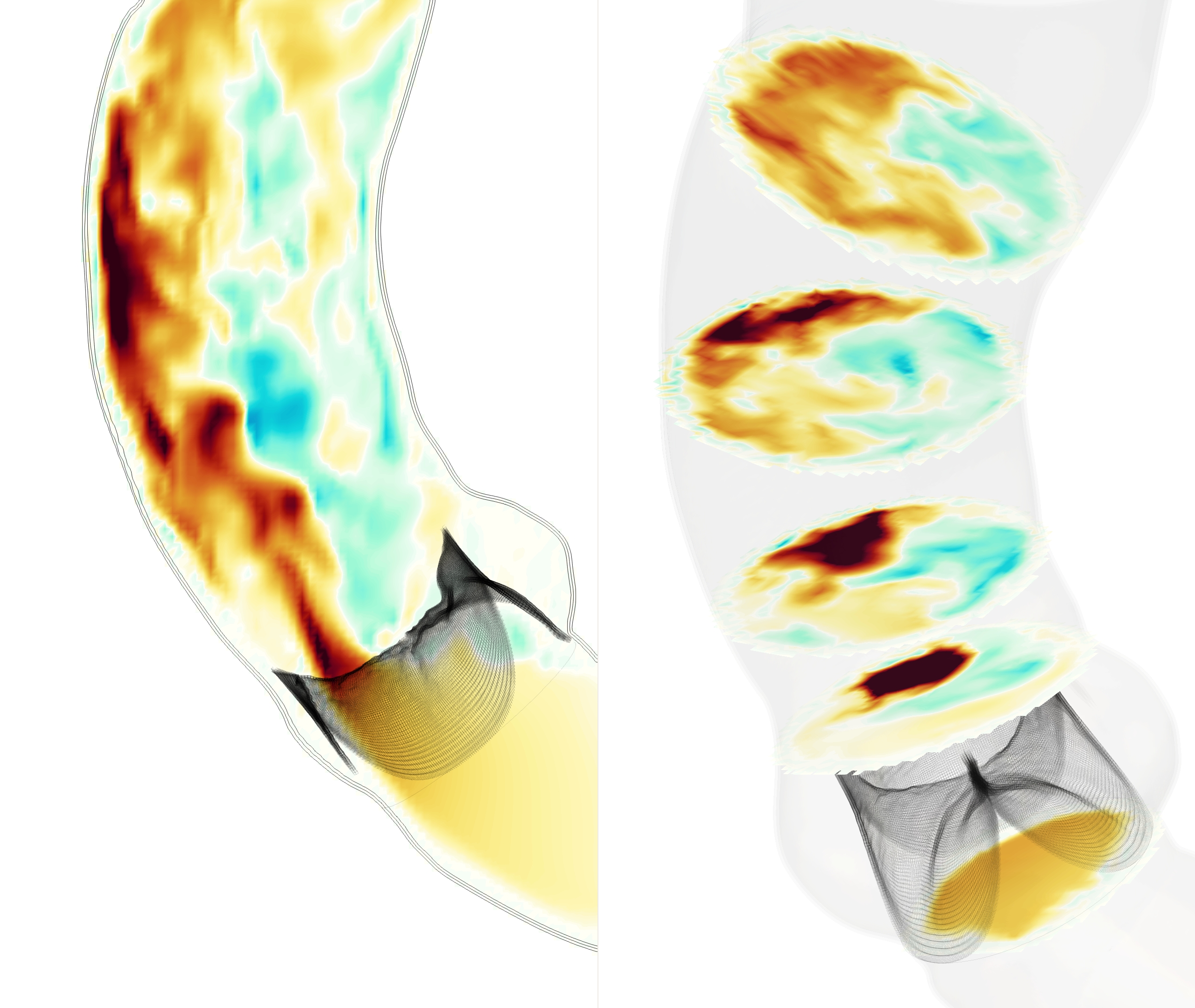}  &
        \includegraphics[width=.24\textwidth]{aortic_9999816_384_ef4cbc_final_setup_2_ee251fa_mesh_comm_2_fused_points_paper1365.jpeg}  & 
        \includegraphics[width=.24\textwidth]{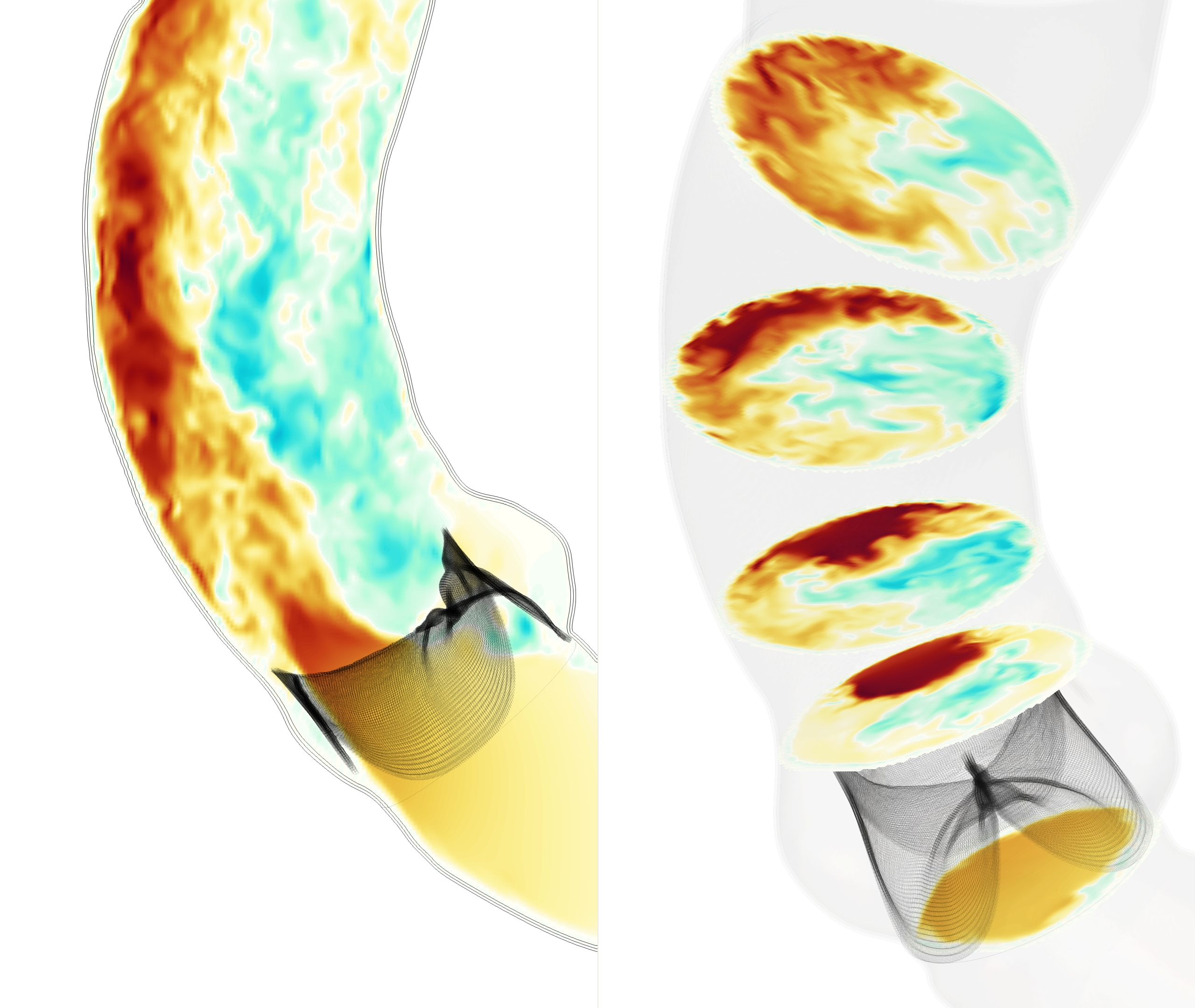} \\ 
        \end{tabular}
	}

	\vspace{15pt}
	{\Large B:}
	\vspace{-20pt}
	
	\begin{center}
	{
    \setlength{\tabcolsep}{0.0pt}
    \begin{tabular}{M{.02\textwidth} M{.03\textwidth}  M{.54\textwidth} M{.22\textwidth}} 
    & \rotatebox[origin=c]{90}{\small pressure (mmHg)} \vfill  & \includegraphics[scale=0.3]{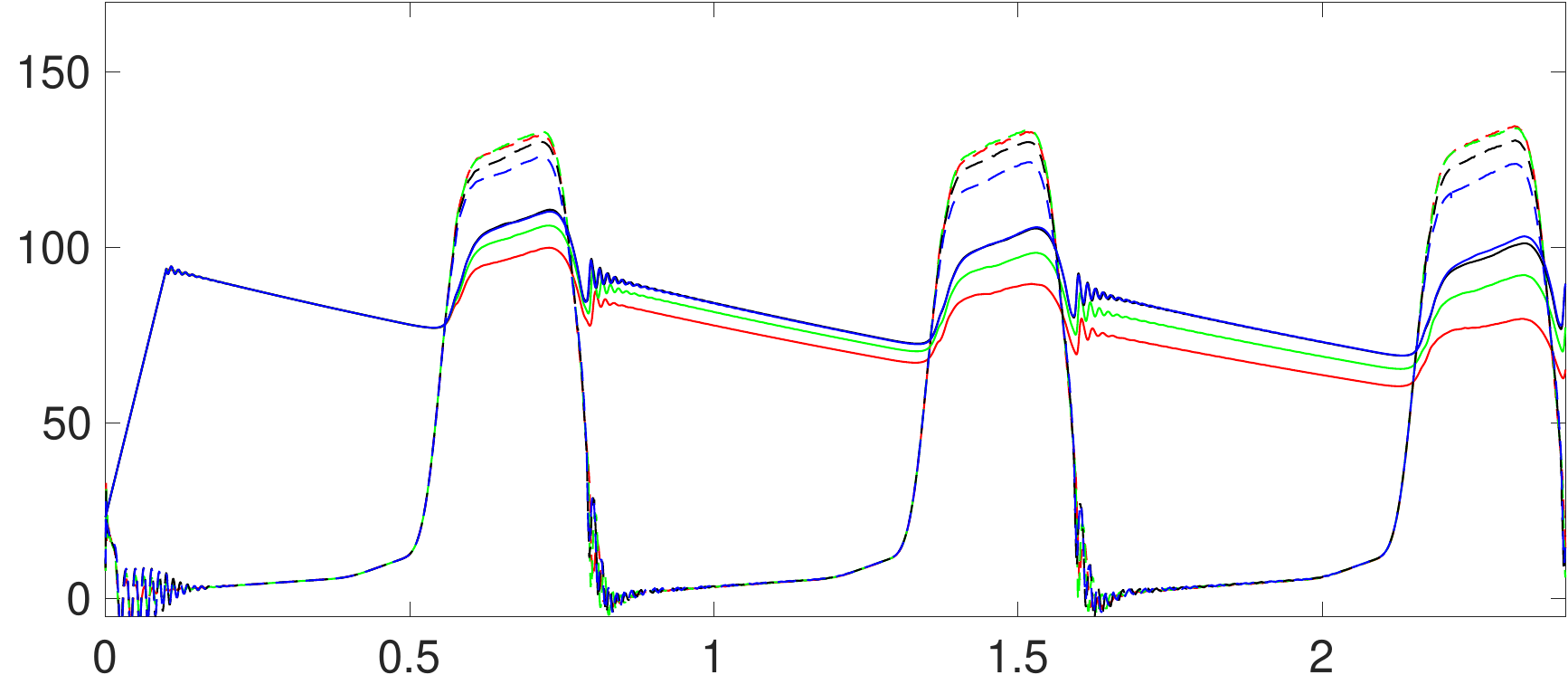} & \includegraphics[scale=0.3]{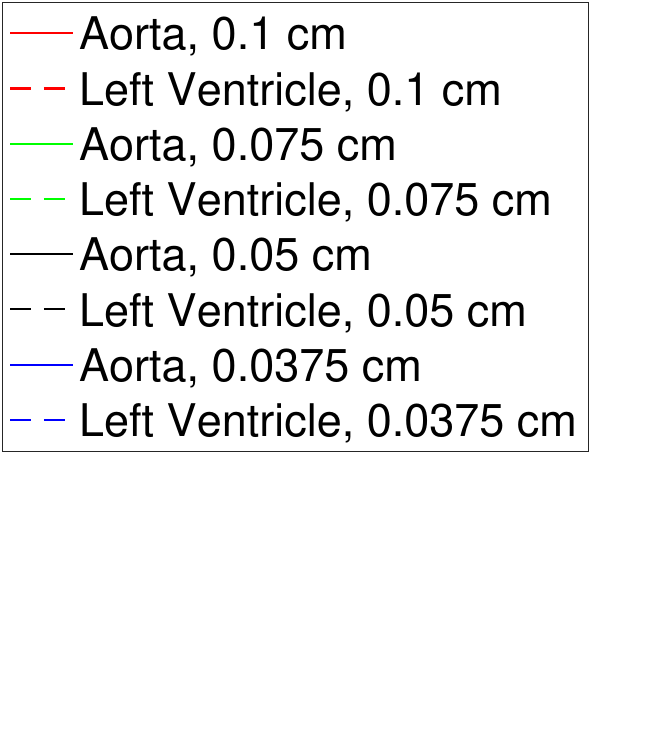}   \\  
     \rotatebox[origin=c]{90}{\small flow (ml/s),} \vfill & \rotatebox[origin=c]{90}{\small cumulative flow (ml)} \vfill &  \includegraphics[scale=0.3]{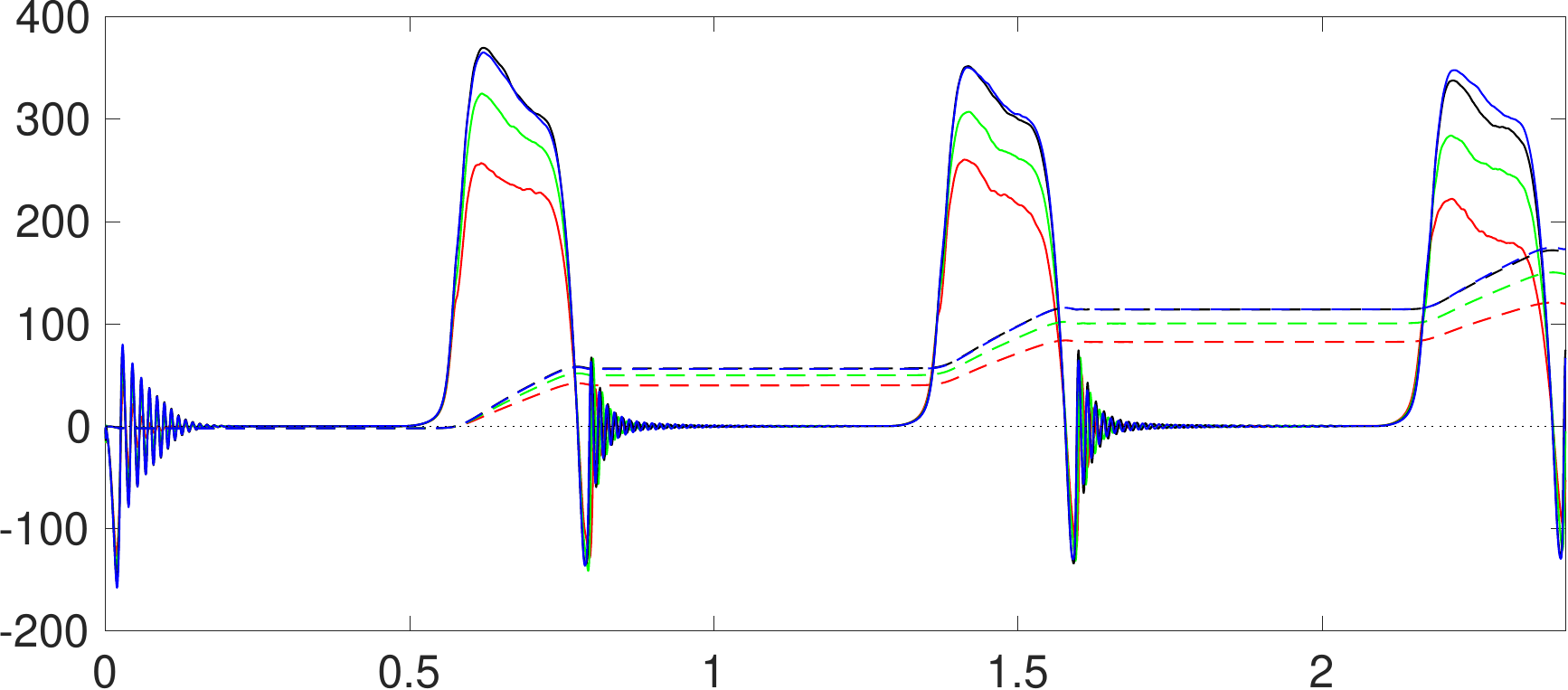} &  \includegraphics[scale=0.3]{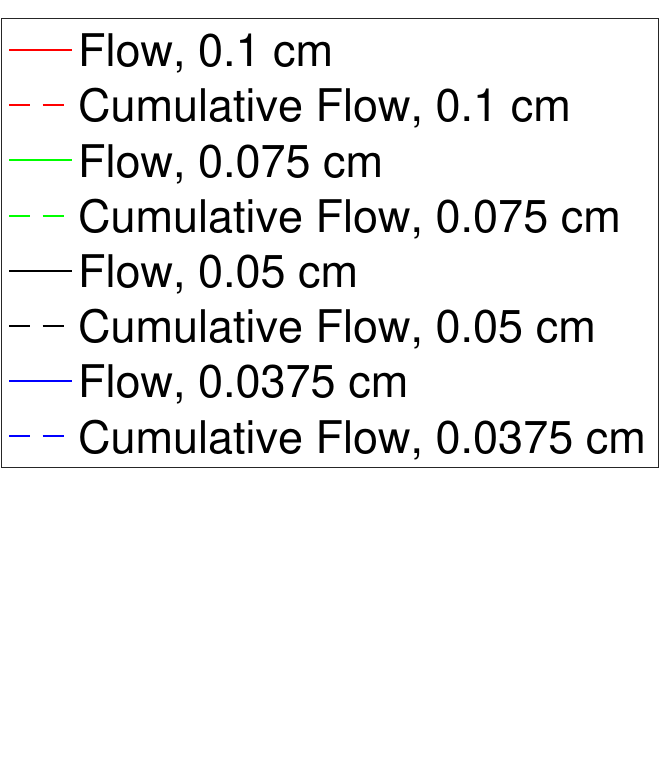} \\      
    & &  time (s) & 
     \end{tabular}   
	}
	\end{center}

	{\Large C:}
	\vspace{-20pt}
	
	\begin{center}
	{
	\setlength{\tabcolsep}{1.0pt} 
	\begin{tabular}{ M{.03\textwidth} M{.16\textwidth} M{.16\textwidth} M{.16\textwidth} M{.16\textwidth}}        
&
         $\Delta x $ = 0.1 cm & 
         $\Delta x $ = 0.075 cm &
         $\Delta x $ = 0.05 cm & 
         $\Delta x $ = 0.0375 cm \\ 
 $I_{1}$ & 
\includegraphics[width=.14\textwidth]{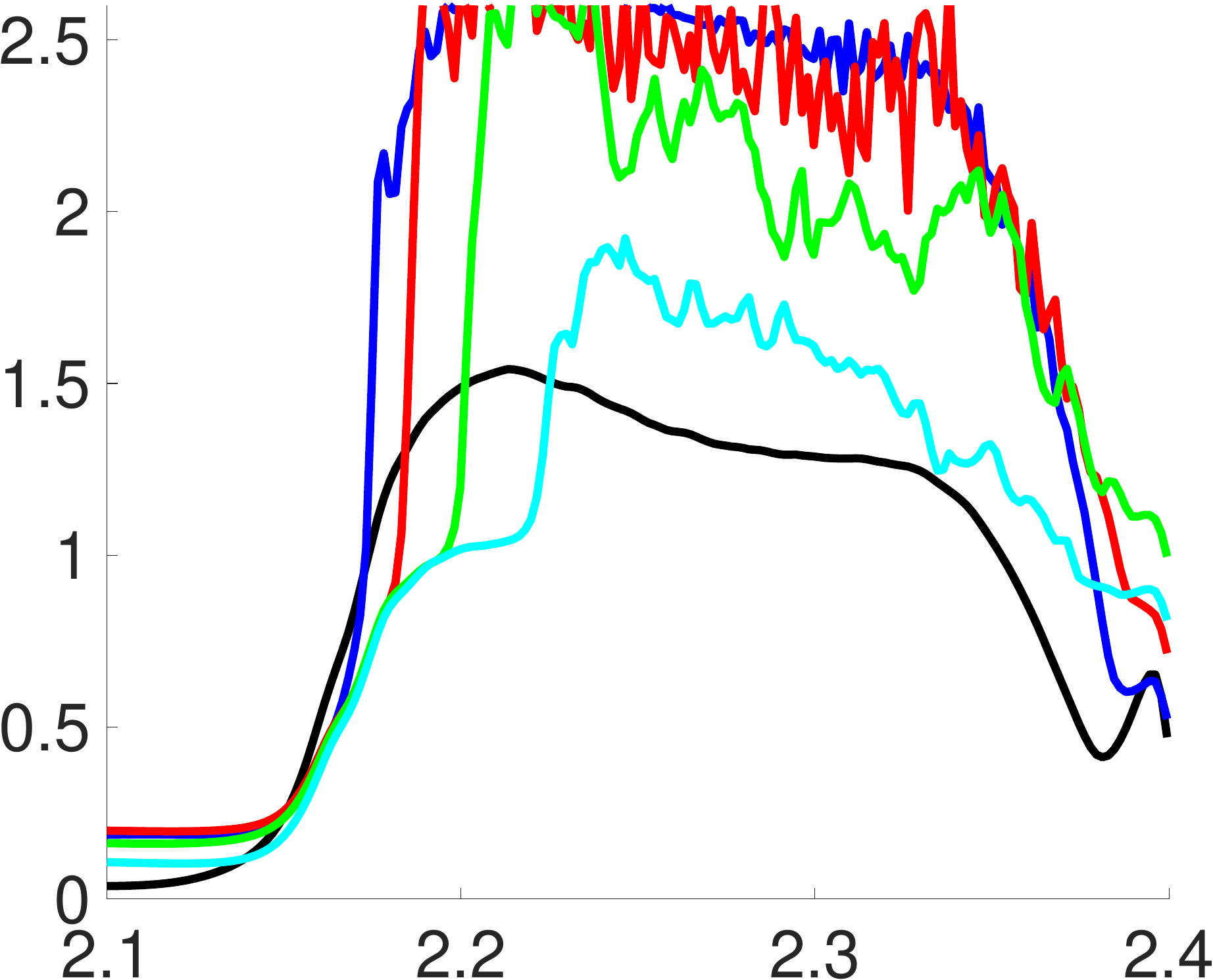} & 
\includegraphics[width=.14\textwidth]{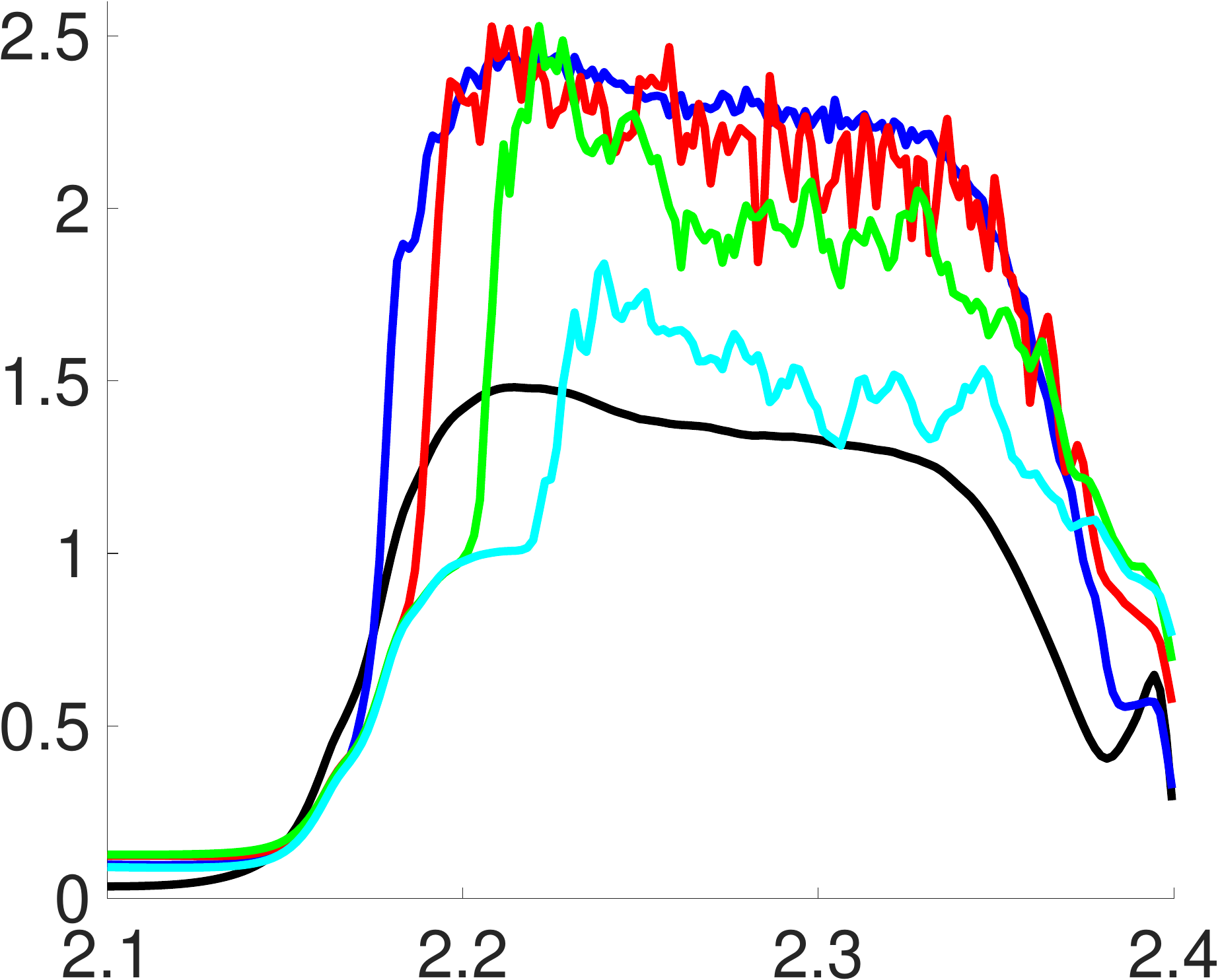} &
\includegraphics[width=.14\textwidth]{IndivRef_I1_Comm2Fused_ThirdCycle_AllContours.pdf} & 
\includegraphics[width=.14\textwidth]{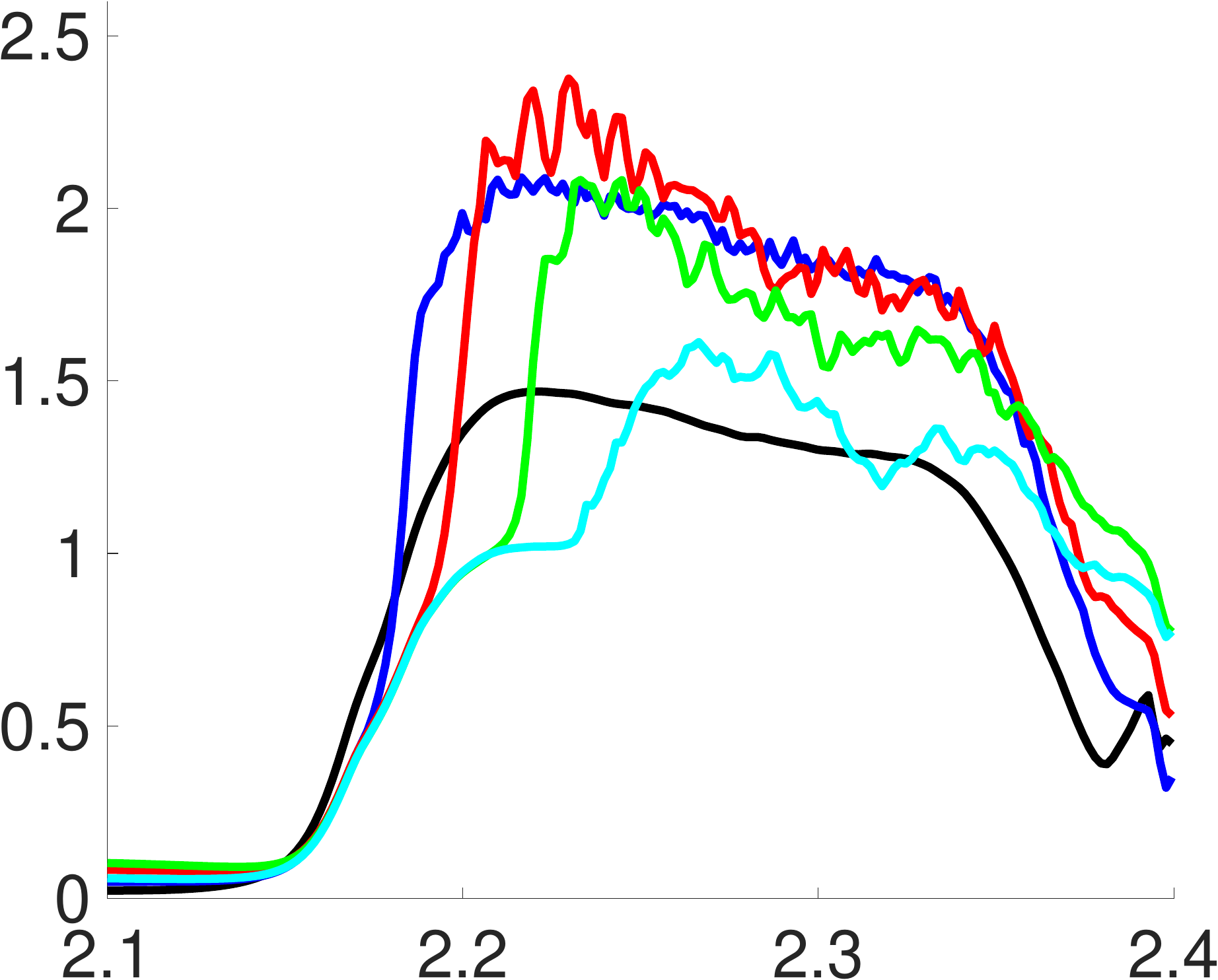} \\ \addlinespace[4.0pt]
$I_{2}$ & 
\includegraphics[width=.14\textwidth]{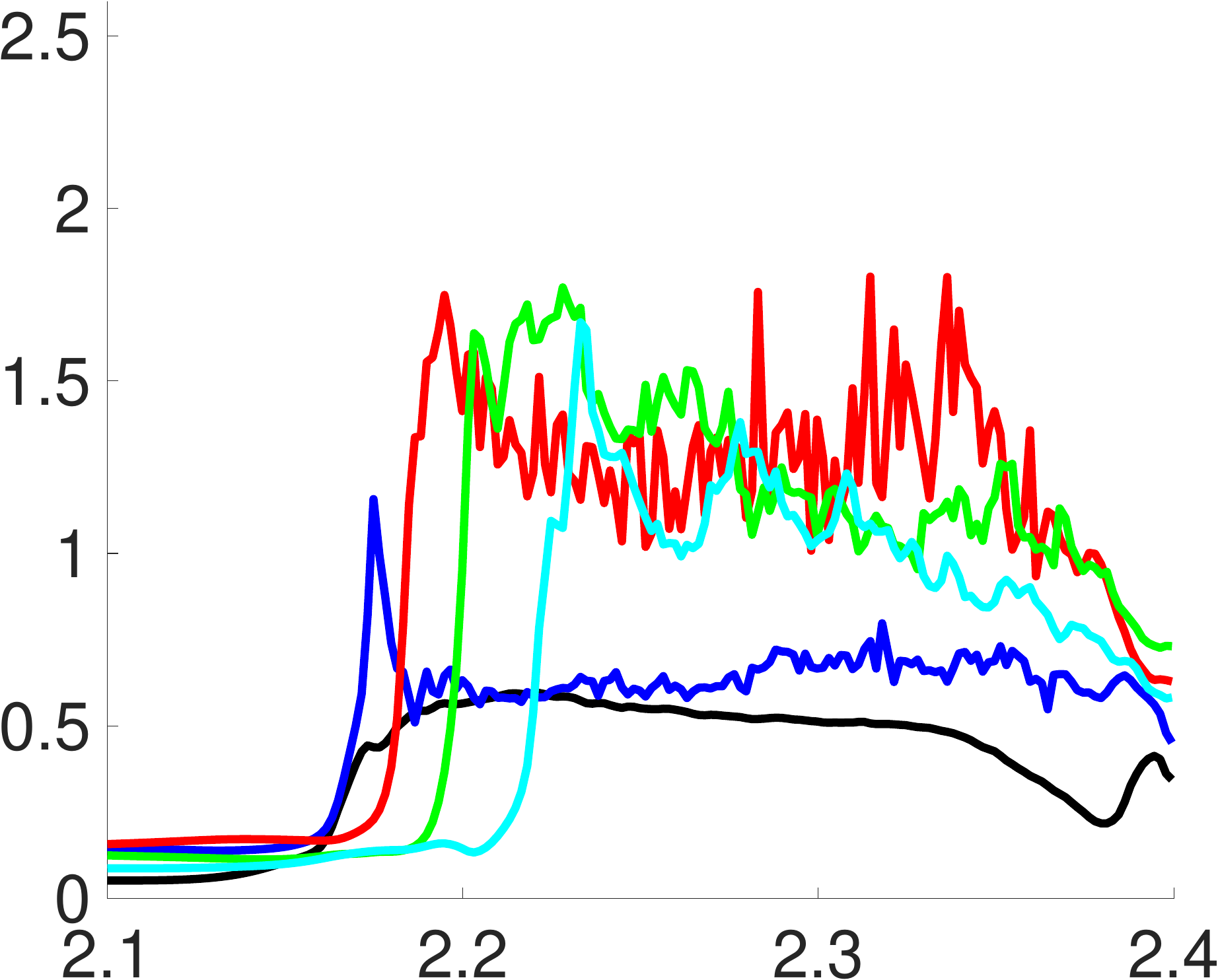} & 
\includegraphics[width=.14\textwidth]{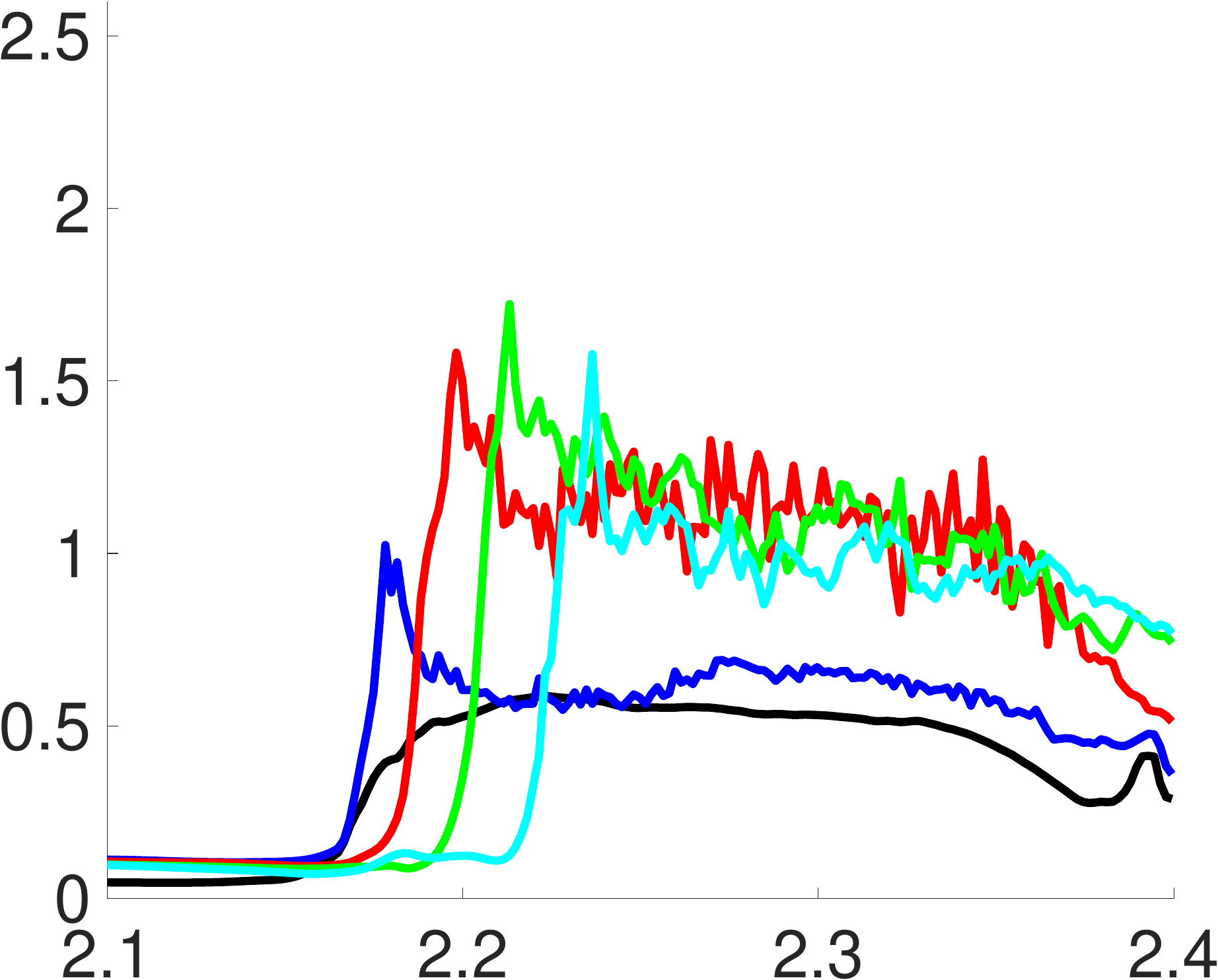} &
\includegraphics[width=.14\textwidth]{IndivRef_I2_Comm2_ThirdCycle_AllContours.pdf} &
\includegraphics[width=.14\textwidth]{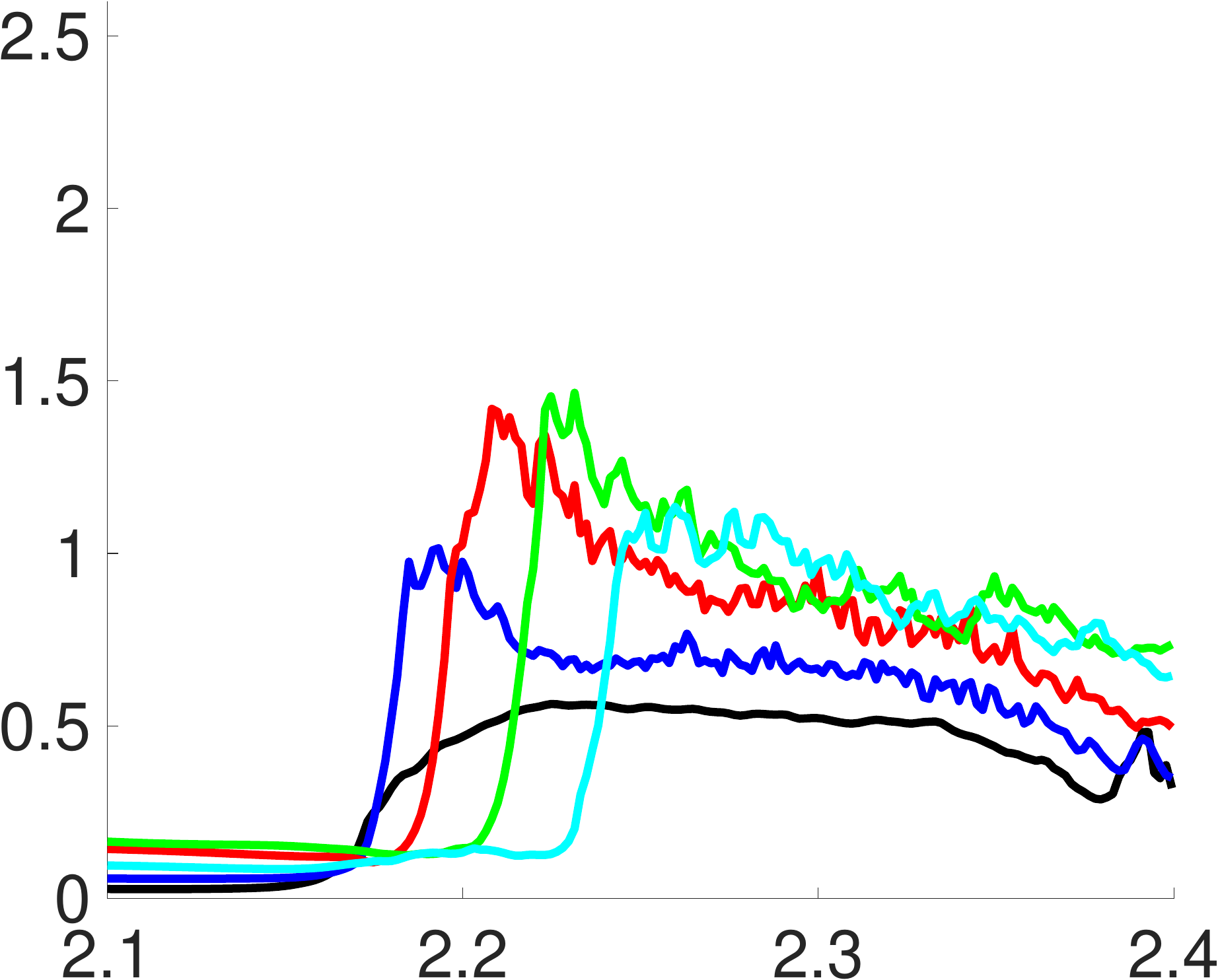} \\ \addlinespace[4.0pt]
$I_{R}$ & 
\includegraphics[width=.14\textwidth]{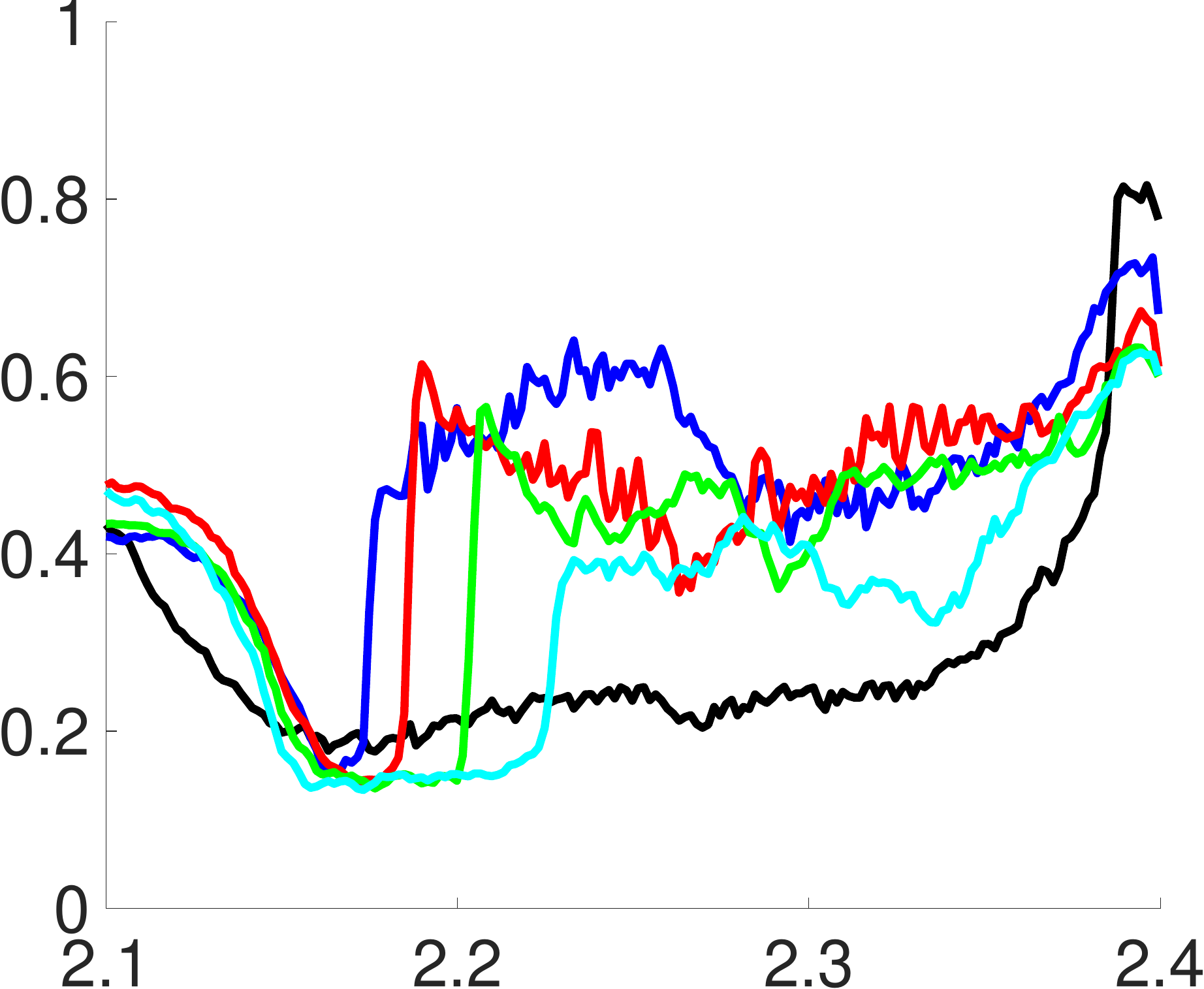} & 
\includegraphics[width=.14\textwidth]{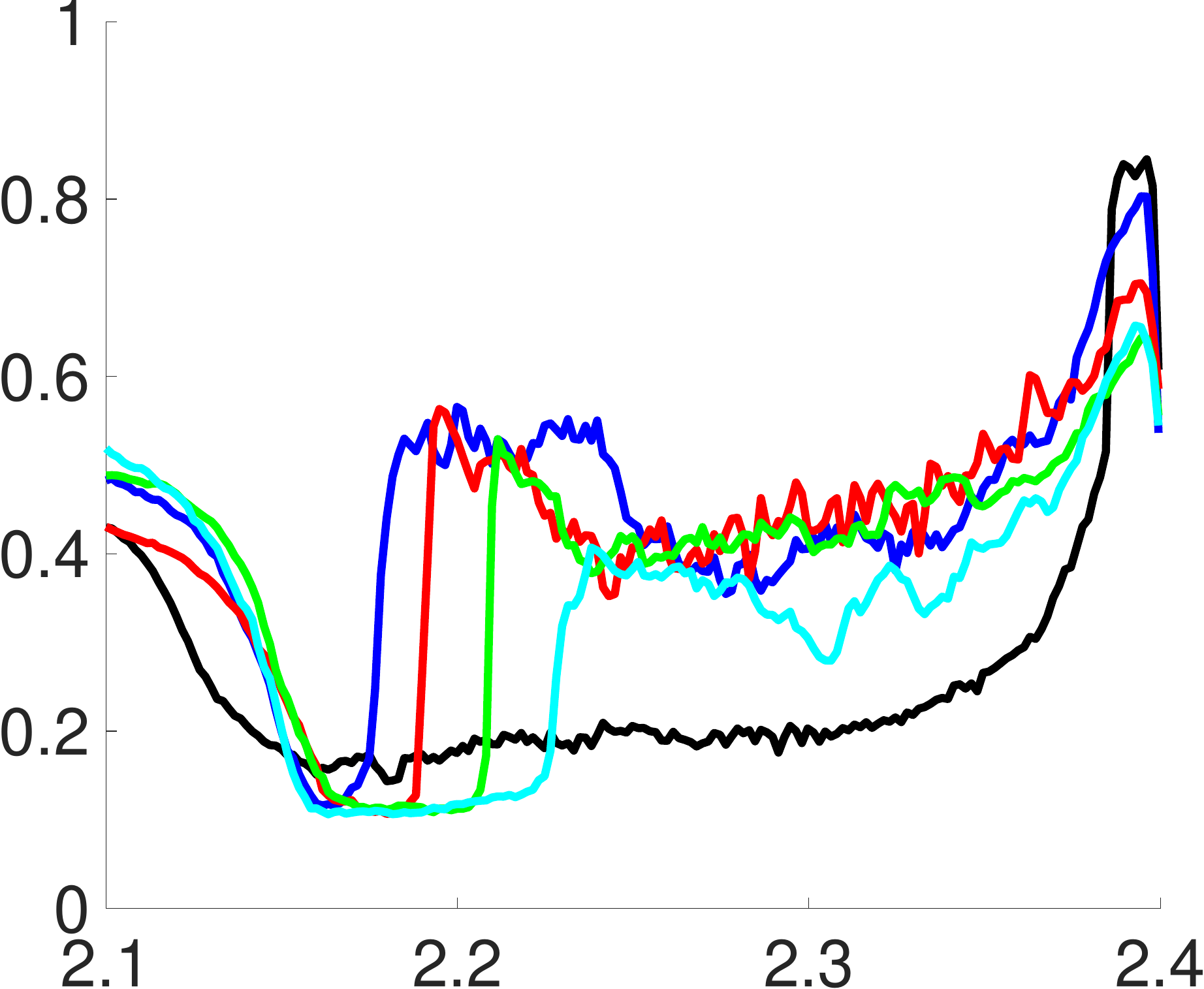} &
\includegraphics[width=.14\textwidth]{ReverseFlow_Comm2_ThirdCycle_AllContours.pdf} & 
\includegraphics[width=.14\textwidth]{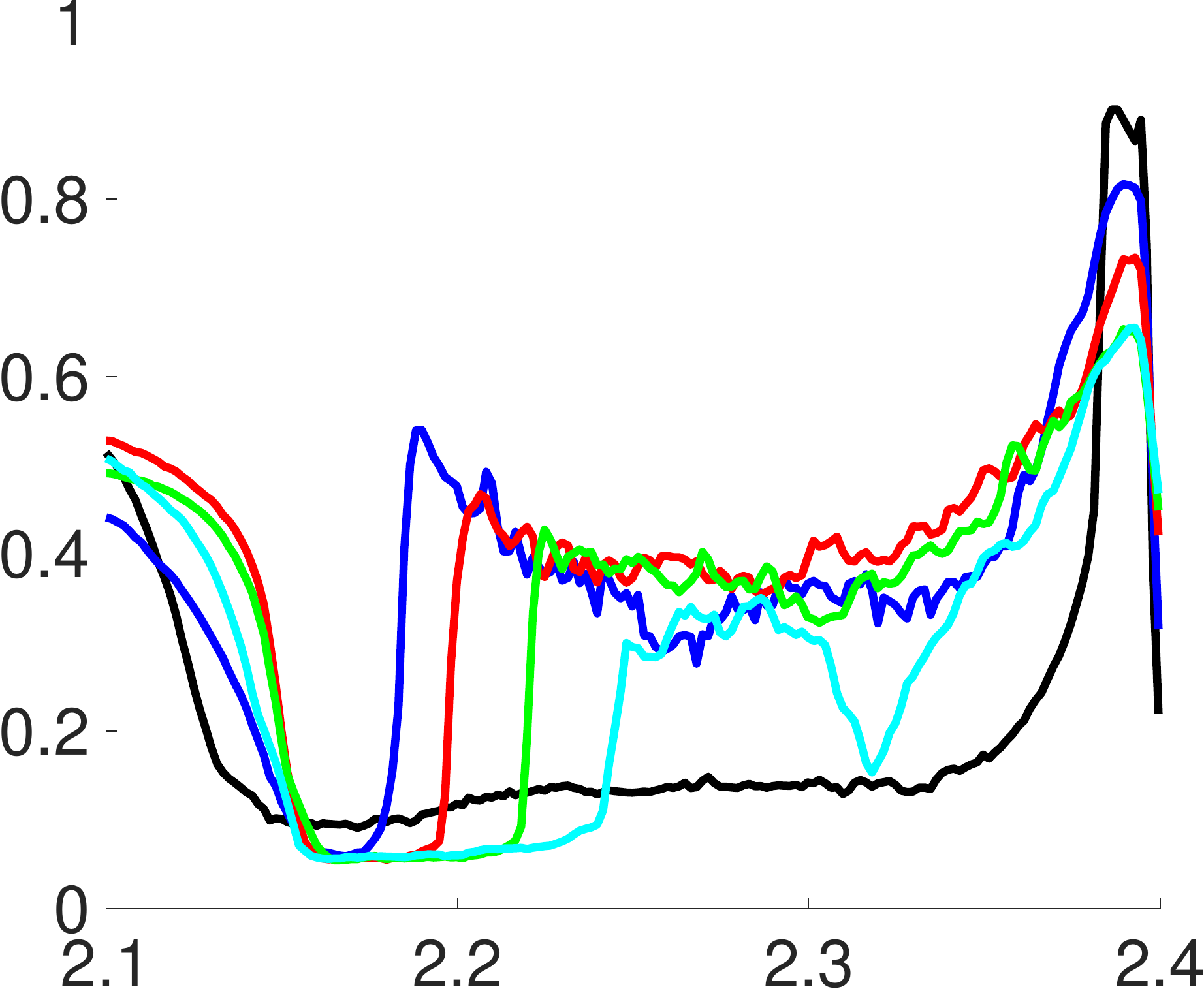} \\
& {\small time (s)} & {\small time (s) } & {\small time (s) } & {\small time (s) }   
\end{tabular}
	}	
	\end{center}

\caption{Convergence study with the bicuspid valve with LC/RC fusion and varying resolution.
A: Fluid velocity, vertical component and velocity normal to five selected slices. 
The color scheme is identical to that of Figure \ref{flow_panels}. 
B: Flow and pressure waveforms.
C: Values of integral metrics. (Values of $I_{1}$ above 2.6 in the $\Delta x = 0.1$ cm case are truncated.)
} 
\label{convergence_figure}
\end{figure*}

We see agreement, despite these limitations, between the resolutions in velocity fields, waveforms and integral metrics, as shown in Figure \ref{convergence_figure}. 
Qualitative trends in the velocity fields are consistent across all resolutions. 
The flow fields show increasing detail with resolution, as expected given the potentially transitional Reynolds number of the flow. 
At the coarsest resolution, the jet appears narrower when leaving the valve orifice, indicating insufficient resolution. 
At the finest two resolutions, the jets, regions of reverse flow and regions of recirculation are qualitatively similar. 

On the two finest resolutions, $\Delta x = 0.05$ and 0.0375 cm, the total cumulative flows are 170.7 and 173.2 ml, respectively, a relative change of 1.4\%, and these simulations show sustained pressure gradients above 18 and 15 mmHg, respectively. 
Thus, the two finest resolutions show similar flow rates and levels of stenosis. 
With $\Delta x = 0.1$ and 0.075 cm the total flow is 148.9 and 119.8 ml, respectively, and these simulations show sustained pressure gradients above 25 and 30 mmHg, respectively. 
We conclude that the two more coarse resolutions are under-resolved.

The integral metric curves show identical trends on the two finest resolutions. 
The values of $I_{1}$, $I_{2}$ and $I_{R}$ are very similar at the two finest scales. 
At the coarsest resolution, $\Delta x = 0.1$ cm, the values of $I_{1}$ are elevated relative to other cases, $I_{2}$ shows subtly different trends and elevated values, especially at slice 3, from finer resolution. 
We conclude that the most coarse resolution, which we do not use elsewhere, is inadequate to evaluate such metrics. 
For all metrics that show agreement at the finer scales, the values do not precisely agree pointwise in time because of the unsteady nature of the flow, which also causes the non-smooth appearance of the curves.

Thus, we believe that our conclusions throughout this work would be consistent with our selected resolution of $\Delta x = 0.05$ cm or increased resolution of $0.0375$ cm. 
We consider results with $\Delta x = 0.05 $ cm to be well-resolved and use this resolution throughout the study.

\begin{small}
\bibliographystyle{acm}
\bibliography{aortic_bicuspid_refs}

\begin{thebibliography}{10}

\bibitem{aggarwal2016vivo}
{\sc Aggarwal, A., Pouch, A.~M., Lai, E., Lesicko, J., Yushkevich, P.~A.,
  Gorman~III, J.~H., Gorman, R.~C., and Sacks, M.~S.}
\newblock In-vivo heterogeneous functional and residual strains in human aortic
  valve leaflets.
\newblock {\em Journal of biomechanics 49}, 12 (2016), 2481--2490.

\bibitem{ahrens2005paraview}
{\sc Ahrens, J., Geveci, B., and Law, C.}
\newblock Paraview: An end-user tool for large data visualization.
\newblock {\em The visualization handbook 717}, 8 (2005).

\bibitem{banko2016oscillatory}
{\sc Banko, A.~J., Coletti, F., Elkins, C.~J., and Eaton, J.~K.}
\newblock Oscillatory flow in the human airways from the mouth through several
  bronchial generations.
\newblock {\em International Journal of Heat and Fluid Flow 61\/} (2016),
  45--57.

\bibitem{IB5_arxiv}
{\sc Bao, Y., Kaiser, A.~D., Kaye, J., and Peskin, C.~S.}
\newblock Gaussian-like immersed boundary kernels with three continuous
  derivatives and improved translational invariance.
\newblock {\em eprint arXiv:1505.07529v3\/} (2017).

\bibitem{5barker2012bicuspid}
{\sc Barker, A.~J., Markl, M., B{\"u}rk, J., Lorenz, R., Bock, J., Bauer, S.,
  Schulz-Menger, J., and von Knobelsdorff-Brenkenhoff, F.}
\newblock Bicuspid aortic valve is associated with altered wall shear stress in
  the ascending aorta.
\newblock {\em Circulation: Cardiovascular Imaging 5}, 4 (2012), 457--466.

\bibitem{bauer2006configuration}
{\sc Bauer, M., Gliech, V., Siniawski, H., and Hetzer, R.}
\newblock Configuration of the ascending aorta in patients with bicuspid and
  tricuspid aortic valve disease undergoing aortic valve replacement with or
  without reduction aortoplasty.
\newblock {\em The Journal of heart valve disease 15}, 5 (2006), 594--600.

\bibitem{bavo2016fluid}
{\sc Bavo, A.~M., Rocatello, G., Iannaccone, F., Degroote, J., Vierendeels, J.,
  and Segers, P.}
\newblock Fluid-structure interaction simulation of prosthetic aortic valves:
  comparison between immersed boundary and arbitrary lagrangian-eulerian
  techniques for the mesh representation.
\newblock {\em PloS one 11}, 4 (2016), e0154517.

\bibitem{CNM:CNM2918}
{\sc Bertoglio, C., Caiazzo, A., Bazilevs, Y., Braack, M., Esmaily, M.,
  Gravemeier, V., Marsden, A., Pironneau, O., Vignon-Clementel, I.~E., and
  Wall, W.~A.}
\newblock Benchmark problems for numerical treatment of backflow at open
  boundaries.
\newblock {\em International Journal for Numerical Methods in Biomedical
  Engineering\/} (2017).
\newblock cnm.2918.

\bibitem{billiar2000biaxial}
{\sc Billiar, K.~L., and Sacks, M.~S.}
\newblock Biaxial mechanical properties of the natural and glutaraldehyde
  treated aortic valve cusp--part i: experimental results.
\newblock {\em J. Biomech. Eng. 122}, 1 (2000), 23--30.

\bibitem{cao2017simulations}
{\sc Cao, K., Atkins, S.~K., McNally, A., Liu, J., and Sucosky, P.}
\newblock Simulations of morphotype-dependent hemodynamics in non-dilated
  bicuspid aortic valve aortas.
\newblock {\em Journal of biomechanics 50\/} (2017), 63--70.

\bibitem{7chen2017loss}
{\sc Chen, J., Peters, A., Papke, C.~L., Villamizar, C., Ringuette, L.-J., Cao,
  J., Wang, S., Ma, S., Gong, L., Byanova, K.~L., Xiong, J., Zhu, M.~X.,
  Madonna, R., Kee, P., Geng, Y.-J., Brasier, A.~R., Davis, E.~C., Prakash, S.,
  Kwartler, C.~S., and Milewicz, D.~M.}
\newblock Loss of smooth muscle $\alpha$-actin leads to nf-$\kappa$b--dependent
  increased sensitivity to angiotensin ii in smooth muscle cells and aortic
  enlargement.
\newblock {\em Circulation research 120}, 12 (2017), 1903--1915.

\bibitem{cotrufo2009association}
{\sc Cotrufo, M., and Della~Corte, A.}
\newblock The association of bicuspid aortic valve disease with asymmetric
  dilatation of the tubular ascending aorta: identification of a definite
  syndrome.
\newblock {\em Journal of Cardiovascular Medicine 10}, 4 (2009), 291--297.

\bibitem{de2020deciphering}
{\sc De~Nisco, G., Tasso, P., Cal{\`o}, K., Mazzi, V., Gallo, D., Condemi, F.,
  Farzaneh, S., Avril, S., and Morbiducci, U.}
\newblock Deciphering ascending thoracic aortic aneurysm hemodynamics in
  relation to biomechanical properties.
\newblock {\em Medical Engineering \& Physics 82\/} (2020), 119--129.

\bibitem{della2006spatial}
{\sc Della~Corte, A., De~Santo, L.~S., Montagnani, S., Quarto, C., Romano, G.,
  Amarelli, C., Scardone, M., De~Feo, M., Cotrufo, M., and Caianiello, G.}
\newblock Spatial patterns of matrix protein expression in dilated ascending
  aorta with aortic regurgitation: congenital bicuspid valve versus marfan's
  syndrome.
\newblock {\em Journal of Heart Valve Disease 15}, 1 (2006), 20--7.

\bibitem{DELLACORTE20088}
{\sc {Della Corte}, A., Quarto, C., Bancone, C., Castaldo, C., {Di Meglio}, F.,
  Nurzynska, D., {De Santo}, L.~S., {De Feo}, M., Scardone, M., Montagnani, S.,
  and Cotrufo, M.}
\newblock Spatiotemporal patterns of smooth muscle cell changes in ascending
  aortic dilatation with bicuspid and tricuspid aortic valve stenosis: Focus on
  cell-matrix signaling.
\newblock {\em The Journal of Thoracic and Cardiovascular Surgery 135}, 1
  (2008), 8--18.e2.

\bibitem{4DuxSantoy_wss_dilation}
{\sc Dux-Santoy, L., Guala, A., Sotelo, J., Uribe, S., Teixid{\'o}-Tur{\`a},
  G., Ruiz-Mu{\~n}oz, A., Hurtado, D.~E., Valente, F., Galian-Gay, L.,
  Guti{\'e}rrez, L., Gonz{\'a}lez-Alujas, T., Johnson, K.~M., Wieben, O.,
  Ferreira-Gonzalez, I., Evangelista, A., and Rodr{\'i}guez-Palomares, J.~F.}
\newblock Low and oscillatory wall shear stress is not related to aortic
  dilation in patients with bicuspid aortic valve.
\newblock {\em Arteriosclerosis, Thrombosis, and Vascular Biology 40}, 1
  (2020), e10--e20.

\bibitem{emendi2021patient}
{\sc Emendi, M., Sturla, F., Ghosh, R.~P., Bianchi, M., Piatti, F.,
  Pluchinotta, F.~R., Giese, D., Lombardi, M., Redaelli, A., and Bluestein, D.}
\newblock Patient-specific bicuspid aortic valve biomechanics: a magnetic
  resonance imaging integrated fluid--structure interaction approach.
\newblock {\em Annals of Biomedical Engineering 49}, 2 (2021), 627--641.

\bibitem{9fedak2003vascular}
{\sc Fedak, P.~W., {de Sa}, M.~P., Verma, S., Nili, N., Kazemian, P., Butany,
  J., Strauss, B.~H., Weisel, R.~D., and David, T.~E.}
\newblock Vascular matrix remodeling in patients with bicuspid aortic valve
  malformations: implications for aortic dilatation.
\newblock {\em The Journal of Thoracic and Cardiovascular Surgery 126}, 3
  (2003), 797--805.

\bibitem{gilmanov2016comparative}
{\sc Gilmanov, A., and Sotiropoulos, F.}
\newblock Comparative hemodynamics in an aorta with bicuspid and trileaflet
  valves.
\newblock {\em Theoretical and Computational Fluid Dynamics 30}, 1-2 (2016),
  67--85.

\bibitem{Girdauskas2011}
{\sc Girdauskas, E., Borger, M.~A., Secknus, M.-A., Girdauskas, G., and Kuntze,
  T.}
\newblock {Is aortopathy in bicuspid aortic valve disease a congenital defect
  or a result of abnormal hemodynamics? A critical reappraisal of a one-sided
  argument}.
\newblock {\em European Journal of Cardio-Thoracic Surgery 39}, 6 (06 2011),
  809--814.

\bibitem{IBAMR}
{\sc Griffith, B.~E.}
\newblock {IBAMR}: Immersed boundary adaptive mesh refinement.
\newblock \url{https://github.com/IBAMR/IBAMR}, 2017.

\bibitem{griffith2010parallel}
{\sc Griffith, B.~E., Hornung, R.~D., McQueen, D.~M., and Peskin, C.~S.}
\newblock Parallel and adaptive simulation of cardiac fluid dynamics.
\newblock {\em Advanced Computational Infrastructures for Parallel and
  Distributed Adaptive Applications\/} (2010), 105.

\bibitem{guala2022wall}
{\sc Guala, A., Dux-Santoy, L., Teixido-Tura, G., Ruiz-Mu{\~n}oz, A.,
  Galian-Gay, L., Servato, M.~L., Valente, F., Guti{\'e}rrez, L.,
  Gonz{\'a}lez-Alujas, T., Johnson, K.~M., Wieben, O., Casas-Masnou, G.,
  Sao~Avil{\'e}s, A., Fernandez-Galera, R., Ferreira-Gonzalez, I., Evangelista,
  J.~F., and Rodr{\'i}guez-Palomares, J.~F.}
\newblock Wall shear stress predicts aortic dilation in patients with bicuspid
  aortic valve.
\newblock {\em JACC: Cardiovascular Imaging 15}, 1 (2022), 46--56.

\bibitem{guzzardi2015valve}
{\sc Guzzardi, D.~G., Barker, A.~J., Van~Ooij, P., Malaisrie, S.~C., Puthumana,
  J.~J., Belke, D.~D., Mewhort, H.~E., Svystonyuk, D.~A., Kang, S., Verma, S.,
  Collins, J., Carr, J., Bonow, R.~O., Markl, M., Thomas, J.~D., McCarthy,
  P.~M., and Fedak, P.~W.}
\newblock Valve-related hemodynamics mediate human bicuspid aortopathy:
  insights from wall shear stress mapping.
\newblock {\em Journal of the American College of Cardiology 66}, 8 (2015),
  892--900.

\bibitem{hammer2011mass}
{\sc Hammer, P.~E., Sacks, M.~S., Pedro, J., and Howe, R.~D.}
\newblock Mass-spring model for simulation of heart valve tissue mechanical
  behavior.
\newblock {\em Annals of biomedical engineering 39}, 6 (2011), 1668--1679.

\bibitem{Humphrey_Remodeling}
{\sc Humphrey, J.~D.}
\newblock Mechanisms of arterial remodeling in hypertension.
\newblock {\em Hypertension 52}, 2 (2008), 195--200.

\bibitem{6ikonomidis2012aortic}
{\sc Ikonomidis, J.~S., Ruddy, J.~M., Benton, S.~M., Arroyo, J., Brinsa, T.~A.,
  Stroud, R.~E., Zeeshan, A., Bavaria, J.~E., Gorman, J.~H., Gorman, R.~C.,
  Spinale, F.~G., and Jones, J.~A.}
\newblock Aortic dilatation with bicuspid aortic valves: cusp fusion correlates
  to matrix metalloproteinases and inhibitors.
\newblock {\em The Annals of thoracic surgery 93}, 2 (2012), 457--463.

\bibitem{jayendiran2020computational}
{\sc Jayendiran, R., Condemi, F., Campisi, S., Viallon, M., Croisille, P., and
  Avril, S.}
\newblock Computational prediction of hemodynamical and biomechanical
  alterations induced by aneurysm dilatation in patient-specific ascending
  thoracic aortas.
\newblock {\em International Journal for Numerical Methods in Biomedical
  Engineering 36}, 6 (2020), e3326.

\bibitem{thesis}
{\sc Kaiser, A.~D.}
\newblock Modeling the mitral valve.
\newblock {\em Ph.D. thesis, Courant Institute of Mathematical Sciences, New
  York University\/} (September 2017).

\bibitem{kaiser2019modeling}
{\sc Kaiser, A.~D., McQueen, D.~M., and Peskin, C.~S.}
\newblock Modeling the mitral valve.
\newblock {\em International journal for numerical methods in biomedical
  engineering\/} (2019), e3240.

\bibitem{kaiser2021validation}
{\sc Kaiser, A.~D., Schiavone, N.~K., Eaton, J.~K., and Marsden, A.~L.}
\newblock Validation of immersed boundary simulations of heart valve
  hemodynamics against in vitro {4D Flow MRI} data.
\newblock {\em eprint arXiv:2111.00720\/} (2021).

\bibitem{kaiser2020designbased}
{\sc Kaiser, A.~D., Shad, R., Hiesinger, W., and Marsden, A.~L.}
\newblock A design-based model of the aortic valve for fluid-structure
  interaction.
\newblock {\em Biomechanics and Modeling in Mechanobiology 20}, 6 (2021),
  2413--2435.

\bibitem{kim2009coupling}
{\sc Kim, H.~J., Vignon-Clementel, I.~E., Figueroa, C.~A., LaDisa, J.~F.,
  Jansen, K.~E., Feinstein, J.~A., and Taylor, C.~A.}
\newblock On coupling a lumped parameter heart model and a three-dimensional
  finite element aorta model.
\newblock {\em Annals of biomedical engineering 37}, 11 (2009), 2153--2169.

\bibitem{kimura2017patient}
{\sc Kimura, N., Nakamura, M., Komiya, K., Nishi, S., Yamaguchi, A., Tanaka,
  O., Misawa, Y., Adachi, H., and Kawahito, K.}
\newblock Patient-specific assessment of hemodynamics by computational fluid
  dynamics in patients with bicuspid aortopathy.
\newblock {\em The Journal of thoracic and cardiovascular surgery 153}, 4
  (2017), S52--S62.

\bibitem{lan2018re}
{\sc Lan, H., Updegrove, A., Wilson, N.~M., Maher, G.~D., Shadden, S.~C., and
  Marsden, A.~L.}
\newblock A re-engineered software interface and workflow for the open-source
  simvascular cardiovascular modeling package.
\newblock {\em Journal of biomechanical engineering 140}, 2 (2018).

\bibitem{laniado1982physiologic}
{\sc Laniado, S., Yellin, E.~L., Yoran, C., Strom, J., Hori, M., Gabbay, S.,
  Terdiman, R., and Frater, R. W.~M.}
\newblock Physiologic mechanisms in aortic insufficiency. i. the effect of
  changing heart rate on flow dynamics. ii. determinants of austin flint
  murmur.
\newblock {\em Circulation 66}, 1 (1982), 226--235.

\bibitem{laskey1990estimation}
{\sc Laskey, W.~K., Parker, H.~G., Ferrari, V.~A., Kussmaul, W.~G., and
  Noordergraaf, A.}
\newblock Estimation of total systemic arterial compliance in humans.
\newblock {\em Journal of Applied Physiology 69}, 1 (1990), 112--119.

\bibitem{Lavon2018}
{\sc Lavon, K., Halevi, R., Marom, G., Ben~Zekry, S., Hamdan, A.,
  Joachim~Sch\"{a}fers, H., Raanani, E., and Haj-Ali, R.}
\newblock {Fluid-Structure Interaction Models of Bicuspid Aortic Valves: The
  Effects of Nonfused Cusp Angles}.
\newblock {\em Journal of Biomechanical Engineering 140}, 3 (01 2018).
\newblock 031010.

\bibitem{3losenno2012bicuspid}
{\sc Losenno, K.~L., Goodman, R.~L., and Chu, M.~W.}
\newblock Bicuspid aortic valve disease and ascending aortic aneurysms: gaps in
  knowledge.
\newblock {\em Cardiology research and practice 2012\/} (2012).

\bibitem{marom2013fully}
{\sc Marom, G., Kim, H.-S., Rosenfeld, M., Raanani, E., and Haj-Ali, R.}
\newblock Fully coupled fluid--structure interaction model of congenital
  bicuspid aortic valves: effect of asymmetry on hemodynamics.
\newblock {\em Medical \& biological engineering \& computing 51}, 8 (2013),
  839--848.

\bibitem{may2009hyperelastic}
{\sc May-Newman, K., Lam, C., and Yin, F.~C.}
\newblock A hyperelastic constitutive law for aortic valve tissue.
\newblock {\em Journal of biomechanical engineering 131}, 8 (2009).

\bibitem{Nishimura_2014_stenosis_guidelines}
{\sc Nishimura, R.~A., Otto, C.~M., Bonow, R.~O., Carabello, B.~A., Erwin,
  J.~P., Guyton, R.~A., O'Gara, P.~T., Ruiz, C.~E., Skubas, N.~J., Sorajja, P.,
  Sundt, T.~M., and Thomas, J.~D.}
\newblock 2014 aha/acc guideline for the management of patients with valvular
  heart disease.
\newblock {\em Journal of the American College of Cardiology 63}, 22 (2014),
  e57--e185.

\bibitem{8pedroza2020single}
{\sc Pedroza, A.~J., Tashima, Y., Shad, R., Cheng, P., Wirka, R., Churovich,
  S., Nakamura, K., Yokoyama, N., Cui, J.~Z., Iosef, C., Hiesinger, W.,
  Quertermous, T., and Fischbein, M.~P.}
\newblock Single-cell transcriptomic profiling of vascular smooth muscle cell
  phenotype modulation in marfan syndrome aortic aneurysm.
\newblock {\em Arteriosclerosis, Thrombosis, and Vascular Biology 40}, 9
  (2020), 2195--2211.

\bibitem{ib_acta_numerica}
{\sc Peskin, C.~S.}
\newblock The immersed boundary method.
\newblock {\em Acta Numerica 11\/} (2002), 479--517.

\bibitem{pham2017quantification}
{\sc Pham, T., Sulejmani, F., Shin, E., Wang, D., and Sun, W.}
\newblock Quantification and comparison of the mechanical properties of four
  human cardiac valves.
\newblock {\em Acta biomaterialia 54\/} (2017), 345--355.

\bibitem{RUSSO2008937}
{\sc Russo, C.~F., Cannata, A., Lanfranconi, M., Vitali, E., Garatti, A., and
  Bonacina, E.}
\newblock Is aortic wall degeneration related to bicuspid aortic valve anatomy
  in patients with valvular disease?
\newblock {\em The Journal of Thoracic and Cardiovascular Surgery 136}, 4
  (2008), 937--942.

\bibitem{sahasakul1988age}
{\sc Sahasakul, Y., Edwards, W.~D., Naessens, J.~M., and Tajik, A.~J.}
\newblock Age-related changes in aortic and mitral valve thickness:
  implications for two-dimensional echocardiography based on an autopsy study
  of 200 normal human hearts.
\newblock {\em The American journal of cardiology 62}, 7 (1988), 424--430.

\bibitem{SCHAEFER2007686}
{\sc Schaefer, B.~M., Lewin, M.~B., Stout, K.~K., Byers, P.~H., and Otto,
  C.~M.}
\newblock Usefulness of bicuspid aortic valve phenotype to predict elastic
  properties of the ascending aorta.
\newblock {\em The American Journal of Cardiology 99}, 5 (2007), 686 -- 690.

\bibitem{2Schaefer1634}
{\sc Schaefer, B.~M., Lewin, M.~B., Stout, K.~K., Gill, E., Prueitt, A., Byers,
  P.~H., and Otto, C.~M.}
\newblock The bicuspid aortic valve: an integrated phenotypic classification of
  leaflet morphology and aortic root shape.
\newblock {\em Heart 94}, 12 (2008), 1634--1638.

\bibitem{schiavone2021vitro}
{\sc Schiavone, N.~K., Elkins, C.~J., McElhinney, D.~B., Eaton, J.~K., and
  Marsden, A.~L.}
\newblock In vitro assessment of right ventricular outflow tract anatomy and
  valve orientation effects on bioprosthetic pulmonary valve hemodynamics.
\newblock {\em Cardiovascular Engineering and Technology\/} (2021), 1--17.

\bibitem{stergiopulos1999use}
{\sc Stergiopulos, N., Segers, P., and Westerhof, N.}
\newblock Use of pulse pressure method for estimating total arterial compliance
  in vivo.
\newblock {\em American Journal of Physiology-Heart and Circulatory Physiology
  276}, 2 (1999), H424--H428.

\bibitem{sullivan2019pyvista}
{\sc Sullivan, C.~B., and Kaszynski, A.}
\newblock {PyVista}: {3D} plotting and mesh analysis through a streamlined
  interface for the {Visualization Toolkit} ({VTK}).
\newblock {\em Journal of Open Source Software 4}, 37 (May 2019), 1450.

\bibitem{valentin2009complementary}
{\sc Valent\'{i}n, A., Cardamone, L., Baek, S., and Humphrey, J.}
\newblock Complementary vasoactivity and matrix remodelling in arterial
  adaptations to altered flow and pressure.
\newblock {\em Journal of The Royal Society Interface 6}, 32 (2009), 293--306.

\bibitem{Verma_nejm_bicuspid_review}
{\sc Verma, S., and Siu, S.~C.}
\newblock Aortic dilatation in patients with bicuspid aortic valve.
\newblock {\em New England Journal of Medicine 370}, 20 (2014), 1920--1929.
\newblock PMID: 24827036.

\bibitem{yap2009dynamic}
{\sc Yap, C.~H., Kim, H.-S., Balachandran, K., Weiler, M., Haj-Ali, R., and
  Yoganathan, A.~P.}
\newblock Dynamic deformation characteristics of porcine aortic valve leaflet
  under normal and hypertensive conditions.
\newblock {\em American Journal of Physiology-Heart and Circulatory Physiology
  298}, 2 (2009), H395--H405.

\bibitem{yellin_book}
{\sc Yellin, E.~L.}
\newblock Dynamics of left ventricular filling.
\newblock In {\em Cardiac Mechanics and Function in the Normal and Diseased
  Heart $\;$}, M.~Hori, H.~Suga, J.~Baan, and E.~L. Yellin, Eds.
  Springer-Verlag, Tokyo, 1989, ch.~C, pp.~225--236.

\bibitem{youssefi2017patient}
{\sc Youssefi, P., Gomez, A., He, T., Anderson, L., Bunce, N., Sharma, R.,
  Figueroa, C.~A., and Jahangiri, M.}
\newblock Patient-specific computational fluid dynamics--assessment of aortic
  hemodynamics in a spectrum of aortic valve pathologies.
\newblock {\em The Journal of thoracic and cardiovascular surgery 153}, 1
  (2017), 8--20.

\end{thebibliography}
\end{small}

\end{document}